\documentclass[12pt]{article}
\usepackage[utf8]{inputenc}
\usepackage{natbib}
\usepackage{graphicx}
\usepackage[export]{adjustbox}
\usepackage{subfigure}
\usepackage{caption}
\usepackage{threeparttable}
\usepackage{tablefootnote}
\usepackage{tabularx}
\usepackage{longtable}
\usepackage{booktabs}
\usepackage{multicol}
\usepackage{multirow}
\usepackage{amsmath}
\usepackage{float}
\usepackage{url}
\usepackage{appendix}
\usepackage{setspace} 
\usepackage{hyperref}
\hypersetup{
   colorlinks = true,
   citecolor = blue,
   linkcolor = blue,
   menucolor = blue,
   filecolor = blue,
   urlcolor  = black,
 }

\usepackage{geometry}
\begin{document}

\title{Seeing Stereotypes
\thanks{
We are deeply grateful to Chiara Ferrando and Marco Madeddu for the expert annotation.
We also thank 
Sule Alan,
Ainoa Aparicio Fenoll,
Michele Belot,
Steven Bosworth,
Deborah Cobb Clark,
Dalit Contini,
Norma De Piccoli, 
Maria Laura Di Tommaso,
Brian Jacobs,
Steven Lee,
Lindsay Macmillan, 
Sandra McNally, 
Rigissa Megalokonomou, 
Viviana Patti,
Daniela Piazzalunga.
Almudena Sevilla,
for their valuable comments and suggestions. Additionally, we appreciate the feedback from participants at the following workshop and seminars:
HERB 2024, IWAEE 2024, CEP Education at LSE, University of Turin, Birmingham, Nanterre, BHEPPE workshop, Tempo Workshop Montepulciano, and LISER Seminar.
All remaining errors are probably Biroli's fault.
} 
} 
\author{
    \begin{tabular}{c@{\hspace{2.5cm}}c}
        Elisa Baldazzi\thanks{University of Bologna. Email: \texttt{elisa.baldazzi3@unibo.it}} &
        Pietro Biroli\thanks{Corresponding author. University of Bologna. Email: \texttt{pietro.biroli@unibo.it}} \\
        Marina Della Giusta\thanks{University of Torino and Collegio Carlo Alberto. Email: \texttt{marina.dellagiusta@unito.it}} & 
        Florent Dubois\thanks{University of Torino and University Paris-Nanterre, EconomiX. Email: \texttt{florentmaximeguillaume.dubois@unito.it}} \\
    \end{tabular}
}

\date{\today}

\maketitle

\begin{abstract}
\singlespace
Reliance on stereotypes is a persistent feature of human decision-making and has been extensively documented in educational settings, where it can shape students' confidence, performance, and long-term human capital accumulation. 
While effective techniques exist to mitigate these negative effects, a crucial first step is to establish whether teachers can recognize stereotypes in their professional environment. 
We introduce the Stereotype Identification Test (SIT), a novel survey tool that asks teachers to evaluate and comment on the presence of stereotypes in images randomly drawn from school textbooks. Their responses are systematically linked to established measures of implicit bias (Implicit Association Test, IAT) and explicit bias (survey scales on teaching stereotypes and social values). 
Our findings demonstrate that the SIT is a valid and reliable measure of stereotype recognition. 
Teachers' ability to recognize stereotypes is linked to trainable traits such as implicit bias awareness and inclusive teaching practices. Moreover, providing personalized feedback on implicit bias improves SIT scores by 0.25 standard deviations, reinforcing the idea that stereotype recognition is malleable and can be enhanced through targeted interventions.


\textit{JEL codes:} I24, J16, J24
\end{abstract}

\newpage
\section{Introduction}\label{sec:intro}
As a social species, humans are wired to automatically generate expectations of other people's behaviours on the basis of their observable traits \citep{lieberman_social_2013}. 
Stereotypes function as mental shortcuts that facilitate rapid judgments in the absence of accurate information \citep{tversky_availability_1973}. 
They lead individuals to assume that a person will behave similarly to the broader group they are perceived to represent, thereby shaping beliefs about individuals based on group-level expectations \citep{kahneman_new_1994}.

Stereotyping and bias have been extensively documented in education, where teachers' beliefs and actions can significantly influence students' cognitive development (e.g., academic achievement), socio-emotional growth (e.g., behavior, character, self-concept, engagement), and their academic and career expectations. 
Teachers shape students’ learning experiences both directly---through instruction, feedback, monitoring, and assessment---and indirectly, by shaping learning environments and selecting teaching materials and books, in which gender and racial stereotypes are often embedded \citep{gorard_cautionary_2016, morris_girls_2017, blumberg2015Eliminating, adukia_what_2023}.
Various strategies can mitigate the effect of stereotypes on decision-making; however, the first and most challenging step is recognizing their presence all around us. How well do teachers fare in this respect?

This paper addresses this essential step by introducing the Stereotype Identification Test (SIT), a novel tool designed to assess teachers' ability to detect stereotypes in educational settings.
The SIT leverages images drawn from school materials to evaluate teachers’ ability to recognize bias in visual representations, following a recent literature in economics that leverages visual cues to understand ideology and culture \citep{caprini_visual_2023,Voth2024Images}. 
We evaluate the validity and reliability of SIT and examine how it relates to teachers’ understanding of bias and other malleable skills that can be enhanced through training.

Our findings show that SIT scores are systematically related to teachable skills such as implicit bias awareness and inclusive teaching practices, as well as attitudes such as growth mindset, beliefs about gender and STEM, locus of control, and social values. 
This suggests that the ability to detect stereotypes, as measured by the SIT, is meaningfully connected to broader cognitive and behavioral constructs that influence decision-making and interactions in educational settings.

Next, we examine whether teachers' ability to recognize stereotypes in images is malleable and responsive to intervention. 
We find that providing personalized feedback on implicit bias--measured by the Implicit Association Test (IAT), which captures differences in reaction time for stereotypical versus counter-stereotypical categorizations--significantly improves SIT scores by a quarter of standard deviation.
However, SIT remains stable despite minor framing variations introduced at the beginning of the survey, suggesting its robustness as a measure of stereotype detection.

By establishing a reliable measure of stereotype detection and demonstrating its malleability, this study provides a foundation for future interventions aimed at reducing bias in educational settings and beyond.

The remainder of the paper is organized as follows. 
Section \ref{sec:stereotypes} provides an overview of stereotypes, their measurement, and their documented impact on decision-making. 
Section \ref{sec:SIT} introduces the Stereotype Identification Test (SIT). 
Section \ref{sec:survey} details the data collection process, while Section \ref{sec:predict} examines how SIT scores relate to individual traits and the effectiveness of our information intervention. 
In Section \ref{sec:reliability_validity}, we assess the reliability and validity of the SIT measure. 
Finally, Section \ref{sec:conclusion} summarizes our findings and discusses broader implications.

\section{Stereotypes}\label{sec:stereotypes}
Stereotypes differ from explicitly held prejudices in that we are not typically conscious of using them in forming our beliefs and are thus a form of implicit belief that we are not necessarily aware of. Stereotypes are operating for instance if upon seeing that men were over-represented in the top tail of the mathematics GPA scores distribution we would form a belief that this also holds true across the whole distribution of scores, making men always better at mathematics than women  \citep{tversky_extensional_1983}.
Given the vast amount of decisions we make in everyday life, a lot of our interactions are indeed driven by simple heuristics that often remain unverified by deliberate reasoning giving rise to biased beliefs \citep{kahneman_thinking_2011}. These are more likely to occur when making decisions under time pressure or when the motivations to be accurate are reduced: for instance, the social perceptions of individuals occupying positions of higher power in social hierarchies, like teachers in a classroom, are often less accurate than those lower in the hierarchy \citep{fiske1993social} who typically allocate more time and energy to social judgment. 

The presence and effect of stereotypes including self stereotypes in shaping beliefs and decisions has been investigated formally and documented empirically in several contexts in the economic literature on the allocation of talent in education and the labor market, voters behavior, and many other domains \citep{gennaioli2023identity, fryer2019updating, coffman2014evidence, bordalo_stereotypes_2016, oxoby2014social}.

The experimental literature has also documented how exposure to negative stereotypes affects effort, self-confidence, and productivity \citep{Carlana2019, bordalo_stereotypes_2016, glover_discrimination_2017}. 
Aspirations are strongly correlated to expectations \citep{la_ferrara_presidential_2019,carlana_goals_2022}, and expectations have been shown to affect performance.
This is particularly important for the case of teachers: optimistic teachers’ expectations have been found to particularly benefit the achievement of students from minorities in the US \citep{jussim_teacher_2005}. More gender egalitarian teachers have been found to increase the performance and uptake of STEM by girls \citep{Alan2018a, Carlana2019, Ash03072024, hawkins_understanding_2023}, and generally to be able to increase the performance of pupils through positive expectations of them \citep{figlio_names_2005, sprietsma_discrimination_2013, campbell_stereotyped_2015, hanna_discrimination_2012}. Research has also shown that teachers’ diminished expectations of children with names associated with low socio-economic status affect student’s cognitive performance \citep{figlio_names_2005}, that essays designated with either German or Turkish names were differently graded in schools in Germany \citep{sprietsma_discrimination_2013}, and that the assessment of children’s behaviour was rated as more disruptive and inattentive by teachers from a different ethnic group \citep{dee_teacher_2005, gilliam_early_2016, blank_unconscious_2016}. \cite{alan_social_2023} have shown the negative effects on children of teachers' ethnic prejudice with teachers who hold prejudicial attitudes creating both socially and spatially segregated classrooms. The consequences are real and long lasting: negative expectations of children are related to educational outcomes independently of previous attainment and parental and other characteristics \citep{jacob_educational_2010}; the consequences of teachers' gender bias persist later on influencing university admissions exams, choice of university field of study, and quality of the enrolled program \citep{Lavy2024}.
\subsection{Measuring Stereotypes}\label{sec:measure}
The use of stereotypes in decision-making are usually measured leveraging techniques that rely on different automatic, and therefore implicit, responses to cues that can be presented in a variety of ways \cite{fazio_automatic_1986,fazio_variability_1995}. 
In sequential priming, for example, bias is measured through the performance in a task such as classifying adjectives into good and bad or by classifying letter strings as words or non-words with or without exposure to social cues. 
Most implicit measures of bias used today follow the same basic strategy: for example, the Affect Misattribution Procedure \citep{payne_affect_2014} relies on visual cues priming emotions before an exercise of evaluating the relative pleasantness of images related to different cultures. The test has been found to predict behaviours and behavioural intentions, including alcohol consumption \citep{friese_control_2009, payne_automatic_2008}, moral decisions \citep{hofmann_immediate_2010}, and behaviours related to health anxiety \citep{jasper_automatic_2013} as well as deliberate behaviours like votes in the 2008 US presidential election \citep{lundberg_decisions_2014, payne_implicit_2010}.
The most widely used test for implicit bias is the Implicit Association Test (IAT).\footnote{\url{https://implicit.harvard.edu/implicit/}} 
The test typically requires assigning words to categories following both stereotypical (congruent) and non-stereotypical (incongruent) associations and measuring the speed with which such associations are made. For example, to test for the presence of a gender and STEM stereotype, the test would provide a measure of the difference in time taken when assigning words like mathematical competence to the female category as opposed to the male category, and thus label bias the difference (if any) between the time taken making a counter-stereotypical (incongruent) vs a stereotypical (congruent) word assignment. Then, the difference in reaction time is standardized, allowing for comparisons between individuals, in the following way: 
\begin{equation}
    IAT_i = \dfrac{\overline{\text{Incongruent reaction time}_i} - \overline{\text{Congruent reaction time}_i}}{\sqrt{\sigma^{2}_{incongruent, i}+\sigma^{2}_{congruent, i}}} \label{eq:IAT}
\end{equation} 
where $\sigma^{2}_{incongruent, i}$ and $\sigma^{2}_{congruent, i}$ are the variance of the incongruent and congruent series for the individual \textit{i} respectively. The test has been performed across a variety of professional contexts to reveal bias in professionals and its correlation with their evaluations of co-workers \citep{reuben_how_2014} or, in the case of doctors, patients and their access to treatment \citep{oxtoby_how_2020}. Results of the test in teachers have been linked to pupil’s performance and self-confidence in math ability \citep{glover_discrimination_2017, Carlana2019} and the revelation of bias to teachers has been shown to lead to subsequent moderating behaviours \citep{alesina_revealing_2018}.
Bias revelation is however also contentious \citep{banks_how_2009} as psychologists worry about the possible negative reactions it can elicit \citep{howell_caught_2015}, most importantly through increasing moral licensing, the process of behaving morally at first, but later being more likely to display immoral behaviours \citep{mazar_green_2010, merritt_strategic_2012, cascio_prospective_2015}. 
In the case of revelation of bias, moral licensing might lead to socially desirable actions at first, but then reversion to previous biased beliefs and even backlash in subsequent longer-term behaviours. 
While the moral licensing effect has been replicated in many studies \citep{blanken_meta-analytic_2015, simbrunner_moral_2017}, \cite{conway_when_2012} point out that the moral licensing effect is at odds with results showing that people strive for consistency with past behaviour. \cite{mullen_consistency_2016} argue that information is likely to elicit consistency when it is more abstract and temporally removed, when it is more relevant to the individual’s identity, and when the individual’s motives underlying their initial behaviour are ambiguous. \cite{della_giusta_bias_2020} have tested experimentally whether bias revelation in the context of the gender and STEM stereotype gives rise to subsequent corrective behaviours and also to further licensing and found that revealing gender bias does not lead to corrective behaviour by male students, but it does on average lead to correction and thereafter to a larger gender biased choice by female students. The experiment also suggests that the effects are different depending on the initial level of bias of the subjects as well as their gender. 

All these tests reveal the recourse to stereotypes in decision-making, however they do not indicate whether people are able to identify stereotypes in their environment. 
We fill this gap by introducing a novel testing procedure.

\section{Stereotype Identification Test}
\label{sec:SIT}
The Stereotype Identification Test (SIT) consists of rating and commenting on images that might appear in one's working environment. 
We develop it specifically for the educational context to measure teachers' responses to sequences of stereotypical images that are likely to be found in educational materials. 
The images are drawn by an educational editor from an educational publisher catalogue who granted us permission for their use, and access to their mailing list of teachers to run the experiment. 
It elicits both a rating of the extent to which respondents find an image stereotypical and a written comment of the reasons why they think it is so. 

\begin{figure}[H]
\caption{Example of task asked to the respondents.\label{fig:example}}
\centering
\subfigure{\includegraphics[scale=0.212]{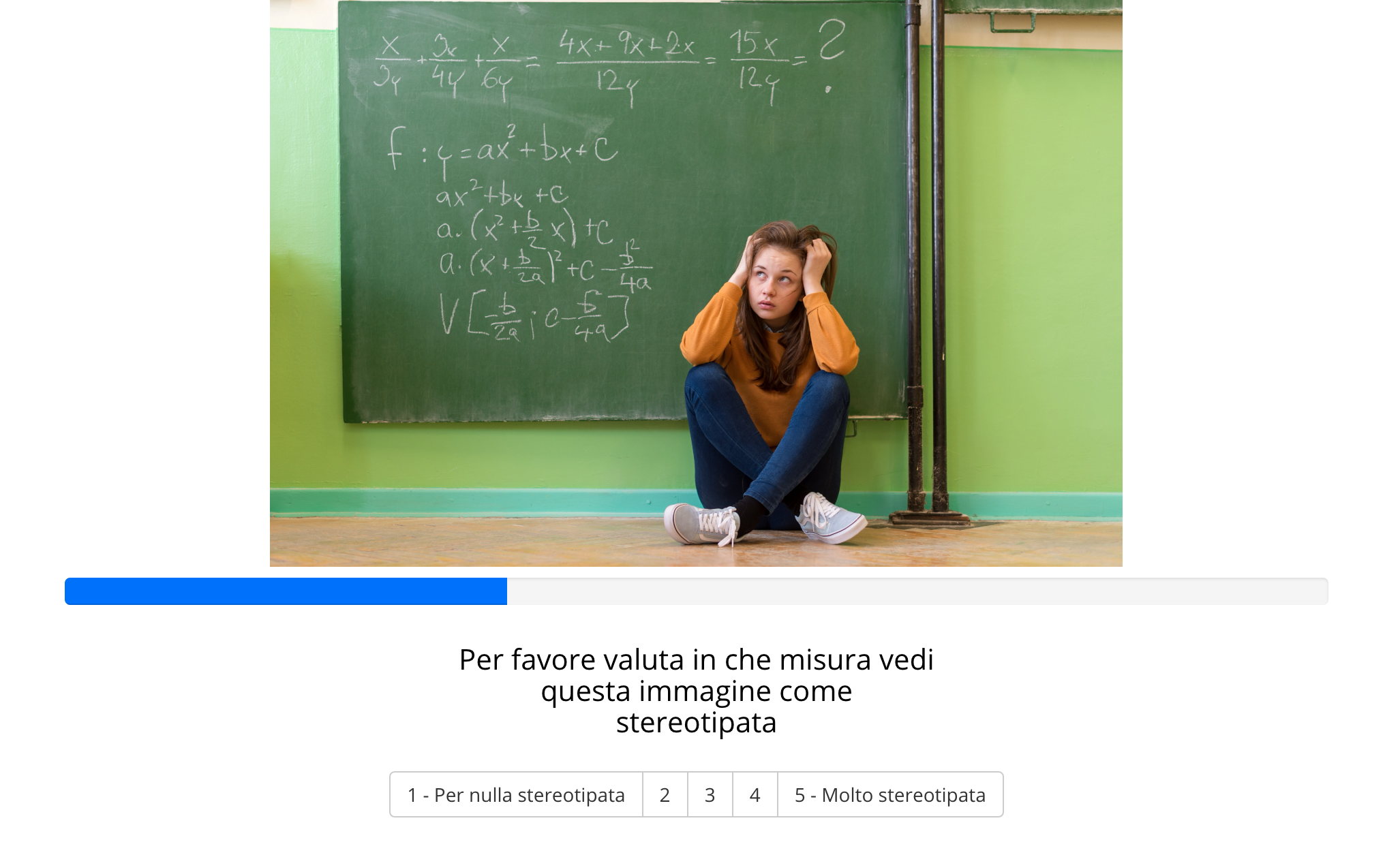}}
\hfill
\subfigure{\includegraphics[scale=0.11]{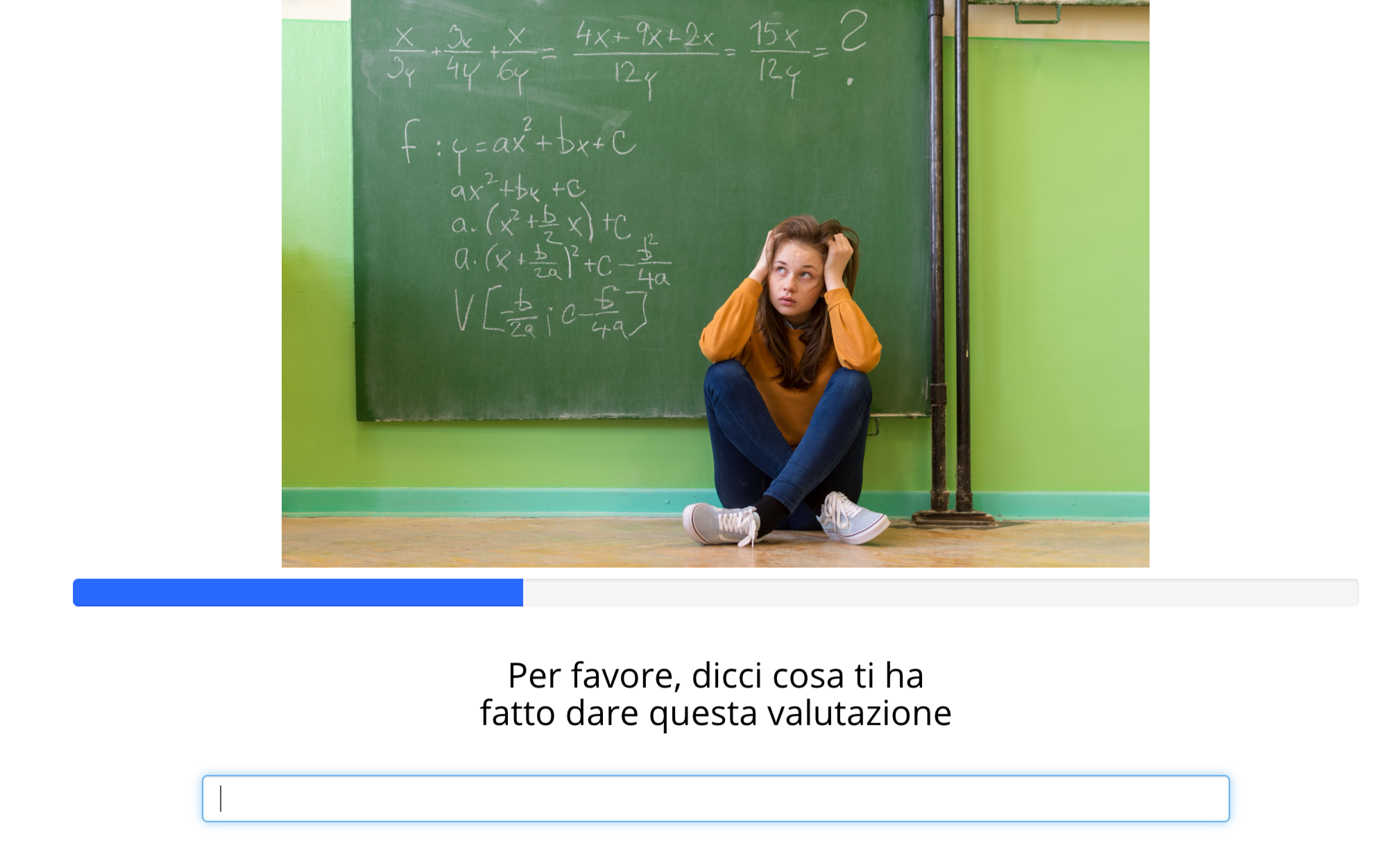}}
\caption*{\footnotesize Example of one SIT prompt. On the left, an example of the rating section. Under the image there was the following text ``Please evaluate the extent to which you see this image as stereotyped'', and then they were asked to give a rate on a scale from 1 (not stereotypical) to 5 (very stereotypical), without the possibility to continue until a rate was given. On the right, an example of the comment section. With the picture still visible, the respondents were asked ``Please tell us what made you give this rate''.} 
\end{figure}

We ask teachers to rate 20 images randomly drawn from a set of 100 that may represent stereotypes of various kinds. 
To ensure comparability with IAT results, six pictures containing representations of a specific stereotype related to Gender and STEM are always administered and rated by all the respondents (see \autoref{sec:gen-stem} for these six pictures); the remaining fourteen are drawn at random and depict other kinds of stereotypes related to ethnicity, ability, and age.
Respondents are asked to rate each picture on a scale from 1 (not stereotypical) to 5 (very stereotypical), as shown in \autoref{fig:example}.

To obtain the score of our test we computed the average by individuals of the differences between the rating and the leave-one-out mean rating of the image: 
        \begin{equation}
            \operatorname{\widetilde{SIT}_i} = \frac{1}{J}\sum_{j=1}^{J}\bigg(\operatorname{rating}{_{ij}} - \frac{1}{N_j-1} \sum_{k=1, k \neq i}^{N_j} \operatorname{rating}{_{kj}}\bigg) \label{eq:SIT}
        \end{equation}
Where \textit{i} and \textit{k} indicate the individual, \textit{j} the image rated and \textit{J} the total number of images rated by individuals, $N_j$ the maximum number of ratings the image received (since the selection of the images was randomized, thus each image has been rated a different amount of times).
We computed one score considering all the images rated by an individual (SIT Score) and a score including only the six Gender-STEM images rated by all the respondents (Gender SIT Score). In the absence of a natural scale for the score, both of them standardized to have zero mean and standard deviation equal one in our sample: $SIT_i = \frac{\widetilde{SIT}_i - \mu_{\widetilde{SIT}}}{\sigma_{\widetilde{SIT}}}$.
Given that we exploit the standardized differences, scores greater than zero indicate that, relative to the average teacher, the individual assigned on average higher ratings to the pictures during the SIT. This implies that the individual may be more adept at identifying stereotypes within the images. Conversely, negative scores suggest that, on average, the individual is less proficient at identifying stereotypes in the images compared to the average teacher (gave lower ratings, on average). 

Respondents are then asked to leave a comment in an open-field box to motivate their rating. 
We do not allow respondents to change the picture rating once they have entered the commentary section. 
For both ratings and comments, we record the time taken to respond. 
We expect that, while the rating of the image may rely more on fast and implicit associations, the written commentary requires activating a more reflective and deliberative mode of thinking. 

\section{Teacher Survey}
\label{sec:survey}

\begin{figure}
    \centering
    \caption{Survey flow diagram. }
    \includegraphics[width=1\linewidth]{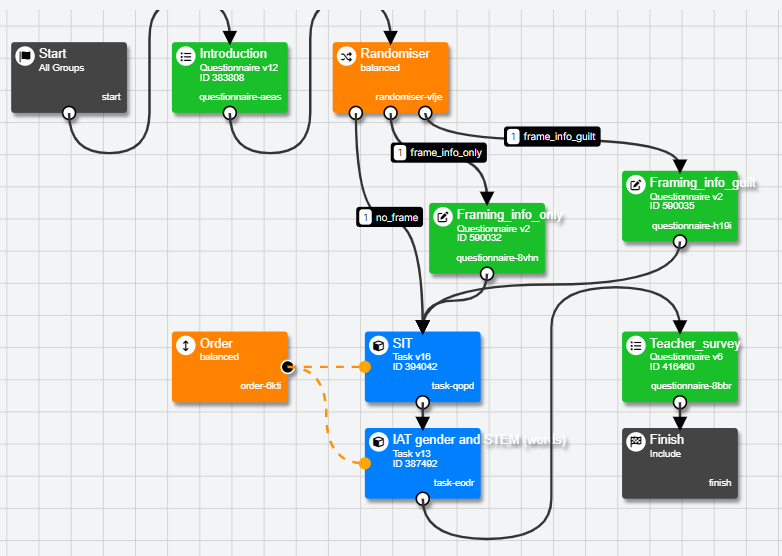}
    \caption*{\footnotesize Arrows represent the flow of the experiment. They describe how participants are moving from one task to another depending on their treatment assignment. Black circles identify the starting point and the related black arrows show the next task. In orange is identified the step of randomization; in blue the step of the tests (SIT and IAT); in dark grey the checkpoints; in green the questionnaire areas.}
    \label{fig:flow}
\end{figure}

The survey structure is outlined in \autoref{fig:flow}. 
Teachers were first randomly assigned to see a brief introductory framing paragraph. They then completed the Stereotype Identification Test (SIT) and the Implicit Association Test (IAT) on Gender-STEM stereotypes, with the order of these two tasks randomized. 
Finally, they answered a questionnaire collecting sociodemographic information and validated measures of six key psychological and behavioral dimensions that may influence their ability to recognize stereotypes.

The questionnaire included items on inclusive teaching practices, assessing the extent to which teachers implement inclusive pedagogical strategies \citep{ewing2018teachers}. 
It also measured Implicit Bias Awareness, capturing teachers’ perceptions of the role of bias in education, their engagement with inequality-related topics, and their confidence in addressing them.
Another section focused on Gender-STEM stereotypes in teaching, evaluating teachers' beliefs about gender differences in academic abilities, learning styles, and discipline, particularly in STEM and humanities subjects \citep{DellaGiusta2022}. 
Additionally, the survey assessed locus of control, referring to the extent to which teachers feel confident in their ability to influence student outcomes \citep{cobb2013two}, as well as growth mindset, which captures beliefs about intelligence as a malleable rather than fixed trait \citep{claro2016growth}. 
Finally, teachers’ social values were measured using validated items from the European Social Survey \citep{Davidov2008}.

All these scales are self-reported using a five-point Likert scale ranging from strongly disagree to strongly agree. To summarize these dimensions, we construct an index using factor analysis.

To check the stability of the response to slight perturbations in the environmental setting, we also conduct a small framing experiment embedded within the survey. As illustrated in the survey flow diagram, this stage is depicted as the preliminary step, occurring prior to any testing or questionnaire administration.
Teachers were randomly shown three different introductory texts selected to evoke a different emotional response to the concept of stereotypes in a school setting. 
The first group was provided with simple and factual information about stereotypes: \textit{``Stereotypes are cognitive shortcuts used by the brain to generate expectations about one's own or others' behaviour. Everyone has them and they don’t always know it.''} We call this the ``info'' group.
The message for the second group was more charged with negative emotions in the school setting, and stated: \textit{``Many studies show that school environments suffer from stereotypes of various kinds. Being exposed to negative stereotypes about one's group has an effect on self-confidence and performance.''} We call this the ``info+guilt" group.
The third group was shown no message. This is the ``no frame" group.

\subsection{Data Collection}\label{sec:data}

We sent our Stereotype Identification Test (SIT) survey to a mailing list of teachers curated by a major educational publisher company in Italy. 
The test data was collected online in two months (from April 16th to June 15th) in 2023.
\footnote{Every member of this mailing list received the following message ``The University of Turin is conducting a research project on the school environment throughout Italy. We are collecting information from people who are experts in the field and who work in direct contact with students. We would appreciate it if you could give us some of your time (about 20 minutes) for this anonymous survey, which we ask you to complete in one session once you start.''}
No incentive was given.
After receiving the email, 1,636 individuals initiated the test, with 645 of them successfully completing it, and 614 valid responses \footnote{We excluded 9 responses who report being 18 years-old; 22 teachers whose response time to the Implicit Association Test was outside of the standard bounds applied for reliability \citep{Carlana2019}.} indicating an approximate completion rate of 40\%. 

\begin{figure}[H]
\caption{\label{fig:place}}
\centering
\subfigure{\includegraphics[scale=0.38]{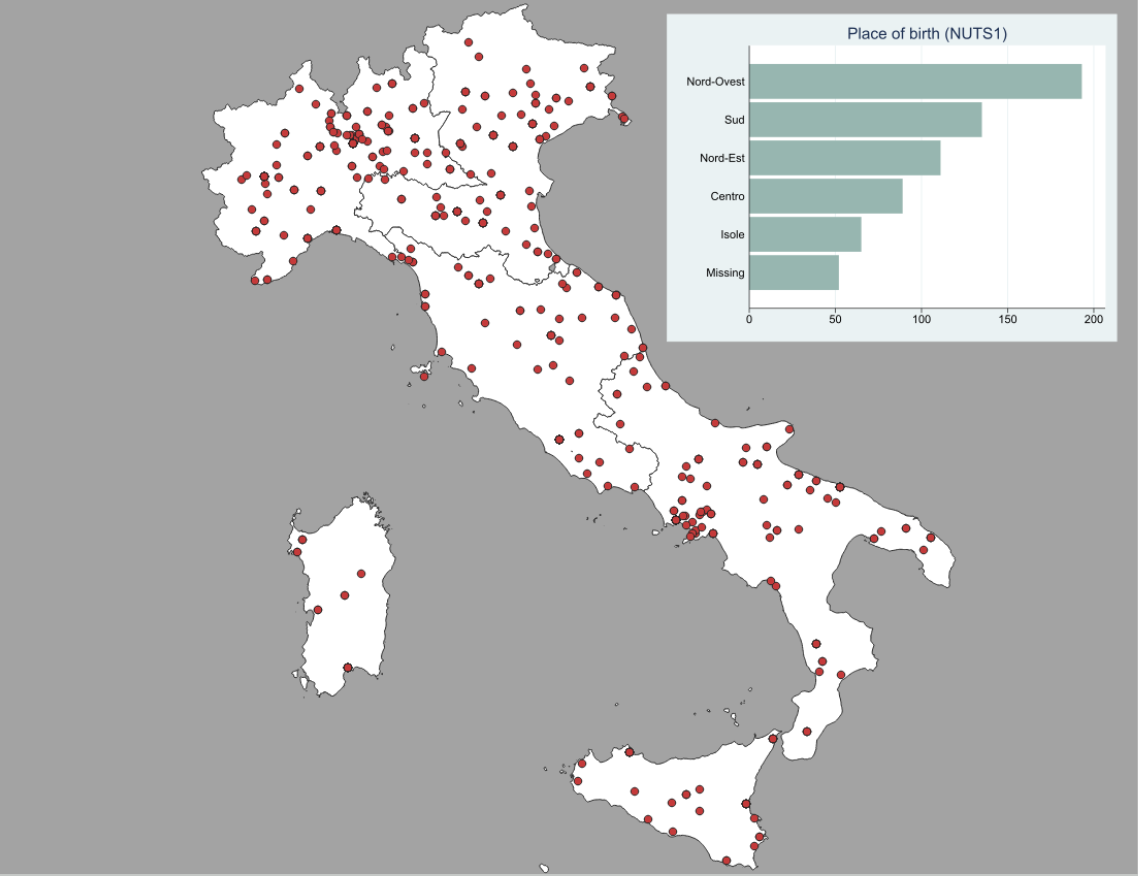}}
\hfill
\subfigure{\includegraphics[scale=0.38]{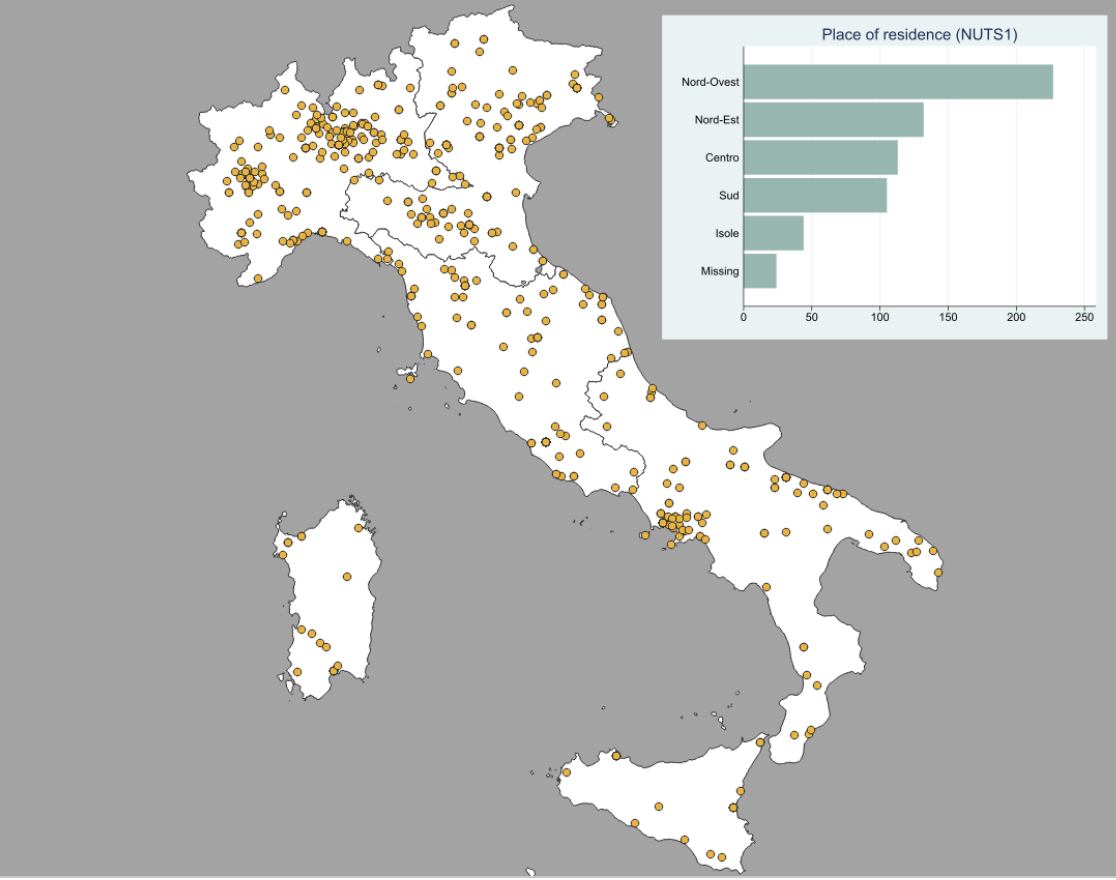}}
\caption*{\footnotesize Map of Italy showing the place of birth (left, red-dots) and place of residence (right, yellow-dots) of each teacher in our sample, together with the distribution broken down by Italian macro areas (North-West, North-East, Center, South and Islands) and the frequency of missing values.}
\end{figure}

Although our respondent sample was self-selected, we were able to collect information from a wide array of demographic backgrounds that are comparable to the overall distribution of Italian teachers, as recorded by the Italian Ministry of Education.\footnote{\url{https://dati.istruzione.it/opendata/}}
Overall, our sample is slightly older and more male than the national level.
Specifically, 84.37\% of teachers in our sample are women, while 96.32\% primary school teachers with a permanent contract are women at the national level.
Considering the age distribution, as shown in  Appendix \autoref{table:agemiur}, 299 respondents in our sample are older than 54 (which is almost 48\%), versus 40.12\% at the national level; 
30.46\% are between 45 and 54 years of age (37.05\% national); 
14.04\% between 35 and 44 years old (19.01\% national);
and 7.81\% are between 18 to 35 years old  (3.82\% national).
Regarding their family status, 66.03\% of our sample respondents are married, and 61.24\% have at least one child. 
In terms of education, 80.06\% of our sample respondents have obtained at least a master's degree, and 93.46\% stated that they like to teach.\footnote{They answered 5 or more on a scale from 1 to 7, where 1 means ``not at all'' and 7 ``completely'' to the question ``How much do you like to teach?''} 
Finally, our sample was also quite geographically widespread and representative in both provinces of birth and residence, as shown in \autoref{fig:place}.
In total, we collected 12,901 ratings and 10,747 comments. Providing comments is optional, but 609 people (97\% of the respondents) commented on at least one picture. The average length of the comments is equal to 58 characters, including spaces, with a minimum of one character and a maximum of 1012 characters.\footnote{Only 108 comments present a unique character and they were left by 15 individuals.}  

\begin{table}[htbp]\centering
{
\def\sym#1{\ifmmode^{#1}\else\(^{#1}\)\fi}
\caption{Summary Statistics}
\begin{tabular}{l*{1}{cccc}}
\toprule
                    &\multicolumn{4}{c}{}                               \\
                    &        mean&          sd&         min&         max\\
\midrule
Gender              &       0.845&       0.362&       0.000&       1.000\\
Age                 &      51.831&       9.536&      25.000&      69.000\\
Like Teaching       &       6.287&       0.945&       1.000&       7.000\\
Master              &       0.813&       0.390&       0.000&       1.000\\
Disability training &       0.235&       0.424&       0.000&       1.000\\
Married             &       0.666&       0.472&       0.000&       1.000\\
Teaching Italian    &       0.427&       0.495&       0.000&       1.000\\
Teaching Maths      &       0.176&       0.381&       0.000&       1.000\\
Centro      &       0.143&       0.351&       0.000&       1.000\\
Isole       &       0.099&       0.299&       0.000&       1.000\\
Missing     &       0.078&       0.269&       0.000&       1.000\\
Nord-Est    &       0.179&       0.384&       0.000&       1.000\\
Nord-Ovest  &       0.301&       0.459&       0.000&       1.000\\
Sud         &       0.199&       0.399&       0.000&       1.000\\
Growth Mindset      &      -0.012&       0.965&      -2.681&       1.147\\
Implicit Bias Awareness&       0.003&       0.880&      -3.462&       1.009\\
Gender-STEM Stereotypes&       0.002&       0.951&      -3.618&       0.952\\
Locus of Control    &      -0.002&       0.795&      -2.570&       1.977\\
Social Values       &       0.015&       0.838&      -3.896&       1.433\\
Inclusive Teaching  &      -0.001&       0.801&      -3.440&       0.861\\
\midrule
Observations        &         614&            &            &            \\
\bottomrule
\end{tabular}
}

\end{table}
\subsection{Ratings}
Analyzing teacher's ratings we identify four defying features: 
individual image ratings tend to be polarized; 
teachers take longer to rate less stereotypical images; 
the overall SIT score shows enough variation and no evidence of floor or ceiling effects;
SIT scores are meaningfully correlated with individual measures of implicit (IAT) and explicit bias (values and beliefs).

\begin{figure}[H]
\caption{Distribution of ratings (ind., whole sample)\label{fig:distribution}}
    \centering
\includegraphics[scale=0.45]{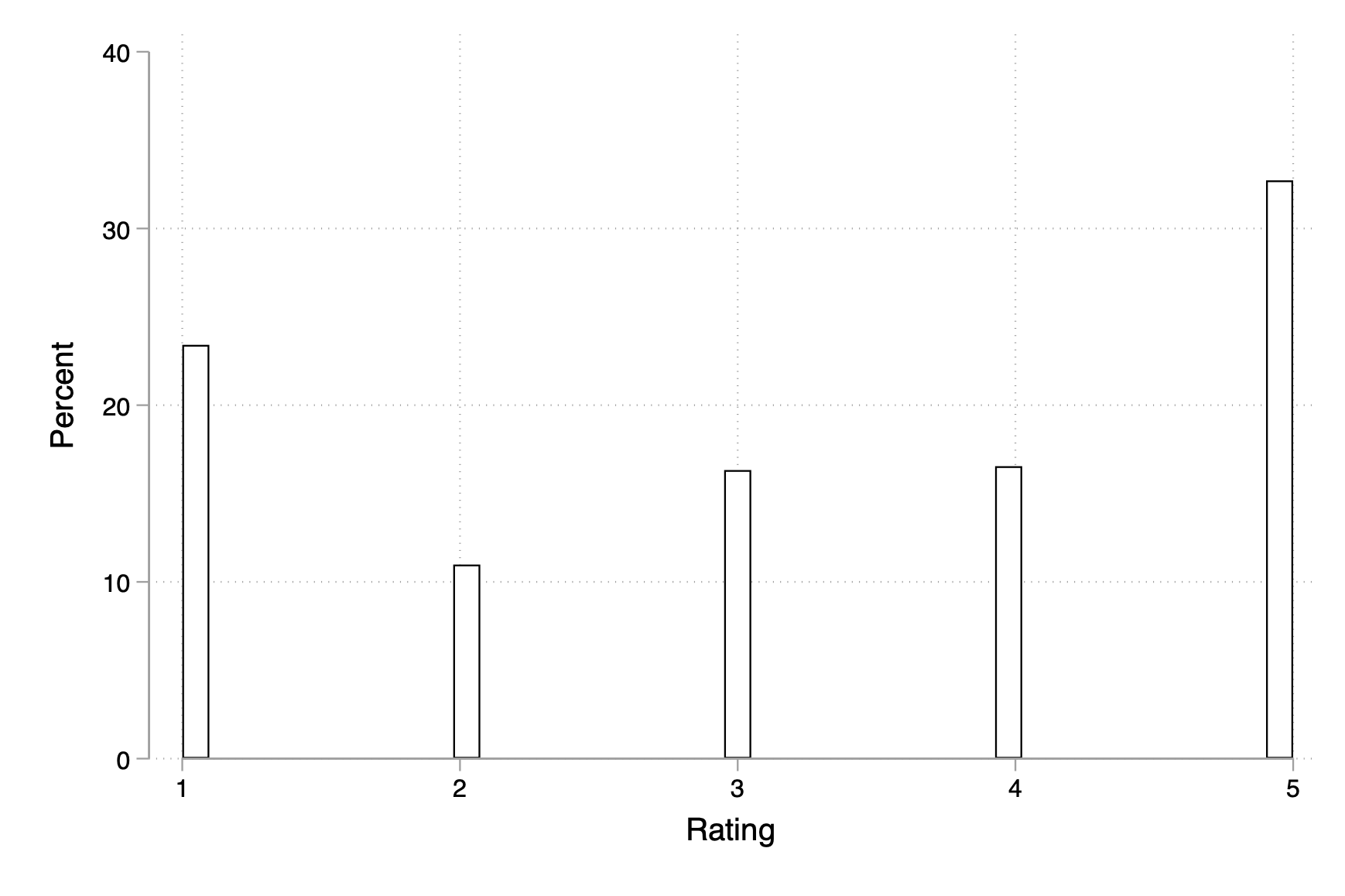}

\caption*{\footnotesize Distribution of all the 12,901 ratings collected in our sample of 614 teachers. 5 means a highly stereotypical image (``Molto stereotipata''); 1 means that the image is not stereotypical at all ``Per nulla stereotipata''.}
\end{figure}

Teacher's ratings are polarized, a result consistent with the literature on online ratings \citep{rating2020}, with more than half of respondents (56,00\%) reporting either a 1 (not stereotypical) or a 5 (very stereotypical), as shown in \autoref{fig:distribution}.
Almost one third of the images (32,68\%) are rated with a 5, while almost one fourth of the ratings (23,32\%) are equal to 1.

\begin{figure}[H]
\caption{Rating times and ratings of images\label{fig:classification}}
    \centering
\includegraphics[scale=0.9]{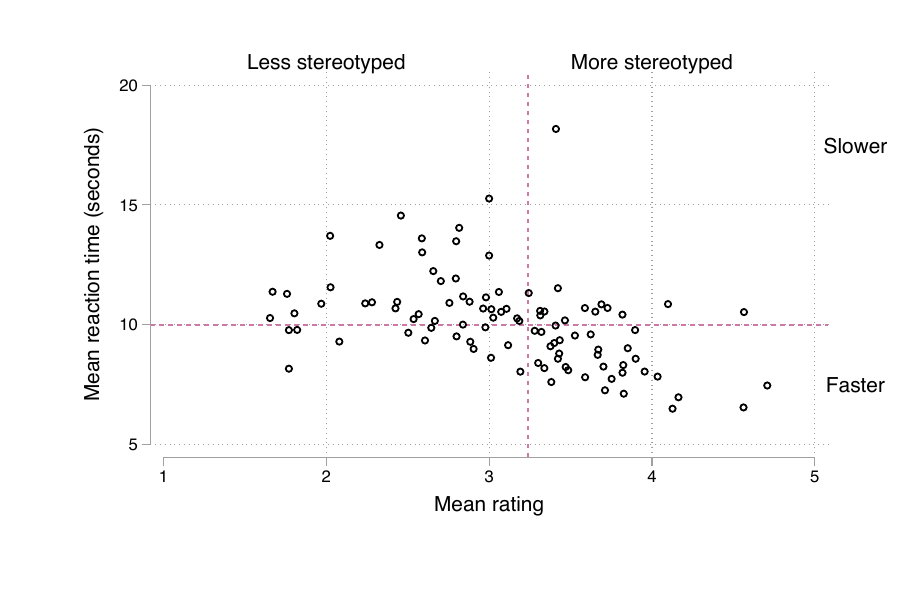} 
\caption*{\footnotesize This graph illustrates the relationship between perceived rating (on a scale from 1, not at all stereotypical, to 5, very stereotypical) and response time, in seconds. Each point represents the average for an image, with the x-axis showing its mean rating and the y-axis indicating the mean time taken to rate it.} 
\end{figure}

Respondents tend to take less time to rate pictures that they perceive as containing more stereotypes, as shown in \autoref{fig:classification}. 
As they are explicitly primed with finding stereotypes, they might be spending more time looking for them in pictures which they eventually rate as less stereotypical.\footnote{We thank Ainoa Aparicio for providing this insight related to the ``Where's Wally'' effect.}

\begin{figure}[H]
\caption{Distribution of the Stereotype Identification Test \label{fig:SIT distribution}}
\centering
\subfigure{\includegraphics[scale=0.25]{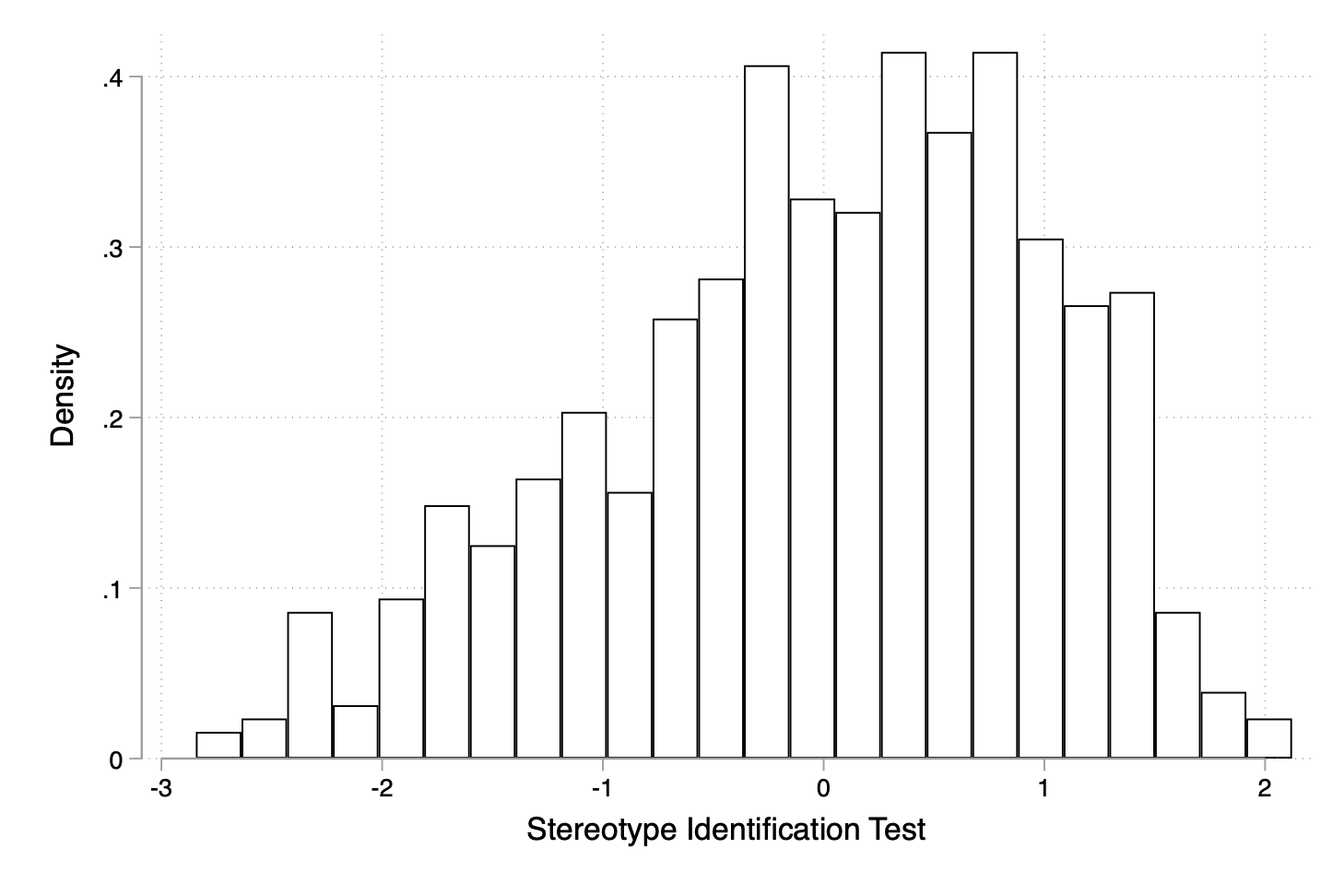}}
\hfill
\subfigure{\includegraphics[scale=0.25]{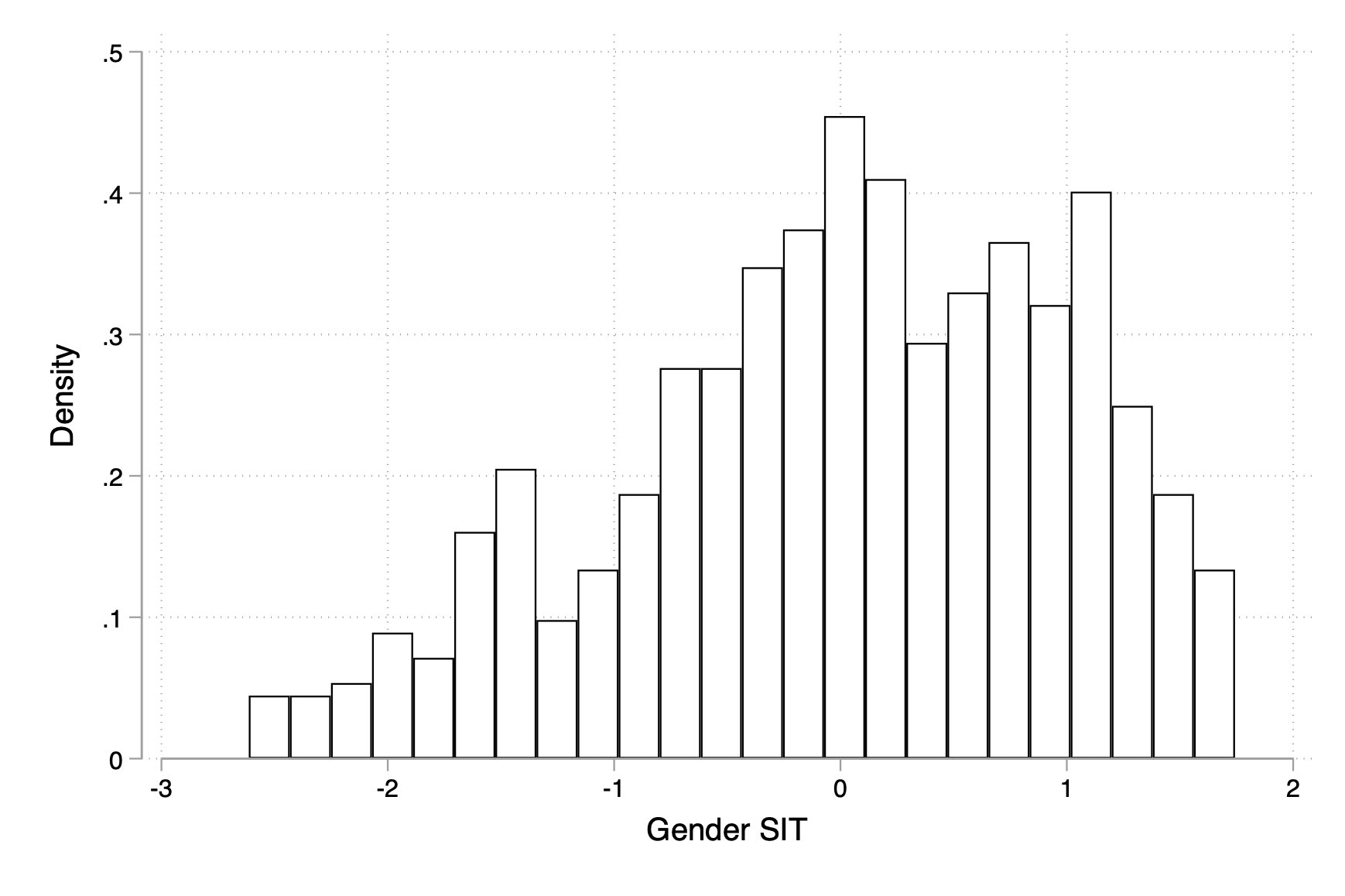}}
\caption*{\footnotesize Distribution of the SIT score and of the Gender SIT score constructed following equation \ref{eq:SIT}. Positive scores indicate a tendency for individuals to rate the images as more biased compared to the average teacher, while negative scores indicate the opposite.}
\end{figure}

In \autoref{fig:SIT distribution} we depict the distributions of the SIT and Gender SIT in our sample.
The left panel shows the distribution of rating differences from the average teacher across the 20 images, which is left-skewed, indicating that only a minority of individuals deviate substantially in a negative direction from this reference point. 
The right panel depicts the same distribution for the Gender SIT scores, showing a similar pattern with left-skewness and peaks at positive values.
Importantly, despite the skewness, there is no excessive bunching at the top or bottom of the distribution, suggesting that responses are well-distributed without strong ceiling or floor effects. 
This indicates that the SIT provides meaningful variation across individuals and can effectively capture differences in stereotype recognition ability.

\begin{figure}[H]
\caption{SIT and  IAT \label{fig:corr_iat}}
    \centering
\includegraphics[scale=0.23]{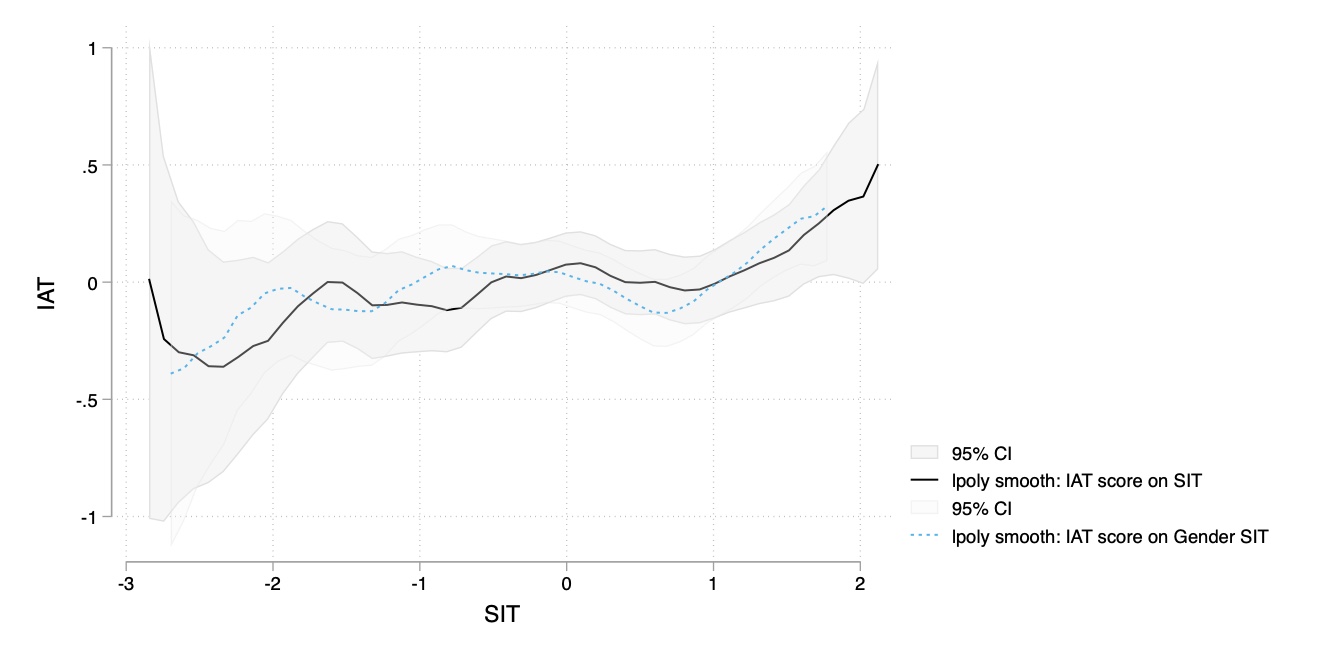}
\caption*{\footnotesize Local polynomial smooth of IAT score on  SIT (the dark line) with the respective confidence interval (95\%) and the local polynomial smooth of the IAT score on the SIT for the six Gender-STEM images (the blue dashed line) with the respective confidence interval (95\%).}
\end{figure}

We expect SIT to be correlated with implicit biases to the extent that these are stereotypical views held but not manifested, and restricting to the set of SIT ratings pertaining to gender STEM stereotypes we can compare SIT to IAT results over the same stereotype.  This relationship is illustrated in \autoref{fig:corr_iat}, where we present a non-linear association of the IAT score against both the overall SIT score (solid black line) and the Gender SIT score (dashed blue line). The results indicate a slight positive trend, suggesting a weak but positive association between SIT scores and the Implicit Association Test (IAT) scores.

\begin{figure}[ht]
\caption{\label{fig:combined_corr}}
    \centering
    \subfigure[Social Values Index]{%
        \includegraphics[width=0.48\linewidth]{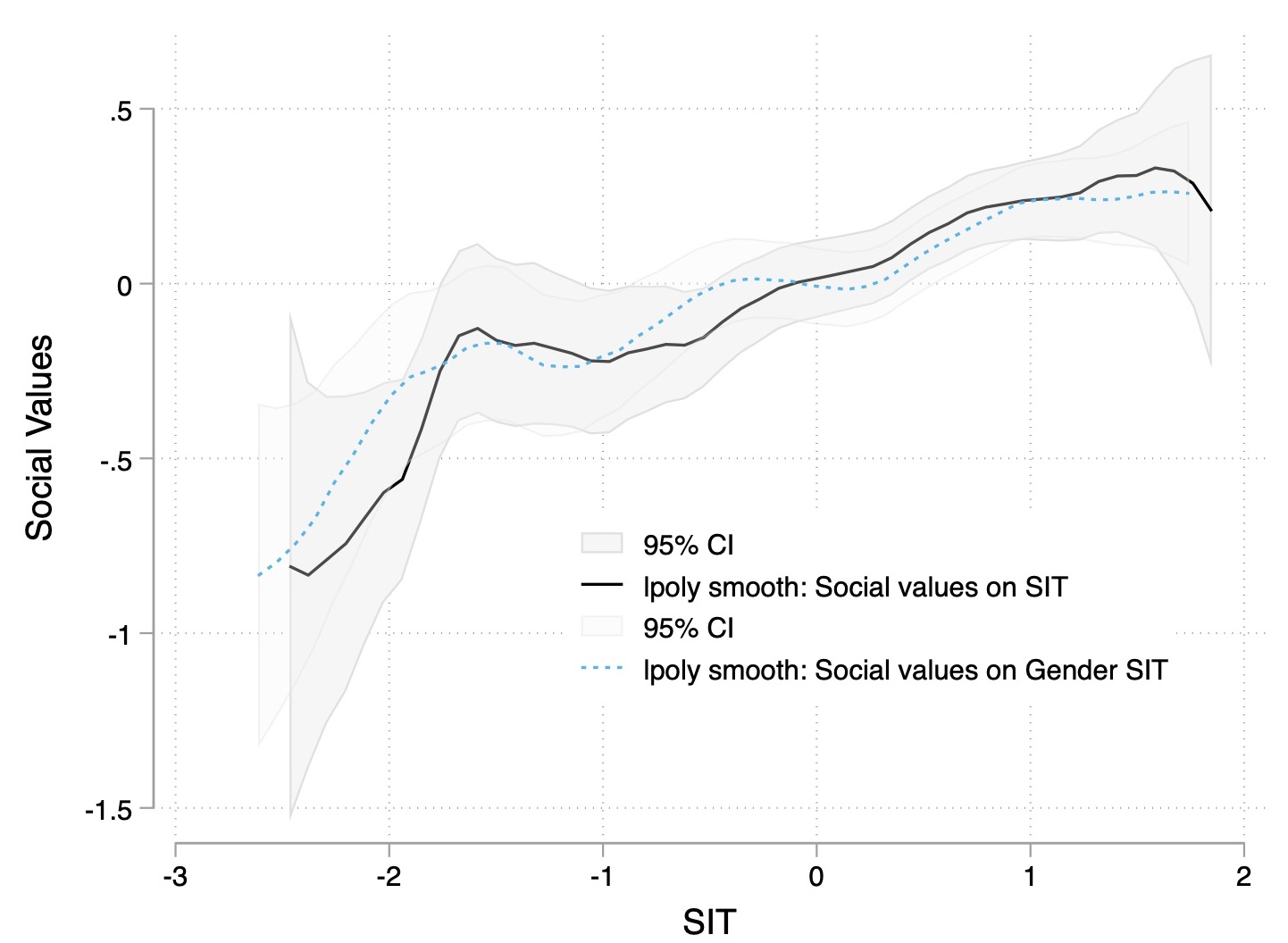}
        \label{fig:corr_soc}
    }%
    \hfill
    \subfigure[Implicit Bias Awareness Index]{
        \includegraphics[width=0.45\linewidth]{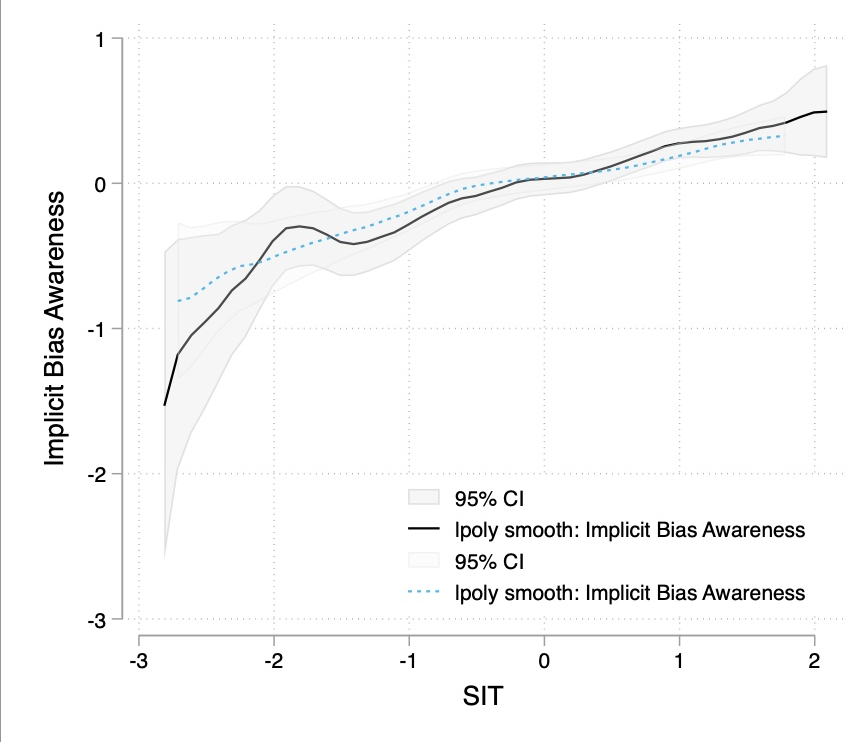}
        \label{fig:corr_ibawareness}
    }%
    \caption*{\footnotesize Local polynomial smooth regression between the SIT score (the dark line) and Gender SIT score (blue dashed line) with the Social Values Index (panel a, left) and the Implicit Bias Awareness Index (panel b, right). Shaded regions represent the 95\% confidence interval.}
   
\end{figure}

Furthermore, we expect SIT to be correlated not only with implicit biases but also with explicitly held stereotypical views, as reflected in progressive social attitudes. 
This expectation is confirmed in \autoref{fig:combined_corr}, where we present a non-linear association of the overall SIT score (solid black line) and the Gender SIT score (dashed blue line) with both the Social Values Index \subref{fig:corr_soc} and the Implicit Bias Awareness Index \subref{fig:corr_ibawareness}. 
All curves exhibit an increasing trend, suggesting that higher SIT scores are associated with more progressive social values and a greater ability to engage with topics related to inequality and bias in education.\footnote{See Appendix \autoref{fig:sit combined} for non-linear associations of the SIT scores with all the other scales measured in the survey.}

%

While the images' ratings are expected to reflect both implicitly and explicitly held stereotypical views, to clearly distinguish the two underlying drivers we can turn to the comments provided after rating each image, in which teachers explain what motivated the rating given. 

\subsection{Comments} 
For the linguistic analysis of the comments, we focused on the six Gender-STEM images.
These were rated and commented by all participants, for a total of 3,295 comments over 3,870 ratings, providing a sufficiently large dataset for meaningful analysis in reference to the core stereotype that our paper focuses on.\footnote{While comments were collected for all images, many had fewer responses and might not allow a reliable and robust linguistic analysis. Focusing on the subset with the most engagement ensures that the analysis captures patterns in participants' reasoning while minimizing noise from sparse data.}
Excluding single characters' comments, we obtain a final sample of 3,265 comments that we can exploit for sentiment and text analysis (84,37\% comment rate). 

We leverage the text to measure respondents' descriptive abilities, specifically lexical variety and lexical density, employing a linguistic profiling tool, Profiling-UD \citep{brunato-etal-2020-profiling}. 
Lexical variety measures the diversity of words used in a text, commonly assessed using the Type/Token Ratio (TTR)— the ratio of unique lexical types to the total number of tokens within a text. Lexical density, on the other hand, reflects the proportion of content words (verbs, nouns, adjectives, and adverbs) to the total word count in a text \citep{brunato-etal-2020-profiling}.
For both lexical variety and density, we calculated the average values across all respondents and all analyzed texts, obtaining identical results for both measures: 0.58 on a scale from 0 to 1, where 0 indicates minimal lexical variety (or density) and 1 represents maximal variety (or density). These findings indicate that respondents exhibit an above-average vocabulary range and lexical density. This aligns with our expectations, given their educational background and professional role in the field of education.
Additionally, we computed individual-level values for these features and incorporated them as predictors in our analysis of test scores.

When considering the most used words, the result is consistent with the theme of those pictures (Gender-STEM stereotypes), hence it is possible to read ``male'' 508 times and ``female'' 199 times, ``man'' 496 times and ``woman'' 445 times. 
Coherently with the type of images (see \autoref{sec:gen-stem}) the following most used words are ``mathematics'' (321), ``dance'' (255), ``construction site'' (201) and ``science'' (185).

Analyzing the use of job titles, we observe a preponderant use of job titles in their male form, with the exception of caring professions. 
Additionally, when trying to use the female form of a job title, often respondents use the modifier ``woman'' alongside the job title (e.g.: woman architect) even when the word exists in Italian in the female gender (\textit{architetta}), indicating that inclusive language is not widely known or used even by teachers, in spite of guidelines existing since the 1980s. 


The final step of our analysis focused on involving linguistic experts to analyze the comments to investigate the reasoned arguments that teachers offer for the ratings given to an image. While the rating itself is driven by the goal of finding stereotypes, the comment is expected to provide a justification that involves both the recognition of a specific type of stereotype and the opinion that the teacher has in relation to it, capturing both what they think and how they feel about it. Two linguistics experts conducted the annotation, assessing two dimensions: subjectivity and stance. Subjectivity refers to whether a subjective opinion is expressed within the comment. Stance, on the other hand, captures the writer’s position on the issue of stereotypes. Specifically, it identifies whether the writer acknowledges stereotypes as a problem and expresses awareness of their implications (pro), remains neutral without taking a clear position (neutral), or recognizes the stereotype but does not perceive it as problematic (against).  In cases of disagreement, a third expert was involved to resolve the conflict and ensure consistency in the annotations. Levels of agreement were measured using Cohen's Kappa, with values of 0.483 for subjectivity (indicating a moderate level of agreement) and 0.649 for stance (indicating substantial agreement). These results suggest that the annotations are reliable and provide a solid foundation for further analysis.

Almost all comments were cataloged as subjective \footnote{An example of a subjective comment is \textit{``Anche le femmine sono portate e interessate alle discipline STEM, basti vedere le iscrizioni al liceo scientifico, eppure nella foto si è scelto di rappresentare un maschio: perché? è uno stereotipo.''} that can be translated as \textit{``Females are also inclined and interested in STEM disciplines, just look at the enrollment in the scientific high school, yet a male was chosen to be represented in the photo: why? it is a stereotype'' }} (96\%) indicating the propensity of our teachers to express their personal opinions and not just a description of what they saw before, letting us know their sentiment. As expected, we observed that ``against'' comments are more frequent following lower ratings, while ``pro'' comments are more common after higher ratings. Specifically, ``against'' comments are associated with an average rating of 1.62, ``neutral'' comments with an average rating of 3.09, and ``pro'' comments with an average rating of 4.34. These findings align with the idea that there is a relationship between the ability to identify stereotypes and the ability to articulate them.
Following the suggestions of linguistic experts, we assigned numeric values to each stance level, consistent with the rating scale: 1 for ``against'', 3 for ``neutral'', and 5 for ``pro''. This allowed us to calculate the mean stance value for each individual. We found a strong correlation (0.80) between individuals' mean stance values and their ratings providing quantitative evidence of consistency.
In the next section, we make use of the annotation data to model the SIT score alongside the other information we collected that can be expected to influence teachers' ability to see stereotypes.

\section{Predicting SIT}\label{sec:predict}

What can influence and shape a teacher's ability to see stereotypes?
We systematically relate teachers' performance on the Stereotype Identification Test to a rich set of demographic characteristics and personal values and beliefs, as well as an information-provision experiment randomly embedded into the survey structure.

As previously explained, the order of the IAT and SIT was randomized. After completing the IAT, teachers received feedback\footnote{An example of the feedback can be found in \autoref{sec:regression}} on how their score was computed, their performance on the test, and the type of stereotype being assessed (i.e., the association between gender and STEM subjects). Thus, to identify the causal impact of information provision about stereotypes on the ability to recognize biases, we estimate the following model:
 \begin{displaymath}
        \begin{aligned}
            SIT_i = \alpha + \beta_1 IATRev_i + \beta_2 IATScore_i + \beta_3 LEX_i + \eta \textbf{W}_i + \gamma \textbf{X}_i + \varepsilon_i 
        \end{aligned}
\end{displaymath}
The dependent variable in our regression model is the SIT score of the individual \textit{i}. The key independent variables include $IATRev_i$, a dummy variable indicating whether the individual completed the IAT test before the SIT, thereby revealing information about their level of implicit stereotypes. The variable $IATScore_i$, represents the score obtained in the IAT, while $LEX_i$ measures the individual's lexical density index. Additionally, the model includes two sets of controls: $\textbf{W}_i$, a vector encompassing indices derived from teachers' questionnaires that capture their values, awareness, and personality traits, and $\textbf{X}_i$, a vector of socio-demographic controls.\footnote{The vector of indices $\textbf{W}_i$ includes: Implicit Bias Awareness, Locus of Control, Social Values, Inclusive Teaching, Growth Mindset, Gender-STEM Stereotypes.
The vector of controls $\textbf{X}_i$ includes: Age (continuous), Gender (0 = male, 1 = female), Like Teaching (7-points Likert Scale), Master degree Disability Training, Married, Teaching Italian, Teaching Mathematics, Place of Birth divided into North-East, North-West, Center, South, Island, Missing (Center as reference category)} 

The results in \autoref{table:IATfirst_paper} show how receiving feedback about individuals' performance on a test about implicit stereotypes increases the level of the SIT score by a quarter of a standard deviation, as teachers who have their IAT revealed see 0.28 standard deviations more stereotypes compared to the ones who completed the Implicit Association Test only after SIT. This effect is similar in size to a one standard deviation increase in implicit bias awareness, and almost twice the size of inclusive teaching practices. 

\begin{table}[htbp]\centering
\def\sym#1{\ifmmode^{#1}\else\(^{#1}\)\fi}
\caption{Predicting SIT}
\begin{tabular}{l*{6}{c}}
\hline\hline
            &\multicolumn{1}{c}{(1)}&\multicolumn{1}{c}{(2)}&\multicolumn{1}{c}{(3)}&\multicolumn{1}{c}{(4)}&\multicolumn{1}{c}{(5)}&\multicolumn{1}{c}{(6)}\\
\hline
IAT Revelation      &       0.285\sym{***}&       0.277\sym{***}&       0.285\sym{***}&       0.243\sym{***}&       0.233\sym{***}&       0.219\sym{***}\\
                    &     (0.085)         &     (0.085)         &     (0.085)         &     (0.080)         &     (0.077)         &     (0.076)         \\
[1em]
IAT score           &                     &                     &       0.069         &       0.043         &       0.052         &       0.049         \\
                    &                     &                     &     (0.043)         &     (0.043)         &     (0.042)         &     (0.042)         \\
[1em]
Growth Mindset      &                     &                     &                     &       0.031         &       0.037         &       0.037         \\
                    &                     &                     &                     &     (0.046)         &     (0.045)         &     (0.044)         \\
[1em]
Implicit Bias Awareness&                     &                     &                     &       0.240\sym{***}&       0.274\sym{***}&       0.266\sym{***}\\
                    &                     &                     &                     &     (0.052)         &     (0.054)         &     (0.053)         \\
[1em]
Gender-STEM Stereotypes&                     &                     &                     &      -0.053         &      -0.042         &      -0.041         \\
                    &                     &                     &                     &     (0.044)         &     (0.044)         &     (0.043)         \\
[1em]
Locus of Control    &                     &                     &                     &       0.098\sym{*}  &       0.104\sym{*}  &       0.110\sym{**} \\
                    &                     &                     &                     &     (0.053)         &     (0.054)         &     (0.053)         \\
[1em]
Social Values       &                     &                     &                     &       0.168\sym{***}&       0.158\sym{***}&       0.144\sym{***}\\
                    &                     &                     &                     &     (0.054)         &     (0.055)         &     (0.053)         \\
[1em]
Inclusive Teaching  &                     &                     &                     &       0.125\sym{**} &       0.110\sym{**} &       0.105\sym{**} \\
                    &                     &                     &                     &     (0.051)         &     (0.051)         &     (0.050)         \\
[1em]
Lexical Density     &                     &                     &                     &                     &                     &       0.673\sym{***}\\
                    &                     &                     &                     &                     &                     &     (0.235)         \\
[1em]
Socio-Demographics     &         NO            &        YES             &         YES            &         NO            &     YES                &      YES        \\
            &                     &                     &                     &                     &                     &             \\
[1em]
\hline
\(N\)      &         614         &         614         &         614         &         614         &         614         &         614         \\
\hline\hline

\end{tabular}
\label{table:IATfirst_paper}
\caption*{\footnotesize Dependent variable: SIT Score. The vector of indices $\textbf{W}_i$ includes: Implicit Bias Awareness, Locus of Control, Social Values, Inclusive Teaching, Growth Mindset, Gender-STEM Stereotypes. The vector of controls $\textbf{X}_i$ includes: Age (continuous), Gender (0 = male, 1 = female), Like Teaching (7-points Likert Scale), Master degree Disability Training, Married, Teaching Italian, Teaching Mathematics, Place of Birth divided into North-East, North-West, Center, South, Island, Missing (Center as reference category). Standard errors in parentheses. \sym{*} \(p<0.10\), \sym{**} \(p<0.05\), \sym{***} \(p<0.01\)}
\end{table}

The results in \autoref{table:IATfirst_paper} support the intuition that, unlike the IAT, our tool measures an ability that depends on traits and knowledge that can be developed and taught. 
As expected, the IAT score is not associated with any specific characteristic or trait, as shown in \autoref{table:reg2}, reinforcing the idea that it captures an implicit bias that individuals may not be consciously aware of.

\begin{table}[htbp]\centering
\def\sym#1{\ifmmode^{#1}\else\(^{#1}\)\fi}
\caption{SIT vs IAT}
\begin{tabular}{l*{2}{c}}
\hline\hline
            &\multicolumn{1}{c}{(1)}&\multicolumn{1}{c}{(2)}\\
            & \multicolumn{1}{c}{SIT Score}&\multicolumn{1}{c}{IAT Score}\\
\hline
IAT score           &       0.042         &                     \\
                    &     (0.042)         &                     \\
[1em]
SIT Score &                     &       0.050         \\
                    &                     &     (0.049)         \\
[1em]
Growth Mindset      &       0.035         &      -0.019         \\
                    &     (0.044)         &     (0.054)         \\
[1em]
Implicit Bias Awareness&       0.277\sym{***}&       0.048         \\
                    &     (0.052)         &     (0.057)         \\
[1em]
Gender-STEM Stereotypes&      -0.045         &       0.018         \\
                    &     (0.043)         &     (0.044)         \\
[1em]
Locus of Control    &       0.102\sym{*}  &       0.053         \\
                    &     (0.053)         &     (0.054)         \\
[1em]
Social Values       &       0.149\sym{***}&      -0.014         \\
                    &     (0.053)         &     (0.060)         \\
[1em]
Inclusive Teaching  &       0.106\sym{**} &      -0.048         \\
                    &     (0.049)         &     (0.064)         \\
[1em]
Lexical Density     &       0.703\sym{***}&       0.081         \\
                    &     (0.236)         &     (0.210)         \\
[1em]
Socio-Demographics              &       YES        &       YES        \\
                    &             &              \\
[1em]
\hline
\(N\)       &         614         &         614         \\
\hline\hline

\end{tabular}
\label{table:reg2}
\caption*{\footnotesize In regression (1) the dependent variable is the SIT score. In regression (2) the dependent variable is the IAT score. The vector of indices $\textbf{W}_i$ includes: Implicit Bias Awareness, Locus of Control, Social Values, Inclusive Teaching, Growth Mindset, Gender-STEM Stereotypes.
The vector of controls $\textbf{X}_i$ includes: Age (continuous), Gender (0 = male, 1 = female), Like Teaching (7-points Likert Scale), Master degree, Disability Training, Married, Teaching Italian, Teaching Mathematics, Place of Birth divided into North-East, North-West, Center, South, Island, Missing (Center as reference category). Standard errors in parentheses. \sym{*} \(p<0.10\), \sym{**} \(p<0.05\), \sym{***} \(p<0.01\)}
\end{table}

\section{Reliability and Validity of the SIT}
\label{sec:reliability_validity}
In the following section, we probe the robustness of our measure by evaluating whether random permutations or sub-group of presented images yield stable and comparable SIT scores; and whether the SIT score actually measures a person's ability to see stereotypes. 
In psychometric terms, we assess the reliability and validity of the SIT.\footnote{More detailed information can be found in \autoref{sec:psychmetrics}.}

\subsection{Reliability}
Reliability refers to the consistency and stability of a test's measurements, ensuring that the results are reproducible across different occasions, forms, or sets of items.

The most commonly used statistic to assess reliability is Cronbach's alpha, which represents the average reliability across all possible item splits \citep{warrens_cronbachs_2015}. The general rule of thumb is that a Cronbach’s alpha greater than 0.7 is acceptable, and greater than 0.9 and above is highly reliable \citep{green1993alpha,taber2018cronbach}. 
Our SIT score has a Cronbach's alpha of 0.98, suggesting extremely strong reliability.\footnote{Given this high level of alpha, fewer than 20 images might have been sufficient for a reliable score.}
The Gender-STEM SIT score, which is constructed using only six images, has a Cronbach's alpha of 0.82.

To further probe reliability, we use a permutation approach based on resampling, which allows us to assess both split-half reliability (when resampling is done \textit{without} replacement) and test-retest reliability (when resampling is done \textit{with} replacement) \citep{williams_reliability_2012,pronk_methods_2022}.\footnote{For greater details, see \autoref{sec:reliability}.} 

\begin{figure}[ht]
\caption{\label{fig:combined_reliability}}
    \centering
    \subfigure[Split-half reliability]{%
        \includegraphics[width=0.48\linewidth]{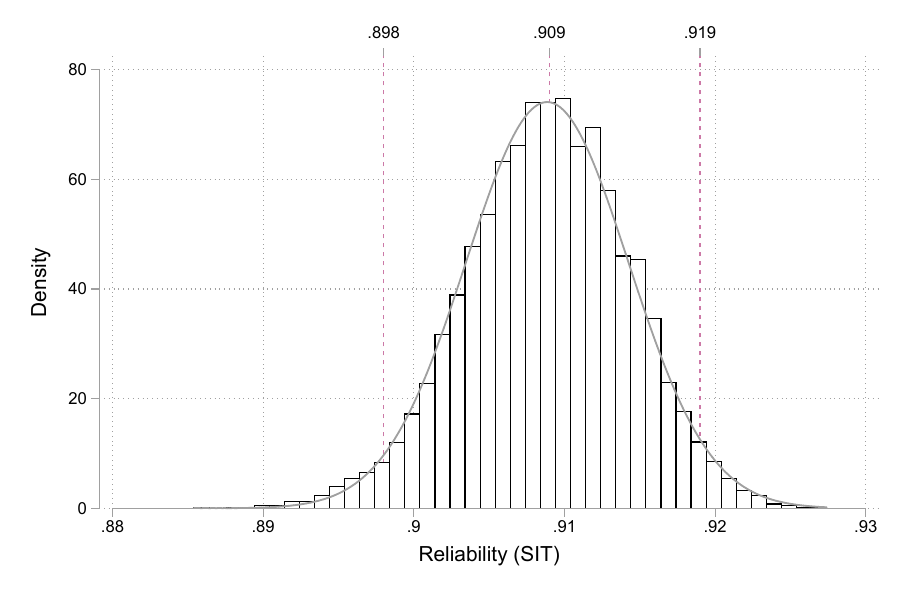}
        \label{fig:simresults_splithalf}
    }%
    \hfill
    \subfigure[Test-retest reliability]{%
        \includegraphics[width=0.48\linewidth]{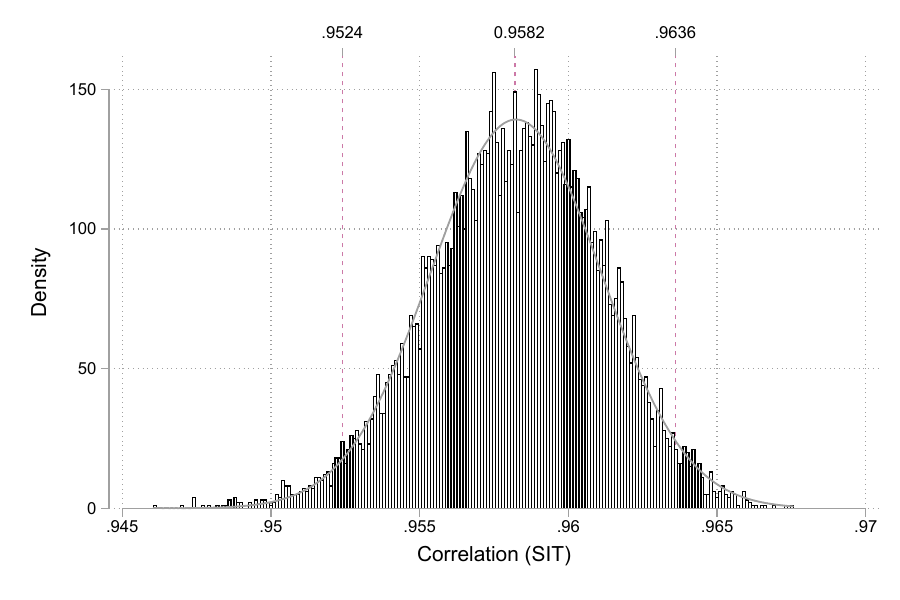}
        \label{fig:simresults_retest}
    }%
    \caption*{Reliability coefficients for SIT. The graph illustrates the mean, along with the 2.5\% and 97.5\% empirical quantiles, of the distribution of Pearson correlation coefficients calculated between the first and second halves of the 20 image ratings when resampled 9,999 times: without replacement for split-half reliability in Figure (a), and with replacement for test-retest reliability in Figure (b).}
    
\end{figure}

 \autoref{fig:combined_reliability} reports the distribution of the simulated reliability coefficients with the associated mean and 2.5\% and 97.5\% empirical quantiles.
The vast majority of the distribution of these permutations are above 0.9, further reinforcing our claim that the SIT test is highly reliable.

To assess the stability of responses to slight environmental variations, we conducted a framing experiment at the beginning of the survey, where teachers were randomly assigned to one of three introductory texts designed to evoke different emotional responses to stereotypes in schools. As mentioned in Section \ref{sec:survey}, the first group (``info") received a neutral definition of stereotypes as cognitive shortcuts. The second group (``info+guilt") was exposed to a more emotionally charged message highlighting the negative effects of stereotypes on self-confidence and performance in school settings. The third group (``no frame") received no introductory message, serving as a control group.

As shown in Appendix Table \ref{tab:Framing}, SIT remains remarkably stable to these framing, suggesting its robustness as a measure of stereotype detection.

\subsection{Validity}
Validity of a test refers to the degree to which the test accurately measures what it is intended to measure; in a somewhat abstract way, it has been defined as ``the degree to which evidence and theory support the interpretations of test scores for proposed uses of tests" \citep[page 11]{AERA2014}. This definition emphasizes that validity is first about interpretation of the test, rather than about the test itself, and, second, that it depends on the context for which the test is relevant. Although validity is still highly debated theoretically,\footnote{Interested readers may refer to \cite{hughes_psychometrics_2018} and \cite{braverman_psychometrics_2022} for more details.} in practice, it is conceived as a cumulative process of different kinds of evidence converging toward a similar interpretation \citep{messick_validity_1989,AERA2014}. In the remaining of the section and the appendix \ref{sec:validity}, we follow standard methods widely recommended for discussing validity \citep{AERA2014, hughes_psychometrics_2018, braverman_psychometrics_2022}

To assess validity of the SIT we take several approaches related to (i) test content; (ii) response processes; and (iii) internal structure. Validity evidence based on relations to other variables was already presented in \autoref{sec:predict} and a complementary discussion can be found in the Appendix \ref{sec:validity}. 


\paragraph{Test content}

First, we evaluate the test content by examining whether the selected images are likely to reflect a broad range of stereotypes. To do so, we utilize the tags provided for each image by the academic publishing company. These tags are meant to facilitate search when book editors need content to illustrate specific situations. Each image has between 18 and 52 tags, averaging 44 tags per image, typically describing aspects such as the individuals, their actions, objects, and context.\footnote{Two images with no tags at all were excluded from this analysis.} We then apply a zero-shot text classification model to categorize these tags into six potentially sensitive characteristics often legally defined as protected characteristics in a broad range of contexts across the world: gender, race, social origin, religion, disability, and age, assigning probability weights to each tag. If the probability of a tag belonging to a particular category exceeds 50\%, it is classified accordingly. After manual review, 347 tags are assigned to these six categories. On average, each image contains 8.9 tags related to protected characteristics, representing 2.7 distinct characteristics. The proportion of tags related to these characteristics ranges from 2\% to 50\%, with an average of 20\%. In summary, each image presented to the teachers has a high probability of featuring potentially stereotypical content.

\paragraph{Response processes}

Second, we evaluate evidence based on response processes, focusing on whether participants' cognitive processes align with the construct being measured. In our SIT test, participants first rate images for stereotypes and then have the option to explain their ratings through written comments, offering insights into their thought processes. We also measure reaction times for both steps, providing an indirect measure of cognitive effort. Of 12,901 ratings, 83\% were accompanied by comments, with high engagement from participants, especially for the Gender-STEM images. Most comments were brief, written in the present tense, and used terms like ``male," ``female," ``math," and ``science," reflecting engagement with stereotypes. Reaction times indicated that less stereotypical images took longer to rate, suggesting participants were thoughtfully evaluating images based on their stereotype knowledge. These patterns, combined with text analysis, demonstrate strong test validity, particularly in how participants engaged with the task.

\paragraph{Internal structure}

Finally, evidence related to the internal structure of a measure examines how the individual items correspond to the underlying construct the test aims to assess. According to the Standards for Educational and Psychological Testing \citep{AERA2014}, the organization of items should align with the construct, in our case, with a unidimensional construct showing high correlations between items. To evaluate the internal structure of SIT, we conducted both exploratory and confirmatory factor analyses using the ratings of 20 images as individual items. The results from both analyses, shown in \autoref{sec:validity}, strongly supported a unidimensional structure, confirming that the SIT accurately measures stereotype detection. This evidence enhances the validity of the test, showing that the ratings align with the intended cognitive process of stereotype recognition.

\section{Conclusion}
\label{sec:conclusion}
Stereotypes influence decision-making across various domains, including education, where they shape both teacher perceptions and student outcomes \citep{Carlana2019}. 
While strategies exist to mitigate the impact of bias, an essential first step is ensuring that individuals can recognize stereotypes in their environment. 
Our study provides an easily implementable measure of this ability through the Stereotype Identification Test (SIT).

We demonstrate that teachers’ ability to detect stereotypes in images commonly found in school textbooks can be reliably measured using the SIT. 
We provide strong evidence that the SIT is a valid and reliable instrument for assessing this skill. 
Furthermore, our findings suggest that the ability to recognize stereotypes may be more malleable than deeper, more implicit cognitive processes, such as those measured by the Implicit Association Test (IAT), making it a promising target for interventions.
Indeed, we find that teachers’ SIT scores are systematically related to trainable traits, such as greater awareness of implicit bias and the use of inclusive teaching practices. 
More strikingly, we find that a simple intervention revealing a person’s own implicit bias can improve their SIT score by a quarter of a standard deviation, highlighting the potential for targeted interventions to enhance stereotype recognition.

These findings suggest that the SIT can serve as a valuable tool for assessing the effectiveness of interventions aimed at reducing the impact of biases in educational settings. 
By tracking improvements in stereotype recognition, the SIT can help evaluate whether training programs or policy changes succeed in fostering more equitable decision-making.

Beyond education, empirical research has documented the influence of stereotypes on decision-making across multiple domains, including labor markets \citep{alexander1992makes,bertrand2004emily,neumark2018experimental},
access to credit \citep{dymski2006discrimination,dobbie2021,macchi2023worth}, 
housing \citep{kain1972housing,ewens2014statistical,edelman2017racial}, 
healthcare services \citep{balsa2001statistical,bridges2018implicit,alsan2019diversity,obermeyer2019dissecting}, 
consumer markets \citep{ayres1995race,yinger1998evidence,list2004nature,doleac2013visible}, 
politics \citep{hooghe2023discrimination}, 
and justice and law enforcement \citep{knowles2001racial,David2018}. 
The SIT framework could be adapted to these settings to better understand how biases affect decision-making and to evaluate interventions aimed at reducing disparities. 
Developing context-specific versions of the SIT could provide policymakers, organizations, and researchers with a practical tool to assess and address bias in hiring, financial decisions, medical treatment, law enforcement practices, and beyond.


\newpage

\appendix

\section{Data from the Italian Ministry of Education (MIUR)} \label{sec:MIUR}
\begin{table}[H]
\centering
\caption{\normalsize Teachers' distribution in Italy according to age and sex.\label{table:agemiur}}
\begin{threeparttable}
\begin{tabular}[H]{lcc}
\toprule
\multicolumn{1}{l}{ Age } & \multicolumn{1}{c}{ National Level } & \multicolumn{1}{c}{ Our Sample }  \\
\midrule
Less than 35 &  9,034 & 40 \\
 &   3.82\% & 6.51\% \\
  & & \\
35 - 44 &  44,979 & 88\\
 &  19.01\% & 14.33\%\\
  & & \\
45 - 54 & 87,674 & 191\\
  & 37.05\% & 31.11\%\\
  & & \\
More than 54 &  94,920 & 295 \\
 & 40.12\% & 48.05\%\\
  & & \\
Total &  236,607 & 614\\ 
 
\bottomrule
\end{tabular}
\caption*{\footnotesize Distribution of teachers with permanent contracts in Italy divided by age. The Italian system counts 236,607 teachers with a permanent contract. Data refers to 2023, do not include information about Val d'Aosta, Trento and Bolzano and are freely available.\tablefootnote{\url{https://dati.istruzione.it/opendata/}}}

\end{threeparttable}
\end{table}

Distribution of teachers in Italy and in our sample according to ann institutionalized division into the five groups: North-West, North-East, Center, South and Islands. 
\begin{figure}[ht]
    \centering
    \subfigure[Place of Birth Italian Teachers]{%
        \includegraphics[width=0.5\linewidth]{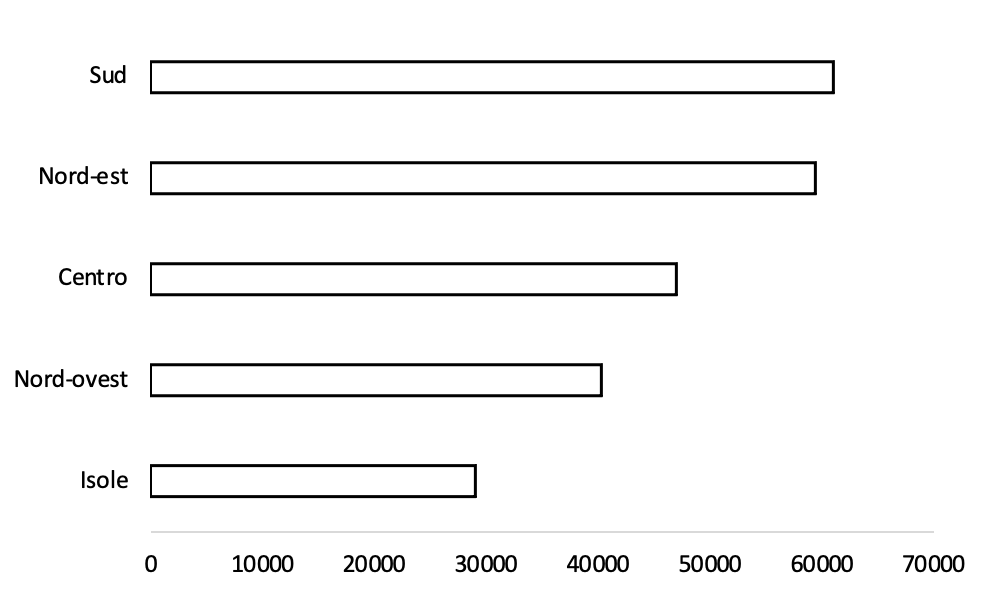}
        \label{fig:prova}
    }%
    \hfill
    \subfigure[Place of Birth Sample Teachers]{%
        \includegraphics[width=0.48\linewidth]{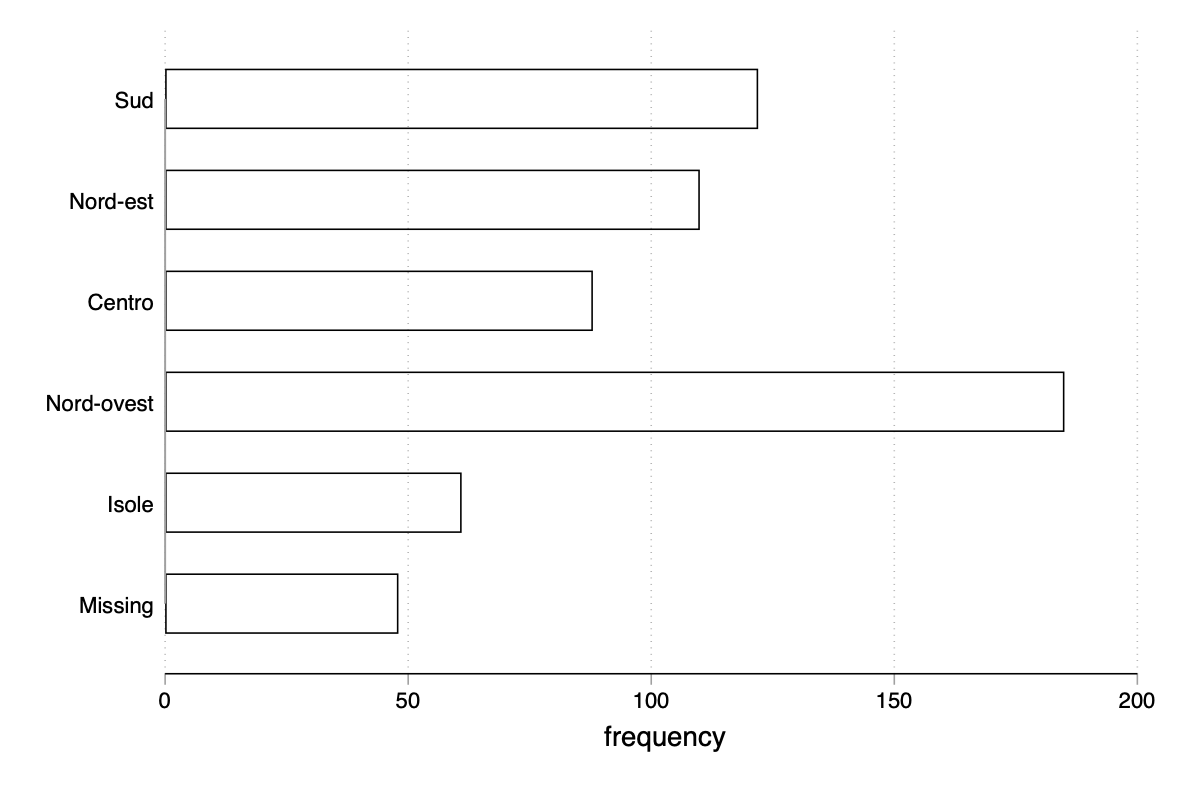}
        \label{fig:prova 2}
    }%
    
    \label{fig:speriamo}
\end{figure}

\section{Gender-STEM Pictures}\label{sec:gen-stem}

\begin{figure}[H]
\centering
\subfigure{\includegraphics[scale=0.04]{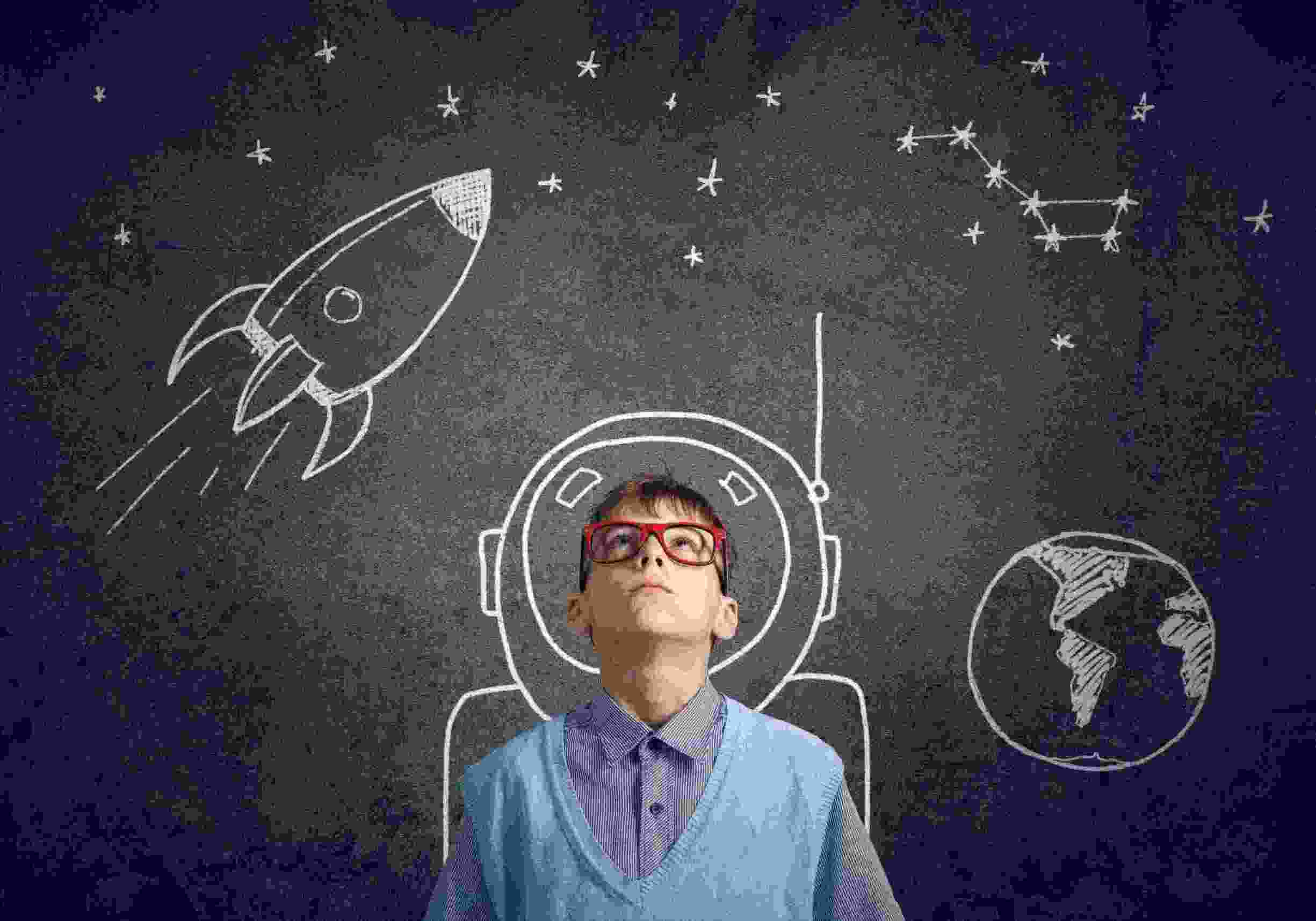}}
\hfill
\subfigure{\includegraphics[scale=0.025]{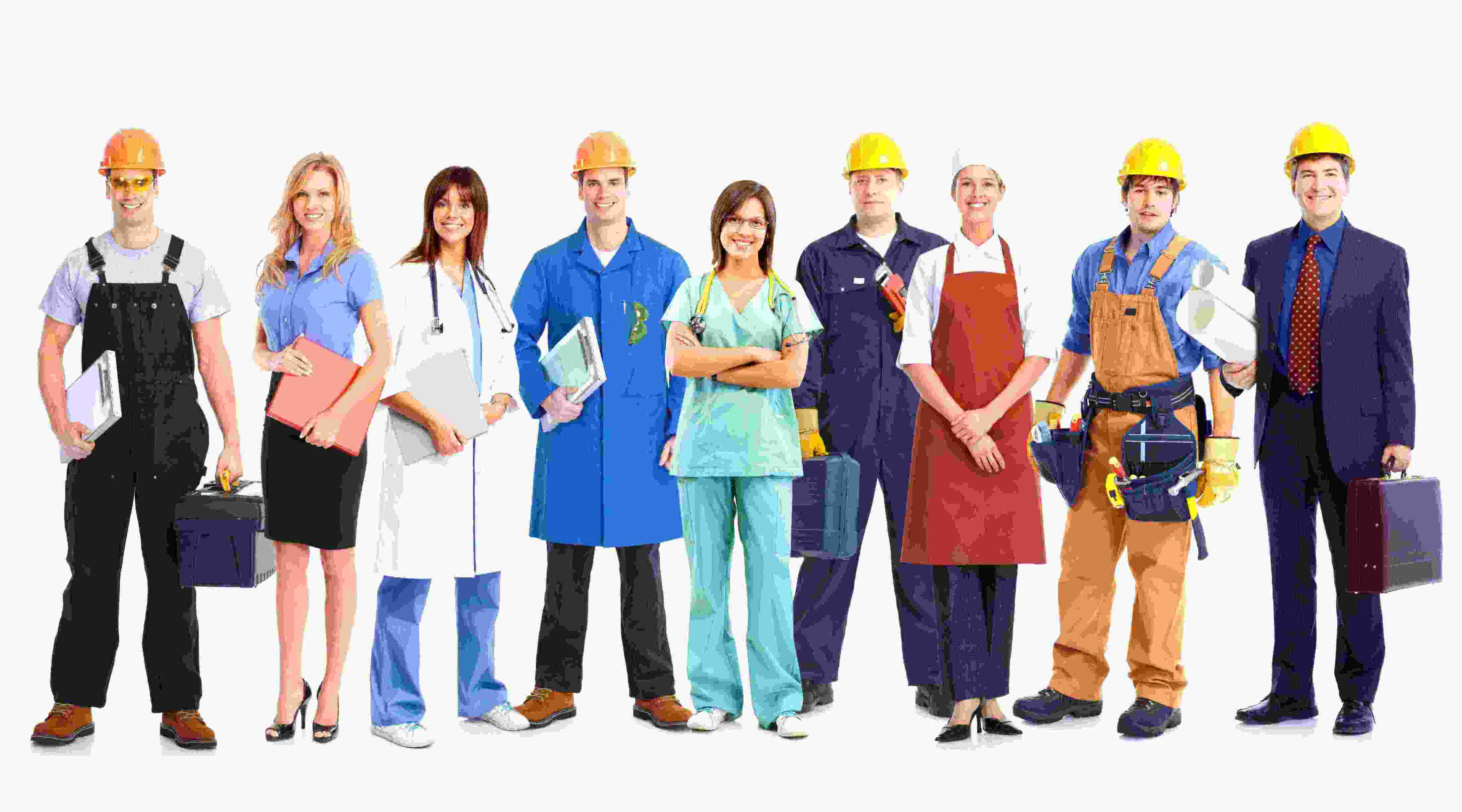}}
\hfill
\subfigure{\includegraphics[scale=0.04]
{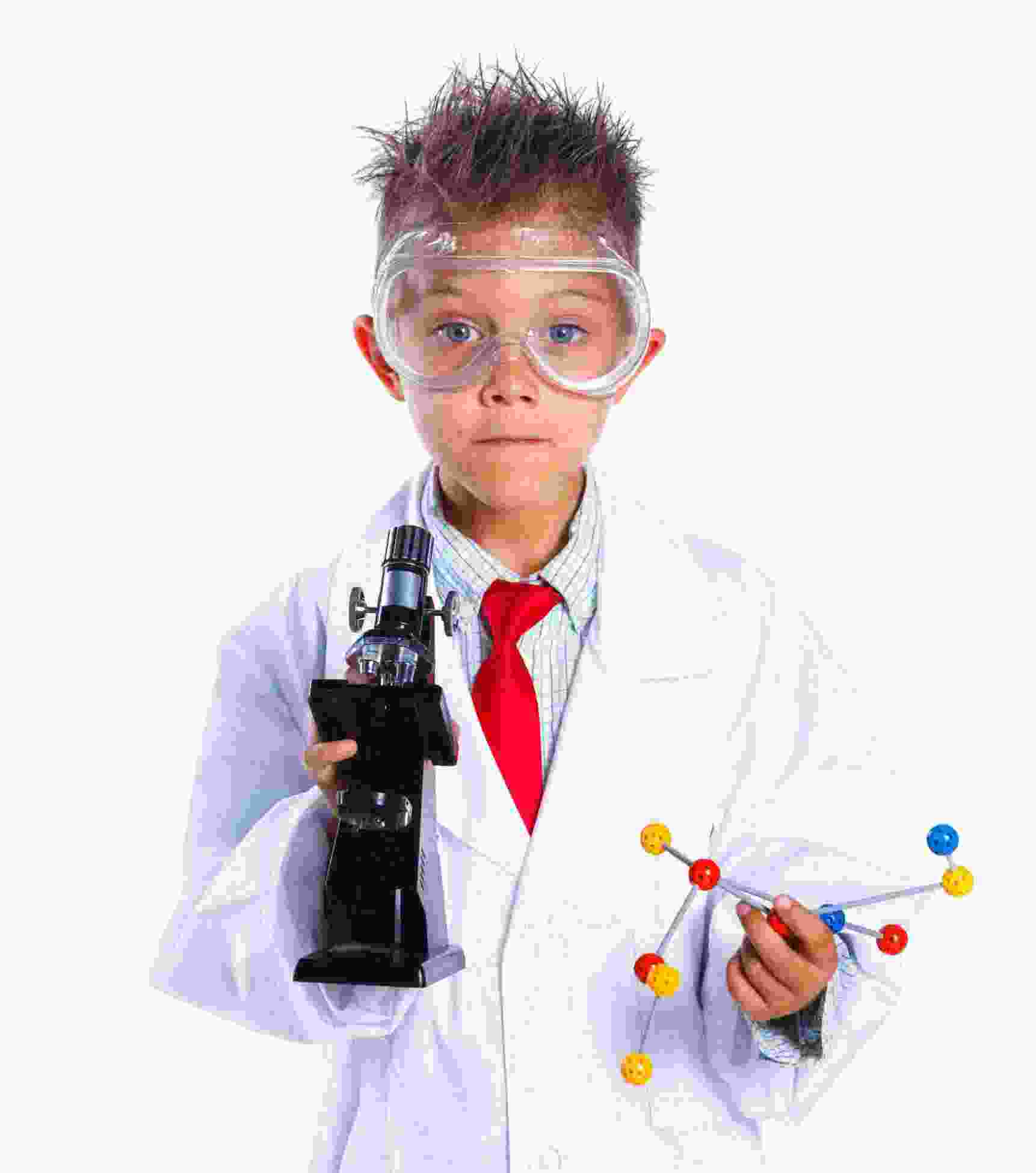}}
\end{figure}

\begin{figure}[H]
\centering
\subfigure{\includegraphics[scale=0.025]
{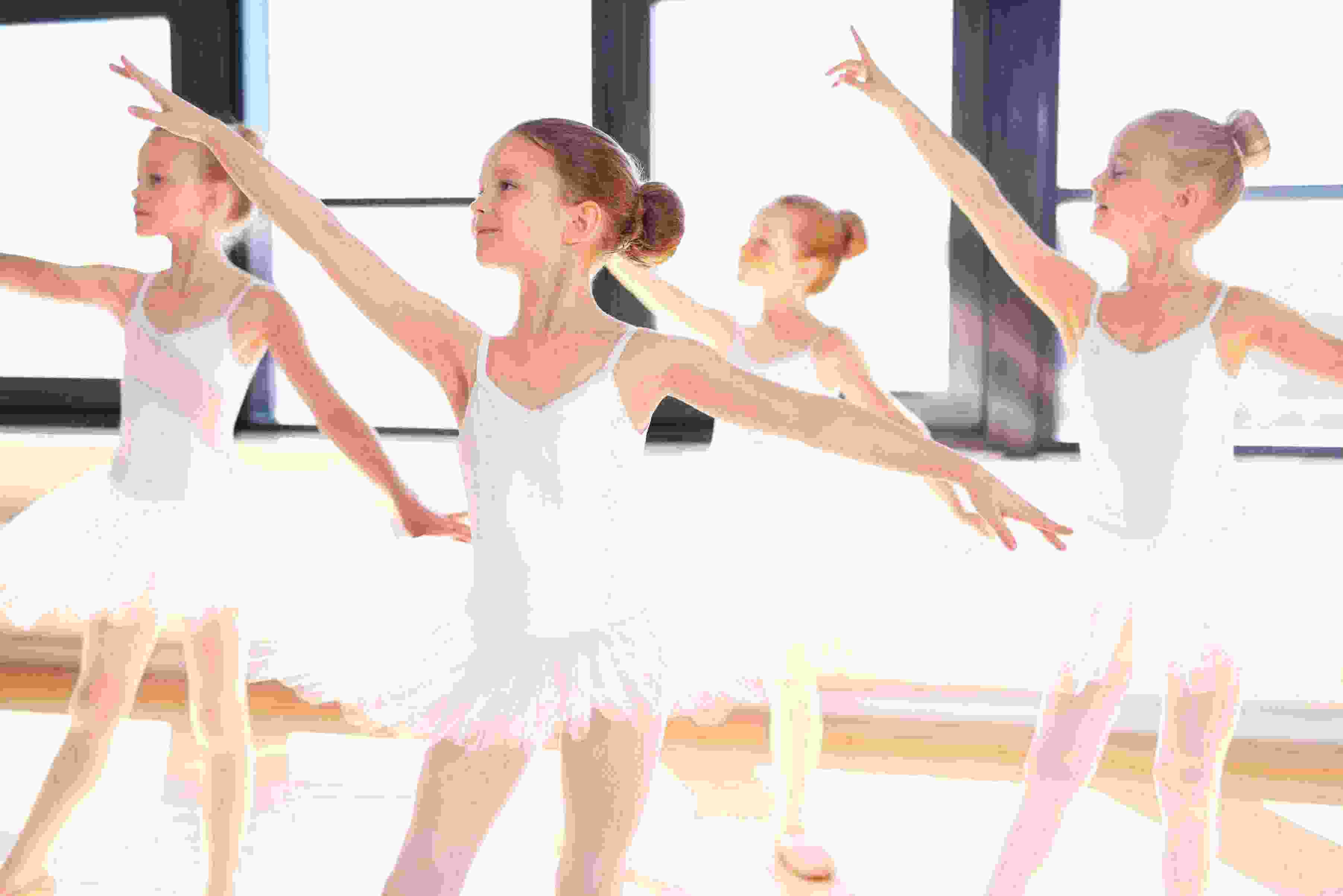}} 
\hfill
\subfigure{\includegraphics[scale=0.04]{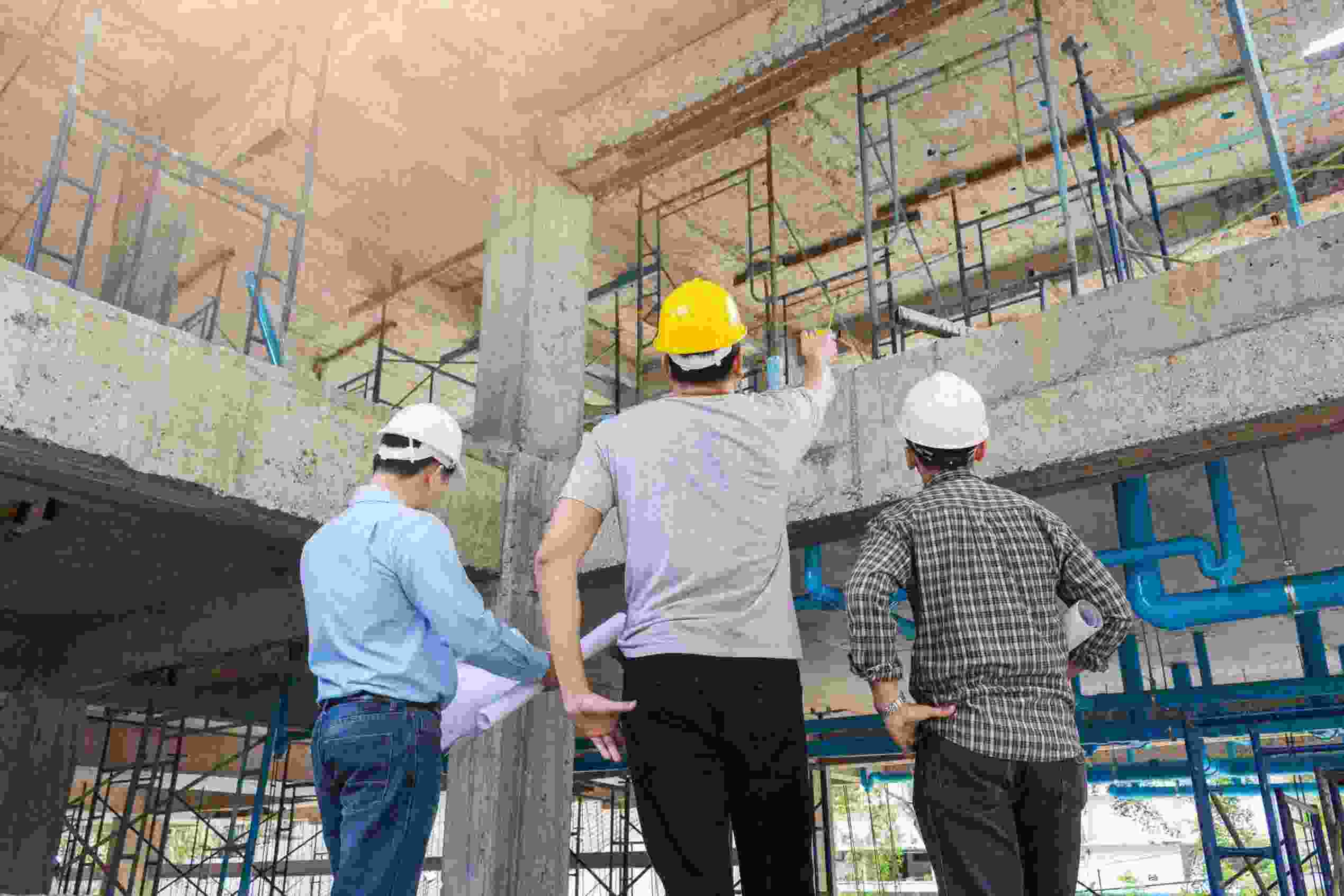}}
\hfill
\subfigure{\includegraphics[scale=0.045]{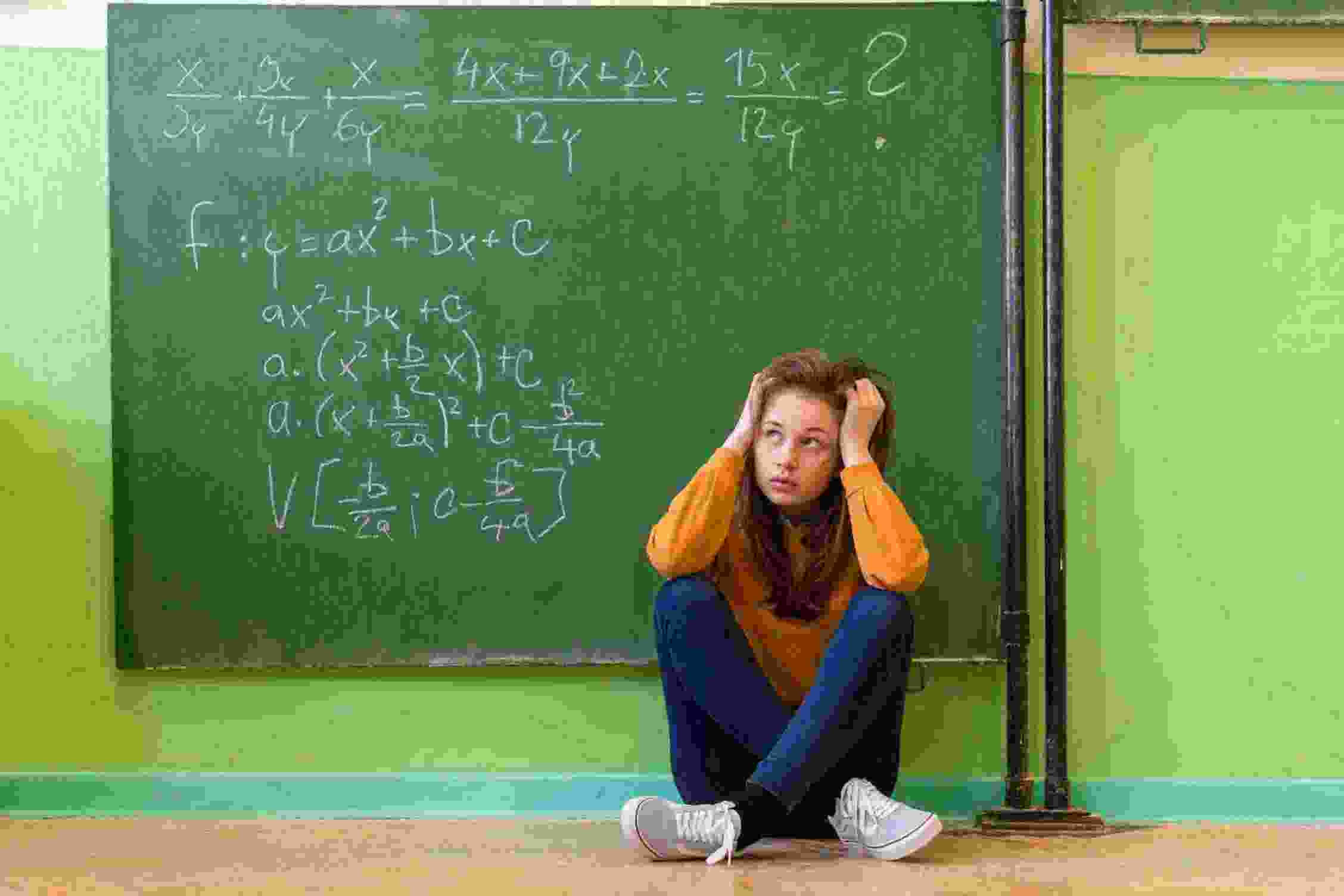}}
\end{figure}

\section{Teacher Survey Data}\label{appsec:survey}
\subsection{Questionnaire}\label{sec:questionnaire}
The specific set of questions used to construct indices on teachers' values, awareness, and personality traits is as follows. Each question was answered on a five-point Likert scale, where 1 indicated ``strongly disagree'' and 5 indicated ``strongly agree''. Below, we provide the English translation of the questions used for each index.
\paragraph{Growth Mindset:}
\begin{enumerate}
    \item Everyone has a certain level of intelligence and cannot do much to change it.
    \item I like challenging training courses so I can learn new things.
    \item Intelligence is a personal trait that cannot be changed much.
    \item I like using what I learn in training courses in my classroom lessons.
    \item You can learn new things, but you cannot change your intelligence.
\end{enumerate}

\paragraph{Implicit Bias Awareness:}
\begin{enumerate}
    \item The way teaching is shaped in schools can contribute to reinforcing conscious or unconscious biases.
    \item It is important to address the issue of biases and inequalities in daily teaching activities.
    \item In my daily teaching activities, I often address issues related to inequality.
    \item I feel that I have the necessary tools to address the issue of biases and inequalities in my teaching practice.
\end{enumerate}

\paragraph{Gender-STEM Stereotypes:}
\begin{enumerate}
    \item Girls are naturally better than male students in humanities subjects.
    \item Male students are naturally better than female students in scientific and mathematical subjects.
    \item Propensity for foreign language communication is typical of girls.
    \item Boys need more time than girls to understand complex or abstract concepts.
    \item More needs to be done to encourage female students to engage in STEM disciplines.
    \item Male students are more undisciplined than female students.
\end{enumerate}

\paragraph{Locus of Control:}
\begin{enumerate}
    \item If I put enough effort and time into it, I can implement a new teaching strategy even with students who are disinclined to learn new things.
    \item When a student gets a better grade than usual generally it is because they have studied more, not because I have tried to explain the lesson better.
    \item If one day I find myself scolding a student more often than usual, it is probably because I was a little less tolerant that day, not because that student was behaving worse than usual.
    \item When a student is able to learn a new concept quickly, it is probably because the student was able to understand it, not because I was able to explain it better.
    \item When a new student fails to make friends with his classmates, it is probably because I have not encouraged other students enough to be nicer to the newcomer.
\end{enumerate}

\paragraph{Social Values:}
\begin{enumerate}
    \item When jobs are scarce, men have more right than women to have jobs.
    \item When work is scarce, employees should give priority to locals over immigrants.
    \item That a woman earns more than her husband can cause problems.
    \item Homosexual parents are just as good as heterosexual parents.
    \item It is a duty to society to have children.
    \item Children in adulthood have an obligation to ensure long-term support for their parents.
    \item People who do not work are lazy.
    \item Work is a duty towards society.
    \item Work should always be a priority, even if it means having less free time.
\end{enumerate}

\paragraph{Inclusive Teaching:} (In this case 1 indicated ``not at all inclusive'' and 5 indicated ``very inclusive''.)
\begin{enumerate}
    \item Critical thinking development activities.
    \item Oral exam.
    \item Cooperative / Collaborative teaching.
    \item Sharing learning objectives before the lesson begins.
    \item Summative assessment tests.
\end{enumerate}

\subsection{Bivariate Associations}\label{sec:correlations}
\begin{figure}[H]
\caption{SIT and Traits \label{fig:sit combined}}
    \centering
\includegraphics[scale=0.55]{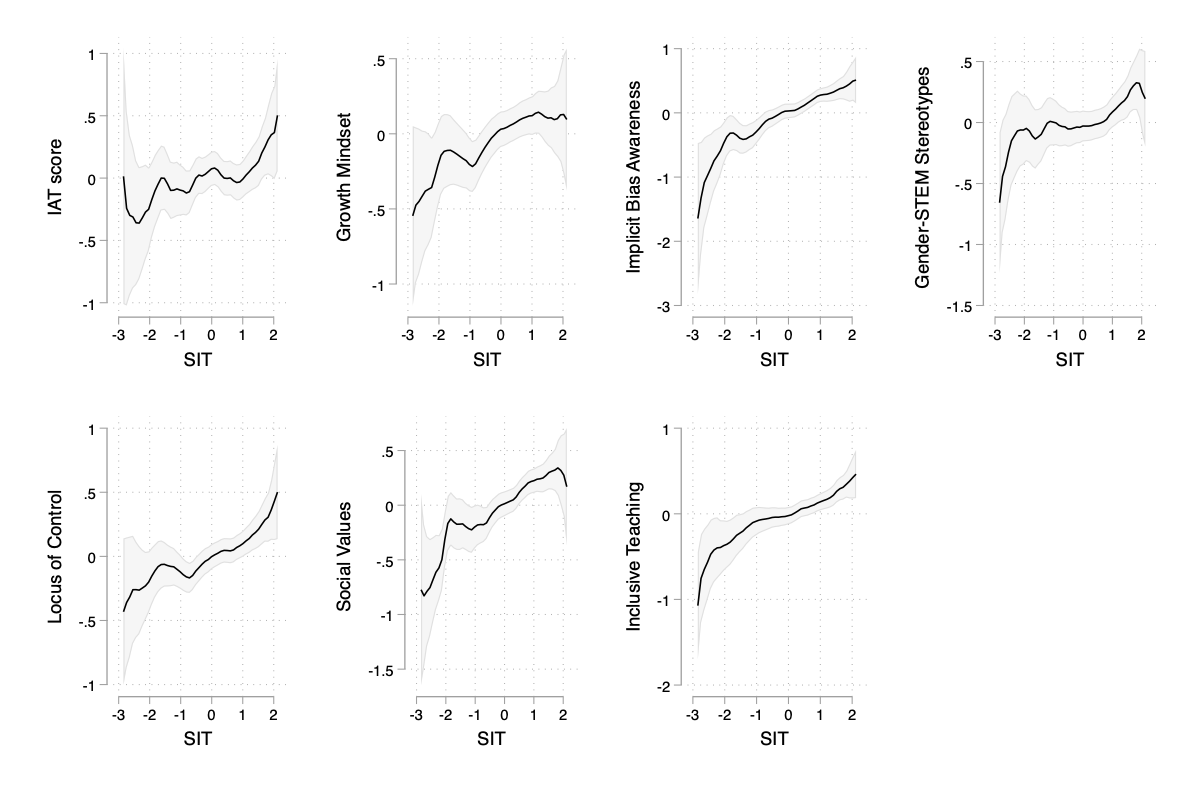}
\caption*{\footnotesize In the graph are depicted the local polynomial smooth of the multiple traits on the SIT (the dark line) with the respective confidence interval (95\%). In particular, in order, we observe: IAT score, growth mindset, implicit bias awareness, Gender-STEM stereotypes, locus of control, social values, and inclusive teaching practices.}
\end{figure}

\autoref{fig:sit combined} presents the non-linear association between various trait indices, derived from our questionnaires, against SIT scores. The indices include growth mindset, implicit bias awareness, Gender-STEM stereotypes, locus of control, social values, and inclusive teaching practices.

\section{Predicting SIT}\label{sec:regression}
\subsection{Feedback IAT}
\begin{figure}[H]
    \centering
\includegraphics[scale=0.65]{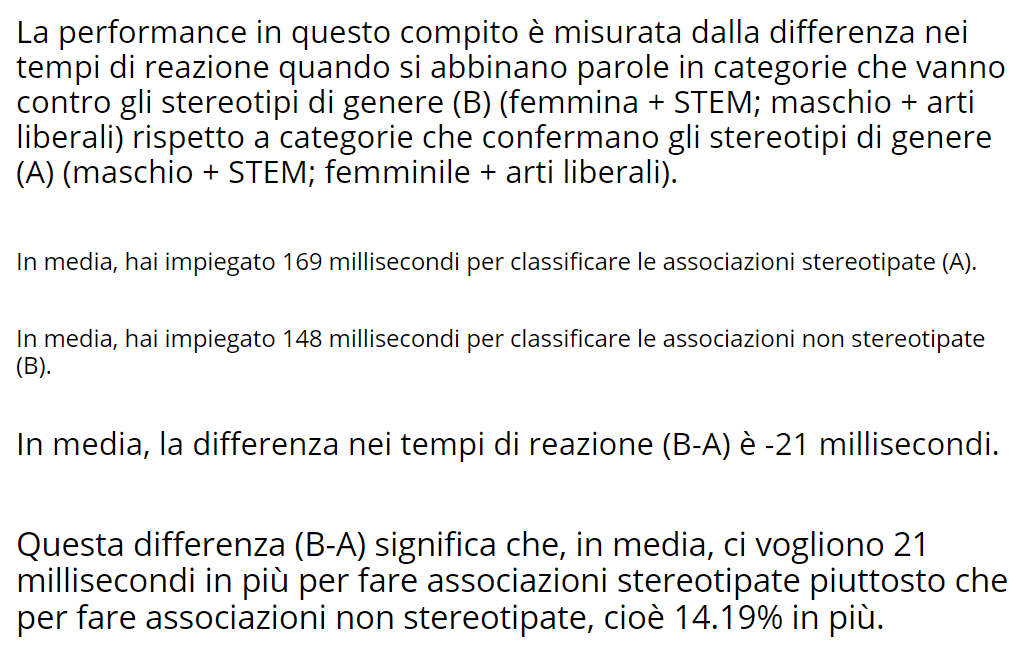}
\end{figure}
\subsection{Predicting SIT - Regression}

\begin{table}[htbp]\centering
\def\sym#1{\ifmmode^{#1}\else\(^{#1}\)\fi}
\caption{Predicting SIT}
\begin{tabular}{l*{6}{c}}
\hline\hline
                    &\multicolumn{1}{c}{(1)}&\multicolumn{1}{c}{(2)}&\multicolumn{1}{c}{(3)}&\multicolumn{1}{c}{(4)}&\multicolumn{1}{c}{(5)}&\multicolumn{1}{c}{(6)}\\
\hline
IAT Revelation      &       0.285\sym{***}&       0.277\sym{***}&       0.285\sym{***}&       0.243\sym{***}&       0.233\sym{***}&       0.219\sym{***}\\
                    &     (0.085)         &     (0.085)         &     (0.085)         &     (0.080)         &     (0.077)         &     (0.076)         \\
[1em]
IAT score           &                     &                     &       0.069         &       0.043         &       0.052         &       0.049         \\
                    &                     &                     &     (0.043)         &     (0.043)         &     (0.042)         &     (0.042)         \\
[1em]
Growth Mindset      &                     &                     &                     &       0.031         &       0.037         &       0.037         \\
                    &                     &                     &                     &     (0.046)         &     (0.045)         &     (0.044)         \\
[1em]
Implicit Bias Awareness&                     &                     &                     &       0.240\sym{***}&       0.274\sym{***}&       0.266\sym{***}\\
                    &                     &                     &                     &     (0.052)         &     (0.054)         &     (0.053)         \\
[1em]
Gender-STEM Stereotypes&                     &                     &                     &      -0.053         &      -0.042         &      -0.041         \\
                    &                     &                     &                     &     (0.044)         &     (0.044)         &     (0.043)         \\
[1em]
Locus of Control    &                     &                     &                     &       0.098\sym{*}  &       0.104\sym{*}  &       0.110\sym{**} \\
                    &                     &                     &                     &     (0.053)         &     (0.054)         &     (0.053)         \\
[1em]
Social Values       &                     &                     &                     &       0.168\sym{***}&       0.158\sym{***}&       0.144\sym{***}\\
                    &                     &                     &                     &     (0.054)         &     (0.055)         &     (0.053)         \\
[1em]
Inclusive Teaching  &                     &                     &                     &       0.125\sym{**} &       0.110\sym{**} &       0.105\sym{**} \\
                    &                     &                     &                     &     (0.051)         &     (0.051)         &     (0.050)         \\
[1em]
Gender              &                     &       0.304\sym{**} &       0.289\sym{**} &                     &       0.122         &       0.129         \\
                    &                     &     (0.121)         &     (0.122)         &                     &     (0.116)         &     (0.114)         \\
[1em]
Age                 &                     &       0.013\sym{***}&       0.013\sym{***}&                     &       0.013\sym{***}&       0.012\sym{***}\\
                    &                     &     (0.004)         &     (0.004)         &                     &     (0.004)         &     (0.004)         \\
[1em]
Like Teaching       &                     &      -0.062         &      -0.061         &                     &      -0.130\sym{***}&      -0.125\sym{***}\\
                    &                     &     (0.041)         &     (0.041)         &                     &     (0.037)         &     (0.037)         \\
[1em]
Master              &                     &       0.059         &       0.053         &                     &       0.077         &       0.041         \\
                    &                     &     (0.105)         &     (0.104)         &                     &     (0.096)         &     (0.096)         \\
[1em]
Disability training &                     &       0.122         &       0.115         &                     &       0.021         &       0.008         \\
                    &                     &     (0.093)         &     (0.093)         &                     &     (0.087)         &     (0.086)         \\
[1em]
Married             &                     &      -0.030         &      -0.033         &                     &      -0.074         &      -0.069         \\
                    &                     &     (0.086)         &     (0.086)         &                     &     (0.082)         &     (0.081)         \\
[1em]
\hline\hline
\end{tabular}

\label{tab:IAT_first_long}
\end{table}

\begin{table}[htbp]\centering
\def\sym#1{\ifmmode^{#1}\else\(^{#1}\)\fi}
\caption*{Predicting SIT - Cont}
\begin{tabular}{l*{6}{c}}
\hline\hline
            &\multicolumn{1}{c}{(1)}&\multicolumn{1}{c}{(2)}&\multicolumn{1}{c}{(3)}&\multicolumn{1}{c}{(4)}&\multicolumn{1}{c}{(5)}&\multicolumn{1}{c}{(6)}\\
\hline
Teaching Italian    &                     &       0.073         &       0.053         &                     &       0.003         &      -0.006         \\
                    &                     &     (0.086)         &     (0.086)         &                     &     (0.082)         &     (0.081)         \\
[1em]
Teaching Maths      &                     &       0.148         &       0.162         &                     &       0.223\sym{**} &       0.218\sym{**} \\
                    &                     &     (0.108)         &     (0.110)         &                     &     (0.104)         &     (0.103)         \\
[1em]
Island             &                     &      -0.024         &      -0.047         &                     &      -0.112         &      -0.129         \\
                    &                     &     (0.159)         &     (0.160)         &                     &     (0.148)         &     (0.145)         \\
[1em]
Missing             &                     &      -0.090         &      -0.109         &                     &      -0.067         &      -0.078         \\
                    &                     &     (0.191)         &     (0.192)         &                     &     (0.171)         &     (0.165)         \\
[1em]
North-East            &                     &      -0.038         &      -0.039         &                     &      -0.049         &      -0.088         \\
                    &                     &     (0.142)         &     (0.142)         &                     &     (0.133)         &     (0.129)         \\
[1em]
North-West         &                     &      -0.099         &      -0.102         &                     &      -0.067         &      -0.118         \\
                    &                     &     (0.129)         &     (0.128)         &                     &     (0.121)         &     (0.119)         \\
[1em]
South                &                     &      -0.200         &      -0.213         &                     &      -0.248\sym{*}  &      -0.266\sym{**} \\
                    &                     &     (0.141)         &     (0.141)         &                     &     (0.135)         &     (0.132)         \\
[1em]
Lexical Density     &                     &                     &                     &                     &                     &       0.673\sym{***}\\
                    &                     &                     &                     &                     &                     &     (0.235)         \\
[1em]
Constant            &      -0.095\sym{*}  &      -0.656\sym{*}  &      -0.634\sym{*}  &      -0.084\sym{*}  &       0.016         &      -0.301         \\
                    &     (0.049)         &     (0.364)         &     (0.364)         &     (0.047)         &     (0.336)         &     (0.358)         \\
\hline
N        &         614         &         614         &         614         &         614         &         614         &         614         \\
\hline\hline
\end{tabular}
\caption*{\footnotesize Dependent variable: SIT Score. The vector of indices $\textbf{W}_i$ includes: Implicit Bias Awareness, Locus of Control, Social Values, Inclusive Teaching, Growth Mindset, Gender-STEM Stereotypes.
The vector of controls $\textbf{X}_i$ includes: Age (continuous), Gender (0 = male, 1 = female), Like Teaching (7-points Likert Scale), Master degree Disability Training, Married, Teaching Italian, Teaching Mathematics, Place of Birth divided into North-East, North-West, Center, South, Island, Missing (Center as reference category). Standard errors in parentheses. \sym{*} \(p<0.10\), \sym{**} \(p<0.05\), \sym{***} \(p<0.01\)}
 \end{table}

\begin{table}[htbp]\centering
\def\sym#1{\ifmmode^{#1}\else\(^{#1}\)\fi}
\caption{SIT vs IAT - Cont}
\begin{tabular}{l*{2}{c}}
\hline\hline
                    &\multicolumn{1}{c}{(1)}&\multicolumn{1}{c}{(2)}\\
\hline
IAT score           &       0.042         &                     \\
                    &     (0.042)         &                     \\
[1em]
SIT Score &                     &       0.050         \\
                    &                     &     (0.049)         \\
[1em]
Growth Mindset      &       0.035         &      -0.019         \\
                    &     (0.044)         &     (0.054)         \\
[1em]
Implicit Bias Awareness&       0.277\sym{***}&       0.048         \\
                    &     (0.052)         &     (0.057)         \\
[1em]
Gender-STEM Stereotypes&      -0.045         &       0.018         \\
                    &     (0.043)         &     (0.044)         \\
[1em]
Locus of Control    &       0.102\sym{*}  &       0.053         \\
                    &     (0.053)         &     (0.054)         \\
[1em]
Social Values       &       0.149\sym{***}&      -0.014         \\
                    &     (0.053)         &     (0.060)         \\
[1em]
Inclusive Teaching  &       0.106\sym{**} &      -0.048         \\
                    &     (0.049)         &     (0.064)         \\
[1em]
Lexical Density     &       0.703\sym{***}&       0.081         \\
                    &     (0.236)         &     (0.210)         \\
[1em]
Gender              &       0.140         &       0.186         \\
                    &     (0.116)         &     (0.124)         \\
[1em]
Age                 &       0.012\sym{***}&      -0.002         \\
                    &     (0.004)         &     (0.005)         \\
[1em]
\hline\hline
\end{tabular}

\label{tab:SIT_vs_IAT_long}
\end{table}

\begin{table}[htbp]\centering
\def\sym#1{\ifmmode^{#1}\else\(^{#1}\)\fi}
\caption*{SIT vs IAT}
\begin{tabular}{l*{2}{c}}
\hline\hline
                    &\multicolumn{1}{c}{(1)}&\multicolumn{1}{c}{(2)}\\
\hline
Like Teaching       &      -0.132\sym{***}&      -0.003         \\
                    &     (0.037)         &     (0.042)         \\
[1em]
Master              &       0.030         &       0.076         \\
                    &     (0.096)         &     (0.098)         \\
[1em]
Disability training &       0.011         &       0.071         \\
                    &     (0.087)         &     (0.094)         \\
[1em]
Married             &      -0.071         &       0.042         \\
                    &     (0.081)         &     (0.087)         \\
[1em]
Teaching Italian    &      -0.001         &       0.261\sym{***}\\
                    &     (0.082)         &     (0.089)         \\
[1em]
Teaching Maths      &       0.225\sym{**} &      -0.197\sym{*}  \\
                    &     (0.102)         &     (0.108)         \\
[1em]
Isole               &      -0.145         &       0.336\sym{*}  \\
                    &     (0.145)         &     (0.173)         \\
[1em]
Missing             &      -0.074         &       0.279         \\
                    &     (0.168)         &     (0.195)         \\
[1em]
Nord-Est            &      -0.070         &       0.003         \\
                    &     (0.130)         &     (0.139)         \\
[1em]
Nord-Ovest          &      -0.113         &       0.042         \\
                    &     (0.120)         &     (0.124)         \\
[1em]
Sud                 &      -0.252\sym{*}  &       0.175         \\
                    &     (0.133)         &     (0.135)         \\
[1em]
Constant            &      -0.187         &      -0.352         \\
                    &     (0.363)         &     (0.387)         \\
\hline
Observations        &         614         &         614         \\
\hline\hline
\end{tabular}
\caption*{\footnotesize In regression (1) the dependent variable is the SIT score. In regression (2) the dependent variable is the IAT score. The vector of indices $\textbf{W}_i$ includes: Implicit Bias Awareness, Locus of Control, Social Values, Inclusive Teaching, Growth Mindset, Gender-STEM Stereotypes. The vector of controls $\textbf{X}_i$ includes: Age (continuous), Gender (0 = male, 1 = female), Like Teaching (7-points Likert Scale), Master degree Disability Training, Married, Teaching Italian, Teaching Mathematics, Place of Birth divided into North-East, North-West, Center, South, Island, Missing (Center as reference category). Standard errors in parentheses. \sym{*} \(p<0.10\), \sym{**} \(p<0.05\), \sym{***} \(p<0.01\)}
\end{table}

\begin{table}[htbp]\centering
\def\sym#1{\ifmmode^{#1}\else\(^{#1}\)\fi}
\caption{Framing on SIT}

\begin{tabular}{l*{3}{c}}
\hline\hline
                    &\multicolumn{1}{c}{(1)}&\multicolumn{1}{c}{(2)}&\multicolumn{1}{c}{(3)}\\
\hline
Info Framing              &       0.065         &       0.035         &       0.091         \\
                    &     (0.099)         &     (0.097)         &     (0.089)         \\
[1em]
No framing          &      -0.051         &      -0.054         &      -0.001         \\
                    &     (0.098)         &     (0.096)         &     (0.089)         \\
[1em]
IAT Revelation      &                     &       0.276\sym{***}&       0.217\sym{***}\\
                    &                     &     (0.085)         &     (0.076)         \\
[1em]
IAT score           &                     &                     &       0.049         \\
                    &                     &                     &     (0.042)         \\
[1em]
Growth Mindset      &                     &                     &       0.036         \\
                    &                     &                     &     (0.043)         \\
[1em]
Implicit Bias Awareness&                     &                     &       0.266\sym{***}\\
                    &                     &                     &     (0.053)         \\
[1em]
Gender-STEM Stereotypes&                     &                     &      -0.042         \\
                    &                     &                     &     (0.043)         \\
[1em]
Locus of Control    &                     &                     &       0.108\sym{**} \\
                    &                     &                     &     (0.053)         \\
[1em]
Social Values       &                     &                     &       0.148\sym{***}\\
                    &                     &                     &     (0.053)         \\
[1em]
Inclusive Teaching  &                     &                     &       0.107\sym{**} \\
                    &                     &                     &     (0.050)         \\
[1em]
Like Teaching       &                     &      -0.057         &      -0.122\sym{***}\\
                    &                     &     (0.042)         &     (0.037)         \\
[1em]
Gender              &                     &       0.305\sym{**} &       0.127         \\
                    &                     &     (0.122)         &     (0.115)         \\
[1em]
Age                 &                     &       0.013\sym{***}&       0.012\sym{***}\\
                    &                     &     (0.004)         &     (0.004)         \\
[1em]
Master              &                     &       0.057         &       0.039         \\
                    &                     &     (0.105)         &     (0.096)         \\
[1em]
Disability training &                     &       0.124         &       0.013         \\
                    &                     &     (0.093)         &     (0.086)         \\
\hline\hline
\end{tabular}

 \label{tab:Framing}
\end{table}

\begin{table}[htbp]\centering
\def\sym#1{\ifmmode^{#1}\else\(^{#1}\)\fi}
\caption*{Framing on SIT - Cont}
\begin{tabular}{l*{3}{c}}
\hline\hline
                    &\multicolumn{1}{c}{(1)}&\multicolumn{1}{c}{(2)}&\multicolumn{1}{c}{(3)}\\
\hline
Married             &                     &      -0.028         &      -0.067         \\
                    &                     &     (0.086)         &     (0.080)         \\
[1em]
Teaching Italian    &                     &       0.073         &      -0.005         \\
                    &                     &     (0.086)         &     (0.081)         \\
[1em]
Teaching Maths      &                     &       0.149         &       0.216\sym{**} \\
                    &                     &     (0.108)         &     (0.103)         \\
[1em]
Island              &                     &      -0.032         &      -0.139         \\
                    &                     &     (0.160)         &     (0.147)         \\
[1em]
Missing             &                     &      -0.096         &      -0.080         \\
                    &                     &     (0.191)         &     (0.165)         \\
[1em]
North-East            &                     &      -0.046         &      -0.098         \\
                    &                     &     (0.142)         &     (0.129)         \\
[1em]
North-West          &                     &      -0.102         &      -0.118         \\
                    &                     &     (0.128)         &     (0.119)         \\
[1em]
South              &                     &      -0.204         &      -0.269\sym{**} \\
                    &                     &     (0.141)         &     (0.133)         \\
[1em]
Lexical Density     &                     &                     &       0.671\sym{***}\\
                    &                     &                     &     (0.236)         \\
[1em]
Constant            &      -0.004         &      -0.667\sym{*}  &      -0.335         \\
                    &     (0.069)         &     (0.366)         &     (0.360)         \\
\hline
Observations        &         614         &         614         &         614         \\
\hline\hline
\end{tabular}
\caption*{\footnotesize Dependent variable: SIT Score. Omitted category in framing: Info + Guilt.The vector of indices $\textbf{W}_i$ includes: Implicit Bias Awareness, Locus of Control, Social Values, Inclusive Teaching, Growth Mindset, Gender-STEM Stereotypes. The vector of controls $\textbf{X}_i$ includes: Age (continuous), Gender (0 = male, 1 = female), Like Teaching (7-points Likert Scale), Master degree Disability Training, Married, Teaching Italian, Teaching Mathematics, Place of Birth divided into North-East, North-West, Center, South, Island, Missing (Center as reference category). Standard errors in parentheses. \sym{*} \(p<0.10\), \sym{**} \(p<0.05\), \sym{***} \(p<0.01\)}
\end{table}

\begin{table}[htbp]\centering
\def\sym#1{\ifmmode^{#1}\else\(^{#1}\)\fi}
\caption{Robustness Check}
\begin{tabular}{l*{3}{c}}
\hline\hline
                    &\multicolumn{1}{c}{(1)}&\multicolumn{1}{c}{(2)}&\multicolumn{1}{c}{(3)}\\
                     & \multicolumn{1}{c}{SIT Score}&\multicolumn{1}{c}{SIT Score Sd}&\multicolumn{1}{c}{SIT Score Factor}\\
\hline
IAT Revelation      &       0.219\sym{***}&       0.214\sym{***}&       0.208\sym{***}\\
                    &     (0.076)         &     (0.077)         &     (0.073)         \\
[1em]
IAT score           &       0.049         &       0.047         &       0.045         \\
                    &     (0.042)         &     (0.042)         &     (0.040)         \\
[1em]
Growth Mindset      &       0.037         &       0.036         &       0.037         \\
                    &     (0.044)         &     (0.043)         &     (0.042)         \\
[1em]
Implicit Bias Awareness&       0.266\sym{***}&       0.268\sym{***}&       0.253\sym{***}\\
                    &     (0.053)         &     (0.053)         &     (0.051)         \\
[1em]
Gender-STEM Stereotypes&      -0.041         &      -0.040         &      -0.037         \\
                    &     (0.043)         &     (0.043)         &     (0.041)         \\
[1em]
Locus of Control    &       0.110\sym{**} &       0.111\sym{**} &       0.103\sym{**} \\
                    &     (0.053)         &     (0.053)         &     (0.051)         \\
[1em]
Social Values       &       0.144\sym{***}&       0.142\sym{***}&       0.135\sym{***}\\
                    &     (0.053)         &     (0.053)         &     (0.050)         \\
[1em]
Inclusive Teaching  &       0.105\sym{**} &       0.106\sym{**} &       0.100\sym{**} \\
                    &     (0.050)         &     (0.050)         &     (0.048)         \\
[1em]
Gender              &       0.129         &       0.126         &       0.125         \\
                    &     (0.114)         &     (0.115)         &     (0.109)         \\
[1em]
Age                 &       0.012\sym{***}&       0.012\sym{***}&       0.012\sym{***}\\
                    &     (0.004)         &     (0.004)         &     (0.004)         \\
[1em]
Like Teaching       &      -0.125\sym{***}&      -0.125\sym{***}&      -0.119\sym{***}\\
                    &     (0.037)         &     (0.037)         &     (0.035)         \\
[1em]
Master              &       0.041         &       0.042         &       0.043         \\
                    &     (0.096)         &     (0.096)         &     (0.092)         \\
[1em]
\hline\hline
\end{tabular}
\label{tab:Robustness}
\end{table}

\begin{table}[htbp]\centering
\def\sym#1{\ifmmode^{#1}\else\(^{#1}\)\fi}
\caption*{Robustness Check - Cont}
\begin{tabular}{l*{4}{c}}
\hline\hline
            &\multicolumn{1}{c}{(1)}&\multicolumn{1}{c}{(2)}&\multicolumn{1}{c}{(3)}\\
\hline
Teaching Italian    &      -0.006         &      -0.007         &      -0.007         \\
                    &     (0.081)         &     (0.081)         &     (0.077)         \\
[1em]
Teaching Maths      &       0.218\sym{**} &       0.220\sym{**} &       0.208\sym{**} \\
                    &     (0.103)         &     (0.102)         &     (0.098)         \\
[1em]
Island               &      -0.129         &      -0.121         &      -0.112         \\
                    &     (0.145)         &     (0.146)         &     (0.139)         \\
[1em]
Missing             &      -0.078         &      -0.065         &      -0.067         \\
                    &     (0.165)         &     (0.167)         &     (0.158)         \\
[1em]
North-East            &      -0.088         &      -0.086         &      -0.086         \\
                    &     (0.129)         &     (0.129)         &     (0.123)         \\
[1em]
North-West         &      -0.118         &      -0.112         &      -0.113         \\
                    &     (0.119)         &     (0.119)         &     (0.113)         \\
[1em]
South                 &      -0.266\sym{**} &      -0.260\sym{*}  &      -0.252\sym{**} \\
                    &     (0.132)         &     (0.133)         &     (0.126)         \\
[1em]
Lexical Density     &       0.673\sym{***}&       0.675\sym{***}&       0.643\sym{***}\\
                    &     (0.235)         &     (0.237)         &     (0.224)         \\
[1em]
Constant            &      -0.301         &      -0.309         &      -0.294         \\
                    &     (0.358)         &     (0.359)         &     (0.343)         \\
\hline
N       &         614         &         614         &         614         \\
\hline\hline
\end{tabular}

\caption*{\footnotesize Dependent variable (1): standard SIT score. Dependent variable (2): SIT score adjusted for the standard deviation of the image. Dependent variable (3): SIT score computed as the main factor of a factor analysis. The vector of indices $\textbf{W}_i$ includes: Implicit Bias Awareness, Locus of Control, Social Values, Inclusive Teaching, Growth Mindset, Gender-STEM Stereotypes.
The vector of controls $\textbf{X}_i$ includes: Age (continuous), Gender (0 = male, 1 = female), Like Teaching (7-points Likert Scale), Master degree Disability Training, Married, Teaching Italian, Teaching Mathematics, Place of Birth divided into North-East, North-West, Center, South, Island, Missing (Center as reference category). Standard errors in parentheses. \sym{*} \(p<0.10\), \sym{**} \(p<0.05\), \sym{***} \(p<0.01\)}
\end{table}

\section{Psychometric Properties of the Stereotype Identification Test}\label{sec:psychmetrics}
In the following section, we probe the robustness of the SIT by evaluating whether it satisfies two of the most important psychometric properties of a test: validity and reliability.

\subsection{Validity}\label{sec:validity} %

In this section, we provide complementary evidence supporting the validity of the SIT. We show that the proposed SIT score constitutes a valid measure of the ability to see stereotypes in an educational setting by providing evidence based on (i) test content, (ii) relations to other variables, (iii) response processes, and (iv) internal structure.

\subsubsection{Evidence based on test content} 
In this section, we describe how the content of our test relates to the main construct we claim can be evaluated with our test, namely the ability to detect stereotypes. The logic of our test exposes people to images that may be stereotypical to different extent. If images are recognized as stereotypical, people can use the 5-point Likert scale to assess how stereotypical it is. For the test to work, the selection of images is crucial and should depict a wide range of different stereotypes. 

Since stereotypes are everywhere, can be about anything, and differ from one culture to another, it is impossible to sample content from all possible stereotypes. However, not all of them have the same importance in terms of diffusion and consequences \citep{bordalo_stereotypes_2016}. Most countries have anti-discrimination legislation based on a list of protected characteristics. For example, Article 21 of the Charter of Fundamental Rights of the European Union lists sex, color, ethnic or social origin, genetic features, language, religion or belief, political or any other opinion, membership of a national minority, property, birth, disability, age, and sexual orientation as protected characteristics for which discrimination is prohibited. 
Most of these characteristics can be inferred from a picture, and are therefore included in the set of 100 images that we chose for our test; special attention has been given to characteristics that are most frequently encountered by primary and secondary school teachers. 


The set of 100 images used for this test were carefully selected by a communication expert drawing from a greater image bank owned by an editorial company that used the images as illustrations for school books. She was given instructions to select images to cover a wide range of situations for the previously mentioned protected characteristics. She based her selection on visual cues and her experience in designing content for children's textbooks. 

\begin{figure}
\caption{Example of a picture having 49 tags.}\label{fig:tag}
    \centering
    \includegraphics[width=0.5\linewidth]{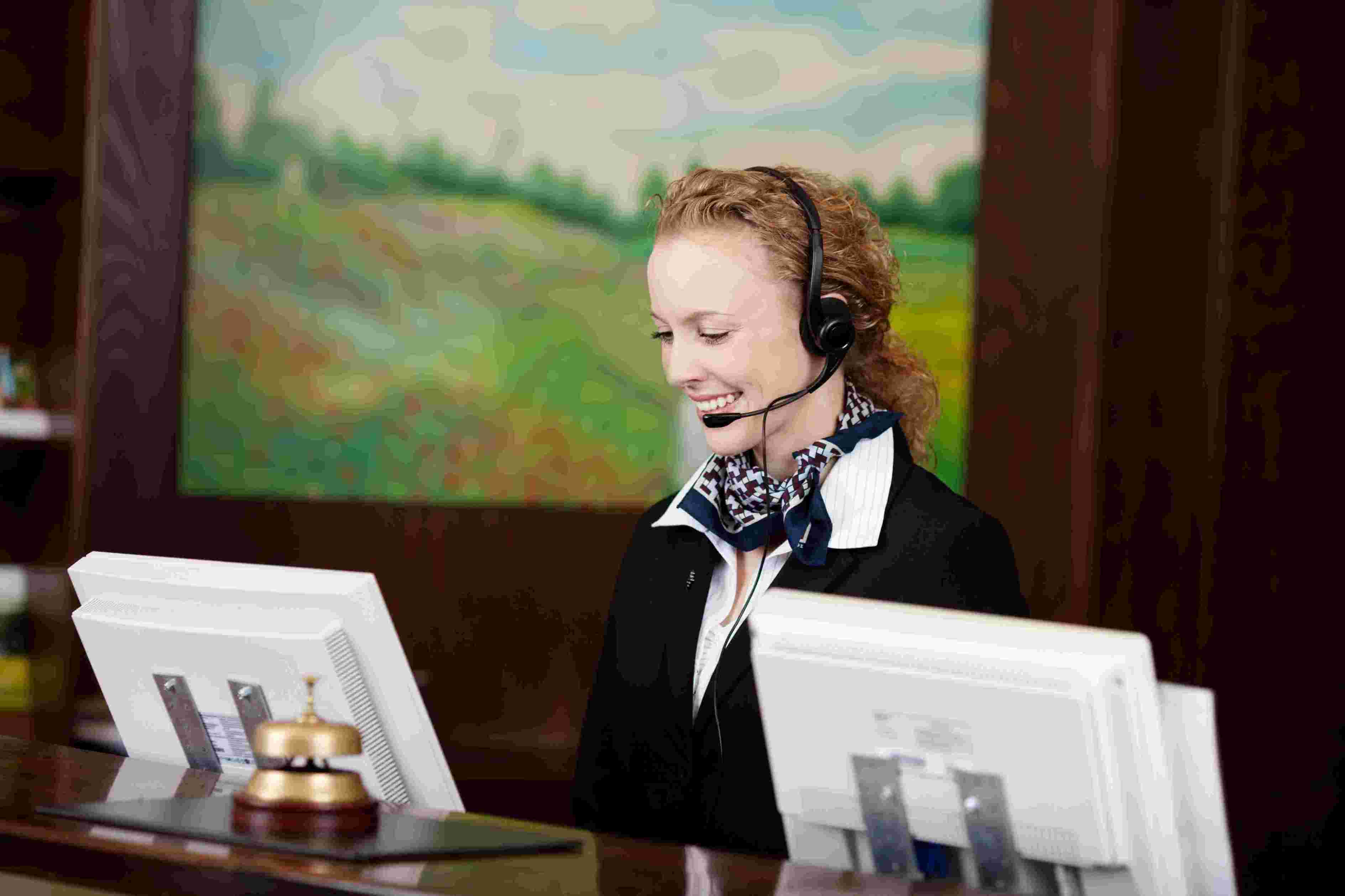}
\end{figure}

To complement this approach, we also exploit the tags associated with each image. 
Each image is associated with a list of tags to facilitate the search of editors when designing books. They are meant to describe the content of the images. For instance, the image in \autoref{fig:tag} has 49 tags associated with it, which describe the individuals involved (caucasian, confident, elegant, employee, female, girl, person, woman, young), what they are doing or their activity (arrival, assistance, call, check-in, desk clerk, executive, holiday, hotel, lobby, reception, receptionist, representative, reservation, resort, service, smile, talking, travel, vacation, welcome, work), objects they might be interacting with or that are present in the scene (bell, computer, costume, counter, desk, front desk, headphone, headset, phone, table, telephone), some other elements of context (corporate, entry, formal, indoor, professional, reception room, workplace), and some other elements related to the picture itself (portrait, technology).
Images have between 18 and 52 tags, with an average of 44 tags.\footnote{These statistics are computed excluding two images that have no tags at all.} 

We use a text classification analysis to connect the tag description to potentially protected characteristics. The goal is to automatically detect the presence of protected characteristics from the tags and to quantify the breadth of protected characteristics that are present in our set of 100 images.
We use a zero-shot text classification approach using the pre-trained Hugging Face model. 
A text classification model associates a text with categories. 
Zero-shot means that the trained model has never been exposed to examples of the associations we want to detect, a very conservative approach.

The list of categories to be predicted is gender, race, social origin, religion, disability, and age. \footnote{Starting from the list of protected characteristics (sex, race, colour, ethnic or social origin, religion, disability, and age), to facilitate the algorithm, we relabel ``sex" into ``gender" and we regroup race, colour, and ethnic origin into a single category called ``race," as skin colour is the most visual defining element that can be depicted in images; other elements separating these concepts, such as language spoken at home or cultural customs, are not easily detectable in images, blurring the frontier between this concepts.}

At the end of the process, the model assigns to every tag probability weights (summing to one) of belonging to each category. We then define a tag as representing a particular category if the probability of belonging to this category is greater than 50\%. If no probability is higher than 50\%, we consider the tag to not reflect any protected characteristic. With such a threshold, the modal category is necessary unique. From a sample of 2189 unique tags, 347 tags are classified into the six categories after reviewing each classification manually.\footnote{This manual review allowed us to exclude from the list tags that were, for instance, incorrectly associated with race due to their association with racing instead of racism.}

\begin{figure}
\caption{ \label{fig:tags_images}}
    \centering
    \includegraphics[width=0.5\linewidth]{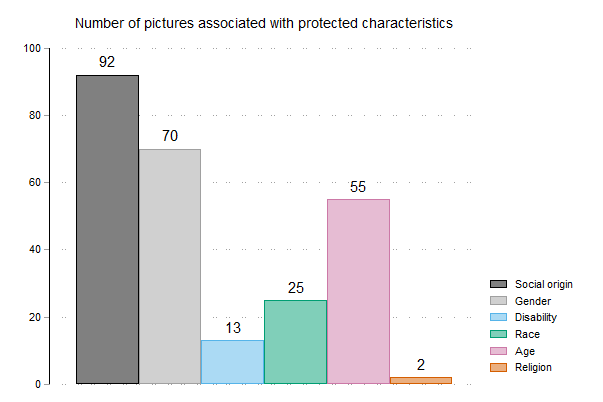}
    \caption*{\footnotesize Number of images containing a tag associated with the six protected characteristics that we consider (gender, race, social origin, religion, disability, and age) based on the output of a zero-shot text classification approach using the pre-trained Hugging Face model.}
   
\end{figure}

When matching these tags for protected characteristics with the specific lists of each image, only five images do not have tags for protected characteristics, among which two do not have tags at all. On average, images have 8.9 tags for protected characteristics (between 1 and 19) covering 2.7 protected characteristics  (between 1 and 5). The tags for protected characteristics represent between 2\% and 50\% of all the tags in an image, with an average of 20\%. 
\autoref{fig:tags_images} shows the number of images associated with each of the six protected characteristics that we consider. 

Virtually all images are associated to the ``social origin'' characteristic, partly because the model classified some tags like ``Caucasian", ``Asians", or ``multiracial group" as social origin.
Racial traits and social conditions are often related when discussing intersectionality. 
Race is represented in 25 images. 
Gender and age are very common characteristics in our images, which is not surprising considering that every image contains at least one person. 
Disability and religion are present but much less frequently.

\subsubsection{Evidence based on relations to other variables} 
The test aims to measure a single construct: individuals' ability to detect stereotypes. 
However, detecting stereotypes requires an understanding of what stereotypes are, a task that is complicated by the stereotypes and personal values that individuals themselves hold. Consequently, the ability to detect stereotypes is likely influenced by other established constructs, such as implicit bias, personal attitudes, and awareness of implicit bias.

In \autoref{sec:predict}, we demonstrate that our SIT measure is systematically related to several self-reported individual traits and attitudes, including growth mindset, beliefs about gender and STEM, locus of control, responses from the social value survey, and inclusive teaching practices. This finding suggests that the ability to detect stereotypes, as measured by the SIT, is meaningfully connected to broader cognitive and behavioral constructs that influence decision-making and interactions in educational settings.  

In contrast, the Implicit Association Test (IAT), while a widely used tool, primarily captures automatic processes that operate just below the level of conscious awareness and often show weak or inconsistent relationships with other self-reported traits. The significant and systematic associations we observe between SIT and these other measures provide further evidence supporting the validity of our construct. This indicates that the SIT captures a skill that is not only relevant to stereotype detection but also reflects an interplay with broader psychological and social dimensions, reinforcing its utility for understanding individual differences in educational contexts.

\subsubsection{Evidence based on response processes} 
Evidence based on response processes is a critical but often overlooked aspect of validity \citep{borsboom2004, cronbach1955, embretson1984, embretson1993, embretson1998, embretson2016}. 
It examines whether the cognitive processes test-takers use to respond align with the construct being measured. For example, answering a long-division question should involve performing the division process as hypothesized. This validation involves analyzing strategies, processes, and knowledge used during responses, ensuring the test accurately measures the intended construct \citep{urbina2004}.

Our SIT test comprises two steps. In the first step, participants rate images based on the degree of stereotype they perceive in them. In the second step, participants have the option to explain their ratings by providing written comments. This qualitative component offers deeper insight into the cognitive processes underlying their evaluations. Additionally, we record reaction times separately for both steps, providing an indirect measure of the thought process and cognitive effort involved in making these judgments.

While comments were optional, participants frequently chose to provide them. Out of 12,901 ratings, 10,747 were accompanied by comments, amounting to responses for over 83\% of the evaluated images. Notably, for the six Gender-STEM images that all participants assessed, teachers provided comments 3,295 times out of 3,870 ratings, corresponding to an 85\% response rate. Most participants commented on all six images, while the remainder was evenly split between those who commented on some and those who provided no comments at all. 
%

On average, the comments consist of 10 words, predominantly written in the present tense (68\%, compared to 30\% in the past tense). They are mostly in the indicative mood (92\%) and typically use the third person (singular or plural) to describe the individuals depicted in the images. The most frequently used words provide insight into participants’ engagement with the task and their appropriate reactions to the test content. Common terms include \textit{male} (508 occurrences), \textit{female} (199), \textit{man} (496), \textit{woman} (445), \textit{math} (321), \textit{dance} (255), \textit{dockyard} (201), \textit{science} (185), \textit{stereotype} (171), \textit{scientific} (144), \textit{astronaut} (99), and \textit{white} (70).

When examining reaction times (\autoref{fig:classification}), we observe that images requiring longer rating times were generally less stereotypical. Conversely, images evaluated more quickly tended to be more stereotypical. This relationship aligns with two possible interpretations. First, when stereotypes are harder to detect, participants may invest more effort into determining the appropriate rating for an image. 
Second, as mentioned above, a ``Where's Wally" effect may occur, where participants spend additional time searching for a stereotype when they perceive none, resulting in extended viewing times. 
In both scenarios, this behavior reflects participants actively engaging with their knowledge of stereotypes and comparing it to the depicted situations. 

Together with the text analysis of the comments, these findings offer compelling evidence of participant engagement and active involvement in the test. The observed patterns in reaction times further suggest that participants were not merely providing superficial responses but were thoughtfully evaluating the images based on their understanding. Taken together, this evidence supports the strong validity of the test, particularly with respect to the response processes involved in stereotype recognition.

\subsubsection{Evidence based on internal structure} 
Evidence relating to the internal structure of a measure refers to the relationships between the individual items and the underlying construct that the test is intended to measure. According to the Standards for Educational and Psychological Testing \citep{AERA2014}, a key aspect of internal structure is the organization of items, which should reflect the construct they are intended to assess. For instance, if a test is designed to measure a unidimensional construct, such as stereotype detection, the items should exhibit high correlations and load onto a single factor. Alternatively, if the test measures multiple subcomponents of a broader construct, we would expect the items to be organized into distinguishable factors that reflect these subcomponents. Examining the internal structure of a measure through statistical techniques, such as factor analysis, allows for the evaluation of whether the test items align with the hypothesized construct.

In our study, we conducted both exploratory and confirmatory factor analyses to assess the internal structure of our Stereotype Identification Test (SIT). We treated the ratings $\widetilde{SIT}_i$ of the 20 images as individual items and examined the factor loadings to determine whether they converged onto a single underlying factor, as predicted by the test's design. The results from the factor analyses, shown in Table \ref{tab:factor}, provided strong evidence for a unidimensional structure, confirming that the SIT successfully captures the construct of stereotype detection. This finding supports the validity of the test, indicating that the ratings of the images reflect a consistent underlying cognitive process, in line with the theoretical framework of stereotype recognition.

\begin{table}
\centering
    \caption{Factor Loadings}
{
\def\sym#1{\ifmmode^{#1}\else\(^{#1}\)\fi}
\begin{tabular}{l*{1}{cc}}
\hline\hline
    Variables        &    Factor 1&         Uniqueness\\
\hline
        $\widetilde{SIT}_1$  & 0.626 & 0.551 \\
        $\widetilde{SIT}_2$  & 0.609 & 0.575 \\
        $\widetilde{SIT}_3$  & 0.556 & 0.622 \\
        $\widetilde{SIT}_4$  & 0.568 & 0.645 \\
        $\widetilde{SIT}_5$  & 0.541 & 0.661 \\
        $\widetilde{SIT}_6$  & 0.497 & 0.654 \\
        $\widetilde{SIT}_7$  & 0.607 & 0.577 \\
        $\widetilde{SIT}_8$  & 0.578 & 0.619 \\
        $\widetilde{SIT}_9$  & 0.588 & 0.601 \\
        $\widetilde{SIT}_{10}$ & 0.579 & 0.590 \\
        $\widetilde{SIT}_{11}$ & 0.604 & 0.595 \\
        $\widetilde{SIT}_{12}$ & 0.605 & 0.552 \\
        $\widetilde{SIT}_{13}$ & 0.604 & 0.581 \\
        $\widetilde{SIT}_{14}$ & 0.572 & 0.617 \\
        $\widetilde{SIT}_{15}$ & 0.576 & 0.624 \\
        $\widetilde{SIT}_{16}$ & 0.562 & 0.589 \\
        $\widetilde{SIT}_{17}$ & 0.571 & 0.615 \\
        $\widetilde{SIT}_{18}$ & 0.581 & 0.619 \\
        $\widetilde{SIT}_{19}$ & 0.547 & 0.627 \\
        $\widetilde{SIT}_{20}$ & 0.545 & 0.661 \\
\hline\hline
\end{tabular}
}

\label{tab:factor}
\caption*{\footnotesize Factor loadings and uniquenesses from a confirmatory factor analysis imposing a single underlying factor. 
Each item $\widetilde{SIT}_i$ refers to the demeaned rating of every image, as describe in equation \ref{eq:SIT}.}
\end{table}

\subsection{Reliability}\label{sec:reliability}

Reliability can be defined as ``the consistency of scores over repeated applications of a test across conditions that can include test forms, items, occasions, and raters.''\citep{kempf-leonard_split-half_2005}. 
The two key forms of reliability are split-half reliability and test-retest reliability. Split-half reliability pertains to the consistency of measurement across different items, while test-retest reliability concerns the consistency of measurement over time. 

Split-half reliability is closely linked to internal consistency and is usually measured using the Cronbach's alpha statistic, which represents the average reliability across all possible item splits \citep{warrens_cronbachs_2015}.

Test-retest reliability involves repeating the same test in the same conditions on the same set of participants on two different occasions. If the test is reliable, participants should perform very similarly in the two tests. 

Performing a test on one occasion may influence performance on the same test later, as participants might improve simply from knowing the task better. To address this, one option is to have participants take two similar tests on the same occasion, with only minor differences between them—this would be a parallel form of the original test. However, creating parallel forms can be challenging and costly. An alternative is split-half reliability, where the test items are randomly divided into two halves, each acting as a pseudo-parallel form. If the test is reliable, participants should score similarly on both halves.

There are several methods to split the halves of a test \citep[for a review, see][]{pronk_methods_2022}. The simplest method, called first-second splitting, assigns the first half of observations to the first group and the second half to the second group based on their order in the trial sequence. However, this approach may be influenced by learning or fatigue effects. To mitigate this, randomization is often used. For instance, one could assign odd trials to one half and even trials to the other (odd-even splitting), or use a simulation approach to generate multiple random splits.

Given the test design, where participants rate different sets of images, calculating Cronbach's alpha directly is challenging. Additionally, test-retest reliability cannot be evaluated on this sample since participants only completed the SIT once.

To address these challenges, we use a simulation approach based on resampling. When resampling is done \textit{without} replacement, the mean of the simulated split-half reliability coefficients converges to Cronbach's alpha. When resampling is done \textit{with} replacement, the mean of the simulated split-half reliability coefficients converges to the test-retest reliability coefficient \citep{williams_reliability_2012,pronk_methods_2022}. 

We also supplement this simulation with traditional test-retest analysis, based on a pilot sample of teachers who completed the SIT twice, with a five-month interval between tests. In the following sections, we further explain the split-half and test-retest reliability assessments and provide additional evidence supporting the simulation approach.

\subsubsection{Split-half reliability}

\begin{figure}
 \caption{Split-half reliability, all images}
    \centering
    \includegraphics[width=0.8\linewidth]{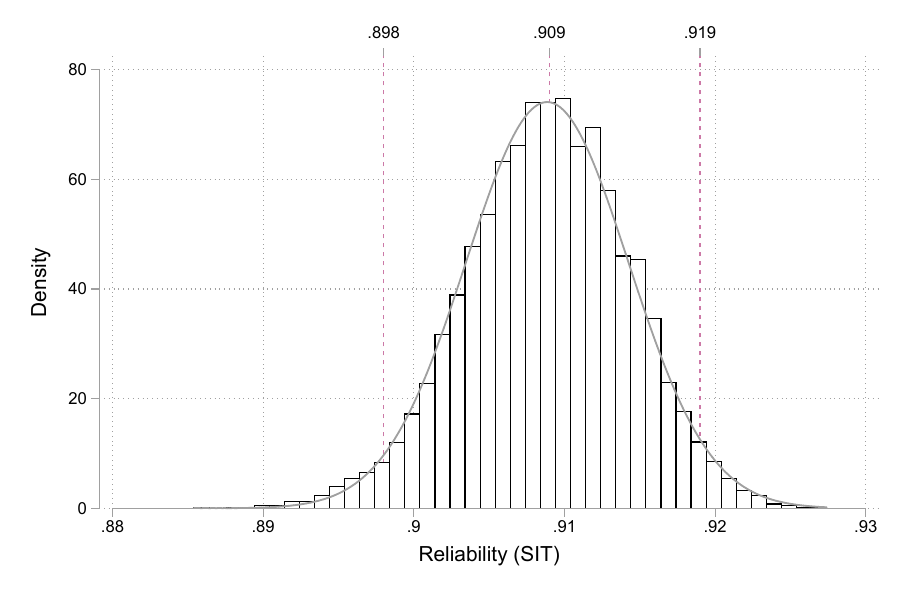}
    \label{fig:simresults_splithalf2}
\end{figure}

\begin{figure}
 \caption{Split-half reliability, Gender-STEM images}
    \centering
    \includegraphics[width=0.8\linewidth]{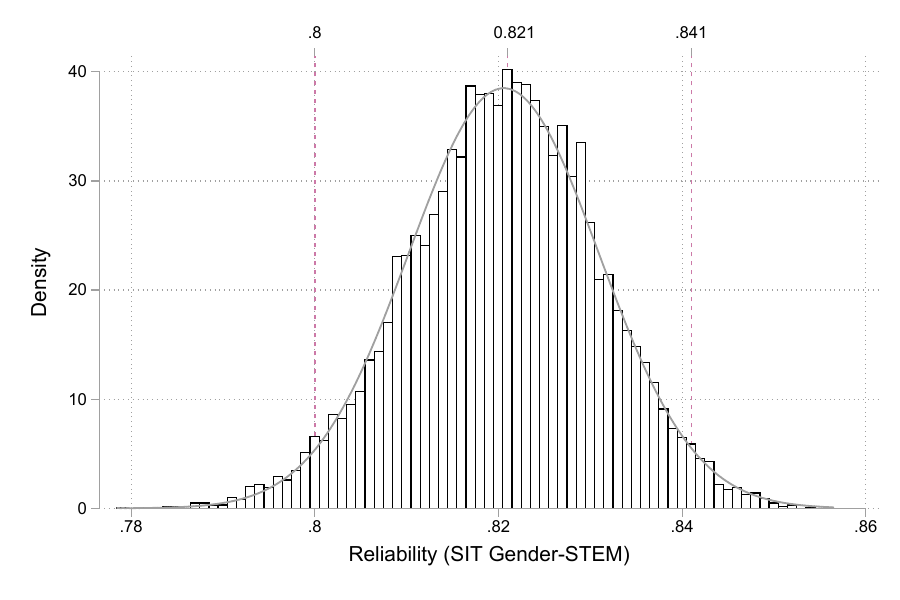}
    \label{fig:simresults_splithalf_genSTEM}
\end{figure}

Following \cite{williams_reliability_2012,pronk_methods_2022}, we implement a simulation approach to estimate split-half reliability in which we consider the series of 20 images rated by the teachers as the pool from which we are resampling without replacement. This is equivalent to randomly generating alternative orders (within teachers) for the 20 pictures that each teacher has already rated. Then, we split the 20 pictures into two halves based on this random order, the first ten images are assigned to the first half and the last ten images to the second half. We measure the Pearson correlation between the first and second halves and adjust it using the Spearman-Brown formula \citep{kempf-leonard_split-half_2005}: 

\begin{equation*}
    R_{adjusted} = \dfrac{2 r_{1,2}}{1+r_{1,2}}
\end{equation*}

where $R_{adjusted}$ is the adjusted correlation coefficient that will act as our measure of reliability, and $r_{1,2}$ is the Pearson correlation between the SIT scores of the first and second halves. We then repeat this process 9999 times. In \autoref{fig:simresults_splithalf2}, we report the distribution of the simulated reliability coefficients with the associated mean and 2.5\% and 97.5\% empirical quantiles. In the same fashion, we report the results of simulating random splits for the six Gender-STEM images that every teacher rated in \autoref{fig:simresults_splithalf_genSTEM}. 

In both cases, reliability is pretty high since we are slightly above 0.9 (0.91) for the series of 20 images and above 0.8 (0.82) for the Gender-STEM images alone. Even the variance is quite small as confidence intervals place these two reliability measures at 0.8 at the lowest with a 95\% probability. Following \cite{rust_modern_2020}, these estimates are very much in line with psychometric standards for a reliable test, which should be between 0.8 and 0.9. In addition, for the Gender-STEM SIT score, the mean value of 0.82 also corresponds (as described above) to the Cronbach's alpha computed for these six images only.  

\subsubsection{Test-retest reliability}

\begin{figure}
\caption{Test-retest reliability, all images}
    \centering
    \includegraphics[width=0.9\linewidth]{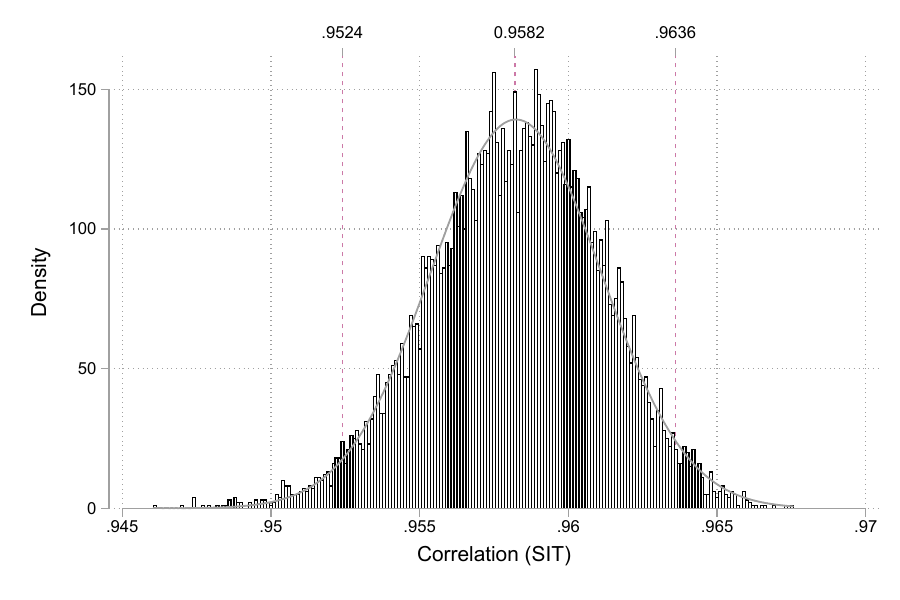}
    \label{fig:simresults_retest2}
\end{figure}

\begin{figure}
\caption{Test-retest reliability, Gender-STEM images}
    \centering
    \includegraphics[width=0.9\linewidth]{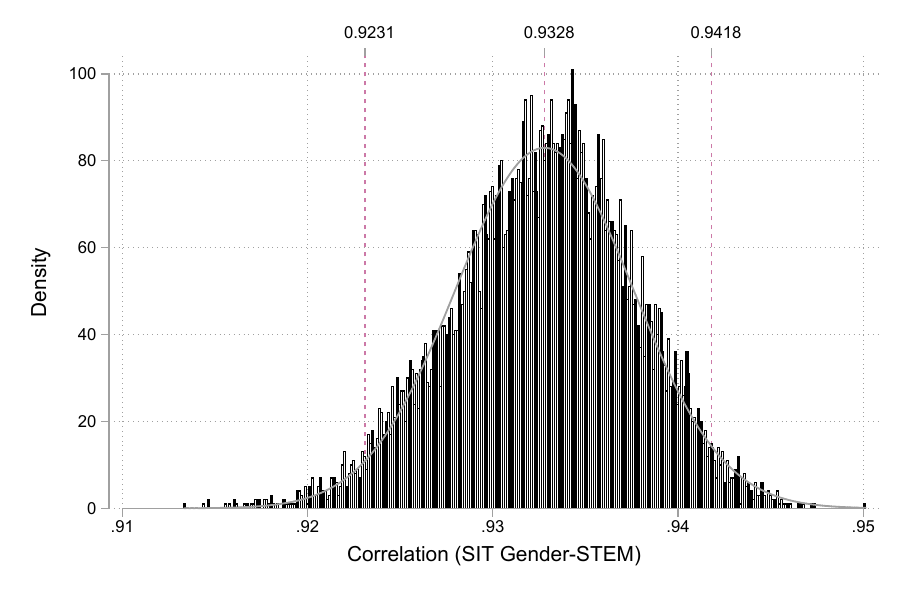}
    \label{fig:simresults_retest_genSTEM}
\end{figure}

For the test-retest reliability, we resample with replacement: some images may be duplicated in the sequences of 20 images \cite{williams_reliability_2012,pronk_methods_2022}. We maintain the same structure as for the main SIT test where 14 images depict non-Gender-STEM stereotypes and six images are about Gender-STEM stereotypes. Then, our measure of reliability is the Pearson correlation between the SIT score from the original series of images and the SIT score from the simulated series. Since we do not have to split series of ratings in two, the simulated series can be considered as an approximation of the retest part, while the original series constitute the test part. As above, we repeat this process 9999 times and report the distributions of test-retest correlations with their means and 2.5 and 97.5 empirical quantiles for both the complete series of images (\autoref{fig:simresults_retest2}) and the subset of Gender-STEM images (\autoref{fig:simresults_retest_genSTEM}).

In both cases, the correlation between the test and retest scores are well above 0.9 (even when considering the width of the confidence intervals) which is considered excellent for a psychometric test \citep{cicchetti_guidelines_1994} (it would be acceptable even for clinical measures, which follow more conservative standards \citep{portney_foundations_2015}).

\paragraph{Test-Retest reliability over time} 

Ideally, we would like to assess the test-retest reliability over time. 
High test-retest reliability for the same person over different time points provides evidence that the test measures a stable construct rather than being influenced by random fluctuations or external factors, and suggests that the test captures a reliable characteristic or skill rather than transient states or momentary variations.
To evaluate this, in 2020, we conducted a small pilot study in which the same teachers completed the SIT questionnaire twice, with a five-month interval between responses.
40 teachers were enrolled in the program, but only 16 completed both pre- and post-test.%
\footnote{The pilot SIT had a slightly different rating scale (from 0 to 5) than the one used in this paper (from 1 to 5).
The text describing the two extrema options is the same in both the online experiment and the pilot, but the pilot included an extra point in the Likert scale. To make it comparable, we rescale the 0-5 scale to a 1-5 scale using the following adjustment formula: 
\begin{equation*}
    \text{Rating}_{Adjusted} = \dfrac{4}{5} \text{Rating}_{Pilot} +1
\end{equation*}
The SIT score computed for the pilot sample is the difference between their (adjusted) rating and the average rating for each picture computed from the online experiment sample.
Additionally, at the end of the SIT pilot, teachers were asked if they wanted to rate more images: they had the opportunity to rate as many sets of 20 images as they wanted. Most of them only rated 20 images in both instances of the test. Four teachers rated more than 40 images (103, 80, 60, 41); one teacher rated less (39). We include all these teachers in the sample.} 

\begin{figure}
\caption{Test-retest reliability, Pilot}
    \centering
    \includegraphics[width=0.8\linewidth]{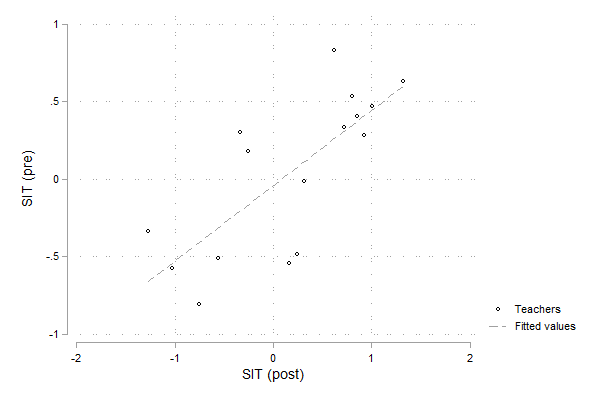}
    \label{fig:retest_pilot}
\end{figure}

Albeit with a very small sample, our SIT measure displays a high rate of test-retest reliability, with a correlation of 0.75 between the two SIT scores.
This clear and positive association is also shown in \autoref{fig:retest_pilot}, plotting teachers' pre and post SIT scores, along with a linear fit.

\newpage
\thispagestyle{empty}
\bibliographystyle{apalike}
\bibliography{UnconciousBias}

\begin{thebibliography}{}

\bibitem[Adukia et~al., 2023]{adukia_what_2023}
Adukia, A., Eble, A., Harrison, E., Runesha, H.~B., and Szasz, T. (2023).
\newblock What {We} {Teach} {About} {Race} and {Gender}: {Representation} in
  {Images} and {Text} of {Children}’s {Books}.
\newblock {\em The Quarterly Journal of Economics}, 138(4):2225--2285.

\bibitem[Alan et~al., 2023]{alan_social_2023}
Alan, S., Duysak, E., Kubilay, E., and Mumcu, I. (2023).
\newblock Social {Exclusion} and {Ethnic} {Segregation} in {Schools}: {The}
  {Role} of {Teachers}' {Ethnic} {Prejudice}.
\newblock {\em The Review of Economics and Statistics}, 105(5):1039--1054.
\newblock \_eprint:
  https://direct.mit.edu/rest/article-pdf/105/5/1039/2158480/rest\_a\_01111.pdf.

\bibitem[Alan et~al., 2018]{Alan2018a}
Alan, S., Ertac, S., and Mumcu, I. (2018).
\newblock Gender {Stereotypes} in the {Classroom} and {Effects} on
  {Achievement}.
\newblock {\em The Review of Economics and Statistics}, 100(5):876--890.

\bibitem[Alesina et~al., 2018]{alesina_revealing_2018}
Alesina, A., Carlana, M., Ferrara, E.~L., and Pinotti, P. (2018).
\newblock Revealing {Stereotypes}: {Evidence} from {Immigrants} in {Schools}.
\newblock {\em NBER Working Paper}.

\bibitem[Alexander, 1992]{alexander1992makes}
Alexander, L. (1992).
\newblock What makes wrongful discrimination wrong? biases, preferences,
  stereotypes, and proxies.
\newblock {\em University of Pennsylvania Law Review}, 141(1):149--219.

\bibitem[Alsan et~al., 2019]{alsan2019diversity}
Alsan, M., Garrick, O., and Graziani, G. (2019).
\newblock Does diversity matter for health? experimental evidence from oakland.
\newblock {\em American Economic Review}, 109(12):4071--4111.

\bibitem[{American Educational Research Association} et~al., 2014]{AERA2014}
{American Educational Research Association}, {American Psychological
  Association}, and {National Council on Measurement in Education} (2014).
\newblock {\em Standards for Educational and Psychological Testing}.
\newblock American Educational Research Association, Washington, DC.
\newblock Prepared by the Joint Committee on Standards for Educational and
  Psychological Testing of the American Educational Research Association, the
  American Psychological Association, and the National Council on Measurement
  in Education. Available at:
  \url{https://www.testingstandards.net/open-access-files.html}.

\bibitem[Arnold et~al., 2018]{David2018}
Arnold, D., Dobbie, W., and Yang, C.~S. (2018).
\newblock Racial bias in bail decisions.
\newblock {\em The Quarterly Journal of Economics}, 133(4):1885--1932.

\bibitem[Ash and Maguire, 2024]{Ash03072024}
Ash, T.~L. and Maguire, S.~C. (2024).
\newblock A scoping review of diversity training for teachers: The potential
  for school psychology.
\newblock {\em School Psychology Review}, 53(4):382--399.

\bibitem[Ayres and Siegelman, 1995]{ayres1995race}
Ayres, I. and Siegelman, P. (1995).
\newblock Race and gender discrimination in bargaining for a new car.
\newblock {\em American Economic Review}, 85(3):304--321.

\bibitem[Balsa and McGuire, 2001]{balsa2001statistical}
Balsa, A.~I. and McGuire, T.~G. (2001).
\newblock Statistical discrimination in health care.
\newblock {\em Journal of health economics}, 20(6):881--907.

\bibitem[Banks and Ford, 2009]{banks_how_2009}
Banks, R.~R. and Ford, R.~T. (2009).
\newblock ({How}) {Does} {Unconscious} {Bias} {Matter}? {Law}, {Politics}, and
  {Racial} {Inequality}.
\newblock {\em Emory Law Journal}, 58.

\bibitem[Bertrand and Mullainathan, 2004]{bertrand2004emily}
Bertrand, M. and Mullainathan, S. (2004).
\newblock Are emily and greg more employable than lakisha and jamal? a field
  experiment on labor market discrimination.
\newblock {\em American Economic Review}, 94(4):991--1013.

\bibitem[Blank et~al., 2016]{blank_unconscious_2016}
Blank, A., Houkamau, C., and Kingi, H. (2016).
\newblock Unconscious bias and education: {A} comparative study of {Māori} and
  {African} {American} students.
\newblock Technical report, Oranui Diversity Leadership.

\bibitem[Blanken et~al., 2015]{blanken_meta-analytic_2015}
Blanken, I., Ven, N. v.~d., and Zeelenberg, M. (2015).
\newblock A {Meta}-{Analytic} {Review} of {Moral} {Licensing}.
\newblock {\em Personality and Social Psychology Bulletin}, 41(4):540--558.

\bibitem[Blumberg, 2015]{blumberg2015Eliminating}
Blumberg, R.~L. (2015).
\newblock Eliminating gender bias in textbooks: pushing for policy reforms that
  promote gender equity in education.
\newblock Technical report, UNESCO, Paris.

\bibitem[Bordalo et~al., 2016]{bordalo_stereotypes_2016}
Bordalo, P., Coffman, K., Gennaioli, N., and Shleifer, A. (2016).
\newblock Stereotypes.
\newblock {\em The Quarterly Journal of Economics}, 131(4):1753--1794.

\bibitem[Borsboom et~al., 2004]{borsboom2004}
Borsboom, D., Mellenbergh, G.~J., and van Heerden, J. (2004).
\newblock The concept of validity.
\newblock {\em Psychological Review}, 111(4):1061--1071.

\bibitem[Braverman, 2022]{braverman_psychometrics_2022}
Braverman, M.~T. (2022).
\newblock Validity in testing and psychometrics.
\newblock In Braverman, M.~T., editor, {\em Evaluating program effectiveness:
  Validity and decision-making in outcome evaluation}, chapter~2, pages 17--33.
  Sage Publications, Inc.

\bibitem[Bridges, 2018]{bridges2018implicit}
Bridges, K. (2018).
\newblock Implicit bias and racial disparities in health care.
\newblock {\em Human Rights Magazine}, 43(3).

\bibitem[Brunato et~al., 2020]{brunato-etal-2020-profiling}
Brunato, D., Cimino, A., Dell{'}Orletta, F., Venturi, G., and Montemagni, S.
  (2020).
\newblock Profiling-{UD}: a tool for linguistic profiling of texts.
\newblock In Calzolari, N., B{\'e}chet, F., Blache, P., Choukri, K., Cieri, C.,
  Declerck, T., Goggi, S., Isahara, H., Maegaard, B., Mariani, J., Mazo, H.,
  Moreno, A., Odijk, J., and Piperidis, S., editors, {\em Proceedings of the
  Twelfth Language Resources and Evaluation Conference}, pages 7145--7151,
  Marseille, France. European Language Resources Association.

\bibitem[Campbell, 2015]{campbell_stereotyped_2015}
Campbell, T. (2015).
\newblock Stereotyped at {Seven}? {Biases} in {Teacher} {Judgement} of
  {Pupils}’ {Ability} and {Attainment}.
\newblock {\em Journal of Social Policy}, 44(3):517--547.

\bibitem[Caprini, 2023]{caprini_visual_2023}
Caprini, G. (2023).
\newblock Visual bias.
\newblock {\em Working Paper}.
\newblock R\&R econometrica.

\bibitem[Carlana, 2019]{Carlana2019}
Carlana, M. (2019).
\newblock Implicit {Stereotypes}: {Evidence} from {Teachers}' {Gender} {Bias}.
\newblock {\em The Quarterly Journal of Economics}, 134(3):1163--1224.

\bibitem[Carlana et~al., 2022]{carlana_goals_2022}
Carlana, M., La~Ferrara, E., and Pinotti, P. (2022).
\newblock Goals and {Gaps}: {Educational} {Careers} of {Immigrant} {Children}.
\newblock {\em Econometrica}, 90(1):1--29.

\bibitem[Cascio and Plant, 2015]{cascio_prospective_2015}
Cascio, J. and Plant, E.~A. (2015).
\newblock Prospective moral licensing: {Does} anticipating doing good later
  allow you to be bad now?
\newblock {\em Journal of Experimental Social Psychology}, 56:110--116.

\bibitem[Cicchetti, 1994]{cicchetti_guidelines_1994}
Cicchetti, D.~V. (1994).
\newblock Guidelines, criteria, and rules of thumb for evaluating normed and
  standardized assessment instruments in psychology.
\newblock {\em Psychological Assessment}, 6(4):284--290.

\bibitem[Claro et~al., 2016]{claro2016growth}
Claro, S., Paunesku, D., and Dweck, C.~S. (2016).
\newblock Growth mindset tempers the effects of poverty on academic
  achievement.
\newblock {\em Proceedings of the National Academy of Sciences},
  113(31):8664--8668.

\bibitem[Cobb-Clark and Schurer, 2013]{cobb2013two}
Cobb-Clark, D.~A. and Schurer, S. (2013).
\newblock Two economists’ musings on the stability of locus of control.
\newblock {\em The Economic Journal}, 123(570):F358--F400.

\bibitem[Coffman, 2014]{coffman2014evidence}
Coffman, K.~B. (2014).
\newblock Evidence on self-stereotyping and the contribution of ideas.
\newblock {\em The Quarterly Journal of Economics}, 129(4):1625--1660.

\bibitem[Conway and Peetz, 2012]{conway_when_2012}
Conway, P. and Peetz, J. (2012).
\newblock When {Does} {Feeling} {Moral} {Actually} {Make} {You} a {Better}
  {Person}? {Conceptual} {Abstraction} {Moderates} {Whether} {Past} {Moral}
  {Deeds} {Motivate} {Consistency} or {Compensatory} {Behavior}.
\newblock {\em Personality and Social Psychology Bulletin}, 38(7):907--919.

\bibitem[Cronbach and Meehl, 1955]{cronbach1955}
Cronbach, L.~J. and Meehl, P.~E. (1955).
\newblock Construct validity in psychological tests.
\newblock {\em Psychological Bulletin}, 52(4):281--302.

\bibitem[Davidov et~al., 2008]{Davidov2008}
Davidov, E., Schmidt, P., and Schwartz, S.~H. (2008).
\newblock Bringing values back in: The adequacy of the european social survey
  to measure values in 20 countries.
\newblock {\em Public Opinion Quarterly}, 72(3):420--445.

\bibitem[Dee, 2005]{dee_teacher_2005}
Dee, T.~S. (2005).
\newblock A {Teacher} like {Me}: {Does} {Race}, {Ethnicity}, or {Gender}
  {Matter}?
\newblock {\em The American Economic Review}, 95(2):158--165.

\bibitem[Della~Giusta and Bosworth, 2020]{della_giusta_bias_2020}
Della~Giusta, M. and Bosworth, S. (2020).
\newblock Bias and {Discrimination}: {What} {Do} {We} {Know}?
\newblock {\em Oxford Review of Economic Policy}, 36(4):925--934.

\bibitem[Della~Giusta and Muratori, 2022]{DellaGiusta2022}
Della~Giusta, M. and Muratori, C. (2022).
\newblock Un’indagine sulle tematiche di genere nella scuola.
\newblock In Della~Giusta, M., Poggio, B., and Spicci, M., editors, {\em
  Educare alla parità: Principi, metodologie didattiche e strategie di azione
  per l'equità e l'inclusione}. Pearson Academy.

\bibitem[Dobbie et~al., 2021]{dobbie2021}
Dobbie, W., Liberman, A., Paravisini, D., and Pathania, V. (2021).
\newblock Measuring bias in consumer lending.
\newblock {\em The Review of Economic Studies}, 88(6):2799--2832.

\bibitem[Doleac and Stein, 2013]{doleac2013visible}
Doleac, J.~L. and Stein, L.~C. (2013).
\newblock The visible hand: Race and online market outcomes.
\newblock {\em Economic Journal}, 123(572):F469--F492.

\bibitem[Dymski, 2006]{dymski2006discrimination}
Dymski, G.~A. (2006).
\newblock Discrimination in the credit and housing markets: findings and
  challenges.
\newblock In {\em Handbook on the Economics of Discrimination}. Edward Elgar
  Publishing.

\bibitem[Edelman et~al., 2017]{edelman2017racial}
Edelman, B., Luca, M., and Svirsky, D. (2017).
\newblock Racial discrimination in the sharing economy: Evidence from a field
  experiment.
\newblock {\em American Economic Journal: Applied Economics}, 9(2):1--22.

\bibitem[Embretson, 1984]{embretson1984}
Embretson, S. (1984).
\newblock A general latent trait model for response processes.
\newblock {\em Psychometrika}, 49(2):175--186.

\bibitem[Embretson, 1993]{embretson1993}
Embretson, S. (1993).
\newblock Psychometric models for learning and cognitive processes.
\newblock In Frederiksen, N., Mislevy, R.~J., and Bejar, I.~I., editors, {\em
  Test Theory for A New Generation of Tests}, chapter~6. Routledge.

\bibitem[Embretson, 1998]{embretson1998}
Embretson, S.~E. (1998).
\newblock A cognitive design system approach to generating valid tests:
  Application to abstract reasoning.
\newblock {\em Psychological Methods}, 3(3):380--396.

\bibitem[Embretson, 2016]{embretson2016}
Embretson, S.~E. (2016).
\newblock Understanding examinees’ responses to items: Implications for
  measurement.
\newblock {\em Educational Measurement: Issues and Practice}, 35(3):6--22.

\bibitem[Ewens et~al., 2014]{ewens2014statistical}
Ewens, M., Tomlin, B., and Wang, L.~C. (2014).
\newblock Statistical discrimination or prejudice? a large sample field
  experiment.
\newblock {\em Review of Economics and Statistics}, 96(1):119--134.

\bibitem[Ewing et~al., 2018]{ewing2018teachers}
Ewing, D.~L., Monsen, J.~J., and Kielblock, S. (2018).
\newblock Teachers’ attitudes towards inclusive education: a critical review
  of published questionnaires.
\newblock {\em Educational Psychology in Practice}, 34(2):150--165.

\bibitem[Fazio et~al., 1995]{fazio_variability_1995}
Fazio, R.~H., Jackson, J.~R., Dunton, B.~C., and Williams, C.~J. (1995).
\newblock Variability in automatic activation as an unobtrusive measure of
  racial attitudes: {A} bona fide pipeline?
\newblock {\em Journal of Personality and Social Psychology}, 69(6):1013--1027.

\bibitem[Fazio et~al., 1986]{fazio_automatic_1986}
Fazio, R.~H., Sanbonmatsu, D.~M., Powell, M.~C., and Kardes, F.~R. (1986).
\newblock On the automatic activation of attitudes.
\newblock {\em Journal of Personality and Social Psychology}, 50(2):229--238.

\bibitem[Figlio, 2005]{figlio_names_2005}
Figlio, D. (2005).
\newblock Names, {Expectations} and the {Black}-{White} {Test} {Score} {Gap}.
\newblock {NBER} {Working} {Papers} w11195, National Bureau of Economic
  Research, Cambridge, MA.

\bibitem[Fiske, 1993]{fiske1993social}
Fiske, A.~P. (1993).
\newblock Social errors in four cultures: Evidence about universal forms of
  social relations.
\newblock {\em Journal of Cross-Cultural Psychology}, 24(4):463--494.

\bibitem[Friese and Hofmann, 2009]{friese_control_2009}
Friese, M. and Hofmann, W. (2009).
\newblock Control me or {I} will control you: {Impulses}, trait self-control,
  and the guidance of behavior.
\newblock {\em Journal of Research in Personality}, 43(5):795--805.

\bibitem[Fryer~Jr et~al., 2019]{fryer2019updating}
Fryer~Jr, R.~G., Harms, P., and Jackson, M.~O. (2019).
\newblock Updating beliefs when evidence is open to interpretation:
  Implications for bias and polarization.
\newblock {\em Journal of the European Economic Association}, 17(5):1470--1501.

\bibitem[Gennaioli and Tabellini, 2023]{gennaioli2023identity}
Gennaioli, N. and Tabellini, G. (2023).
\newblock Identity politics.
\newblock {\em Available at SSRN 4395173}.

\bibitem[Gilliam et~al., 2016]{gilliam_early_2016}
Gilliam, W.~S., Maupin, A.~N., Reyes, C.~R., Accavitti, M., and Shic, F.
  (2016).
\newblock Do {Early} {Educators}’ {Implicit} {Biases} {Regarding} {Sex} and
  {Race} {Relate} to {Behavior} {Expectations} and {Recommendations} of
  {Preschool} {Expulsions} and {Suspensions}?
\newblock Research {Study} {Brief}, Yale Child Study Center.

\bibitem[Glover et~al., 2017]{glover_discrimination_2017}
Glover, D., Pallais, A., and Pariente, W. (2017).
\newblock Discrimination as a {Self}-{Fulfilling} {Prophecy}: {Evidence} from
  {French} {Grocery} {Stores}.
\newblock {\em The Quarterly Journal of Economics}, 132(3):1219--1260.

\bibitem[Gorard, 2016]{gorard_cautionary_2016}
Gorard, S. (2016).
\newblock A {Cautionary} {Note} on {Measuring} the {Pupil} {Premium}
  {Attainment} {Gap} in {England}.
\newblock {\em Journal of Education, Society and Behavioural Science},
  14(2):1--8.

\bibitem[Green, 1993]{green1993alpha}
Green, R.~M. (1993).
\newblock What is coefficient alpha? an examination of theory and applications.
\newblock {\em Journal of Applied Psychology}, 78(1):98--104.

\bibitem[Hanna and Linden, 2012]{hanna_discrimination_2012}
Hanna, R.~N. and Linden, L.~L. (2012).
\newblock Discrimination in {Grading}.
\newblock {\em American Economic Journal: Economic Policy}, 4(4):146--168.

\bibitem[Hawkins et~al., 2023]{hawkins_understanding_2023}
Hawkins, C.~B., Lofaro, N., Umansky, E., and Ratliff, K.~A. (2023).
\newblock Understanding implicit bias ({UIB}): Experimental evaluation of an
  online bias education program.
\newblock {\em Journal of Experimental Psychology: Applied}, 29(4):887--902.

\bibitem[Hofmann and Baumert, 2010]{hofmann_immediate_2010}
Hofmann, W. and Baumert, A. (2010).
\newblock Immediate affect as a basis for intuitive moral judgement: {An}
  adaptation of the affect misattribution procedure.
\newblock {\em Cognition and Emotion}, 24(3):522--535.

\bibitem[Hooghe and Quintelier, 2023]{hooghe2023discrimination}
Hooghe, M. and Quintelier, E. (2023).
\newblock Discrimination and political engagement: A cross-national test.
\newblock {\em Journal of Race, Ethnicity, and Politics}, 8(3):1--23.

\bibitem[Howell et~al., 2015]{howell_caught_2015}
Howell, J.~L., Gaither, S.~E., and Ratliff, K.~A. (2015).
\newblock Caught in the {Middle}: {Defensive} {Responses} to {IAT} {Feedback}
  {Among} {Whites}, {Blacks}, and {Biracial} {Black}/{Whites}.
\newblock {\em Social Psychological and Personality Science}, 6(4):373--381.

\bibitem[Hughes, 2018]{hughes_psychometrics_2018}
Hughes, D.~J. (2018).
\newblock Psychometric validity.
\newblock In Irwing, P., Booth, T., and Hughes, D.~J., editors, {\em The Wiley
  Handbook of Psychometric Testing}, volume~2, chapter~24, pages 751--779. John
  Wiley \& Sons, Ltd.

\bibitem[Jacob and Wilder, 2010]{jacob_educational_2010}
Jacob, B.~A. and Wilder, T. (2010).
\newblock Educational {Expectations} and {Attainment}.
\newblock {NBER} {Working} {Papers} 15683, National Bureau of Economic
  Research.

\bibitem[Jasper and Witthöft, 2013]{jasper_automatic_2013}
Jasper, F. and Witthöft, M. (2013).
\newblock Automatic evaluative processes in health anxiety and their relations
  to emotion regulation.
\newblock {\em Cognitive Therapy and Research}, 37(3):521--533.

\bibitem[Johnson and Penny, 2005]{kempf-leonard_split-half_2005}
Johnson, R.~L. and Penny, J. (2005).
\newblock Split-{Half} {Reliability}.
\newblock In Kempf-Leonard, K., editor, {\em Encyclopedia of {Social}
  {Measurement}}, pages 649--654. Elsevier, New York.

\bibitem[Jussim and Harber, 2005]{jussim_teacher_2005}
Jussim, L. and Harber, K.~D. (2005).
\newblock Teacher {Expectations} and {Self}-{Fulfilling} {Prophecies}: {Knowns}
  and {Unknowns}, {Resolved} and {Unresolved} {Controversies}.
\newblock {\em Personality and Social Psychology Review}, 9(2):131--155.

\bibitem[Kahneman, 1994]{kahneman_new_1994}
Kahneman, D. (1994).
\newblock New {Challenges} to the {Rationality} {Assumption}.
\newblock {\em Journal of Institutional and Theoretical Economics (JITE) /
  Zeitschrift für die gesamte Staatswissenschaft}, 150(1):18--36.

\bibitem[Kahneman, 2011]{kahneman_thinking_2011}
Kahneman, D. (2011).
\newblock {\em Thinking, fast and slow}.
\newblock Thinking, fast and slow. Farrar, Straus and Giroux, New York, NY, US.
\newblock Pages: 499.

\bibitem[Kain and Quigley, 1972]{kain1972housing}
Kain, J.~F. and Quigley, J.~M. (1972).
\newblock Housing market discrimination, home-ownership, and savings behavior.
\newblock {\em The American Economic Review}, 62(3):263--277.

\bibitem[Knowles et~al., 2001]{knowles2001racial}
Knowles, J., Persico, N., and Todd, P. (2001).
\newblock Racial bias in motor vehicle searches: Theory and evidence.
\newblock {\em Journal of Political Economy}, 109(1):203--229.

\bibitem[La~Ferrara, 2019]{la_ferrara_presidential_2019}
La~Ferrara, E. (2019).
\newblock Presidential {Address}: {Aspirations}, {Social} {Norms}, and
  {Development}.
\newblock {\em Journal of the European Economic Association}, 17(6):1687--1722.

\bibitem[Lavy and Megalokonomou, 2024]{Lavy2024}
Lavy, V. and Megalokonomou, R. (2024).
\newblock The short- and the long-run impact of gender-biased teachers.
\newblock {\em American Economic Journal: Applied Economics}, 16(2):176–218.

\bibitem[Lieberman, 2013]{lieberman_social_2013}
Lieberman, M.~D. (2013).
\newblock {\em Social: {Why} our brains are wired to connect}.
\newblock Social: {Why} our brains are wired to connect. Crown
  Publishers/Random House, New York, NY, US.

\bibitem[List, 2004]{list2004nature}
List, J.~A. (2004).
\newblock The nature and extent of discrimination in the marketplace: Evidence
  from the field.
\newblock {\em Quarterly Journal of Economics}, 119(1):49--89.

\bibitem[Lundberg and Payne, 2014]{lundberg_decisions_2014}
Lundberg, K.~B. and Payne, B.~K. (2014).
\newblock Decisions among the {Undecided}: {Implicit} {Attitudes} {Predict}
  {Future} {Voting} {Behavior} of {Undecided} {Voters}.
\newblock {\em PLOS ONE}, 9(1):1--11.

\bibitem[Macchi, 2023]{macchi2023worth}
Macchi, E. (2023).
\newblock Worth your weight: experimental evidence on the benefits of obesity
  in low-income countries.
\newblock {\em American Economic Review}, 113(9):2287--2322.

\bibitem[Mazar and Zhong, 2010]{mazar_green_2010}
Mazar, N. and Zhong, C.-B. (2010).
\newblock Do {Green} {Products} {Make} {Us} {Better} {People}?
\newblock {\em Psychological Science}, 21(4):494--498.

\bibitem[Merritt et~al., 2012]{merritt_strategic_2012}
Merritt, A.~C., Effron, D.~A., Fein, S., Savitsky, K.~K., Tuller, D.~M., and
  Monin, B. (2012).
\newblock The strategic pursuit of moral credentials.
\newblock {\em Journal of Experimental Social Psychology}, 48(3):774--777.

\bibitem[Messick, 1989]{messick_validity_1989}
Messick, S. (1989).
\newblock Validity.
\newblock In Linn, R.~L., editor, {\em Educational measurement}, chapter~2,
  pages 13--104. American Council on education and Macmillan.

\bibitem[Morris and Perry, 2017]{morris_girls_2017}
Morris, E.~W. and Perry, B.~L. (2017).
\newblock Girls {Behaving} {Badly}? {Race}, {Gender}, and {Subjective}
  {Evaluation} in the {Discipline} of {African} {American} {Girls}.
\newblock {\em Sociology of Education}, 90(2):127--148.

\bibitem[Mullen and Monin, 2016]{mullen_consistency_2016}
Mullen, E. and Monin, B. (2016).
\newblock Consistency {Versus} {Licensing} {Effects} of {Past} {Moral}
  {Behavior}.
\newblock {\em Annual Review of Psychology}, 67(1):363--385.

\bibitem[Neumark, 2018]{neumark2018experimental}
Neumark, D. (2018).
\newblock Experimental research on labor market discrimination.
\newblock {\em Journal of Economic Literature}, 56(3):799--866.

\bibitem[Obermeyer et~al., 2019]{obermeyer2019dissecting}
Obermeyer, Z., Powers, B., Vogeli, C., and Mullainathan, S. (2019).
\newblock Dissecting racial bias in an algorithm used to manage the health of
  populations.
\newblock {\em Science}, 366(6464):447--453.

\bibitem[Oxoby, 2014]{oxoby2014social}
Oxoby, R.~J. (2014).
\newblock Social inference and occupational choice: Type-based beliefs in a
  bayesian model of class formation.
\newblock {\em Journal of Behavioral and Experimental Economics}, 51:30--37.

\bibitem[Oxtoby, 2020]{oxtoby_how_2020}
Oxtoby, K. (2020).
\newblock How unconscious bias can discriminate against patients and affect
  their care.
\newblock {\em BMJ}, 371:m4152.

\bibitem[Payne et~al., 2008]{payne_automatic_2008}
Payne, B.~K., Govorun, O., and Arbuckle, N.~L. (2008).
\newblock Automatic attitudes and alcohol: {Does} implicit liking predict
  drinking?
\newblock {\em Cognition and Emotion}, 22(2):238--271.

\bibitem[Payne et~al., 2010]{payne_implicit_2010}
Payne, B.~K., Krosnick, J.~A., Pasek, J., Lelkes, Y., Akhtar, O., and Tompson,
  T. (2010).
\newblock Implicit and explicit prejudice in the 2008 {American} presidential
  election.
\newblock {\em Journal of Experimental Social Psychology}, 46(2):367--374.

\bibitem[Payne and Lundberg, 2014]{payne_affect_2014}
Payne, K. and Lundberg, K. (2014).
\newblock The affect misattribution procedure: {Ten} years of evidence on
  reliability, validity, and mechanisms.
\newblock {\em Social and Personality Psychology Compass}, 8(12):672--686.

\bibitem[Portney and Watkins, 2015]{portney_foundations_2015}
Portney, L.~G. and Watkins, M.~P. (2015).
\newblock {\em Foundations of clinical research: applications to practice},
  volume 892.
\newblock Pearson/prentice hall Upper Saddle River, NJ.

\bibitem[Pronk et~al., 2022]{pronk_methods_2022}
Pronk, T., Molenaar, D., Wiers, R.~W., and Murre, J. (2022).
\newblock Methods to split cognitive task data for estimating split-half
  reliability: {A} comprehensive review and systematic assessment.
\newblock {\em Psychonomic Bulletin \& Review}, 29(1):44--54.

\bibitem[Reuben et~al., 2014]{reuben_how_2014}
Reuben, E., Sapienza, P., and Zingales, L. (2014).
\newblock How stereotypes impair women’s careers in science.
\newblock {\em Proceedings of the National Academy of Sciences},
  111(12):4403--4408.

\bibitem[Rust et~al., 2020]{rust_modern_2020}
Rust, J., Kosinski, M., and Stillwell, D. (2020).
\newblock {\em Modern {Psychometrics}: {The} {Science} of {Psychological}
  {Assessment}}.
\newblock Routledge, 4th edition.

\bibitem[Schoenmueller et~al., 2020]{rating2020}
Schoenmueller, V., Netzer, O., and Stahl, F. (2020).
\newblock The polarity of online reviews: Prevalence, drivers and implications.
\newblock {\em Journal of Marketing Research}, 57(5):853--877.

\bibitem[Simbrunner and Schlegelmilch, 2017]{simbrunner_moral_2017}
Simbrunner, P. and Schlegelmilch, B.~B. (2017).
\newblock Moral licensing: a culture-moderated meta-analysis.
\newblock {\em Management Review Quarterly}, 67(4):201--225.

\bibitem[Sprietsma, 2013]{sprietsma_discrimination_2013}
Sprietsma, M. (2013).
\newblock Discrimination in grading: experimental evidence from primary school
  teachers.
\newblock {\em Empirical Economics}, 45(1):523--538.

\bibitem[Taber, 2018]{taber2018cronbach}
Taber, K.~S. (2018).
\newblock The use of cronbach’s alpha when developing and reporting research
  instruments in science education.
\newblock {\em Research in Science Education}, 48(6):1273--1296.

\bibitem[Tversky and Kahneman, 1973]{tversky_availability_1973}
Tversky, A. and Kahneman, D. (1973).
\newblock Availability: {A} heuristic for judging frequency and probability.
\newblock {\em Cognitive Psychology}, 5(2):207--232.

\bibitem[Tversky and Kahneman, 1983]{tversky_extensional_1983}
Tversky, A. and Kahneman, D. (1983).
\newblock Extensional versus intuitive reasoning: {The} conjunction fallacy in
  probability judgment.
\newblock {\em Psychological Review}, 90(4):293--315.

\bibitem[Urbina, 2004]{urbina2004}
Urbina, S. (2004).
\newblock {\em Essentials of psychological testing}.
\newblock John Wiley \& Sons Inc.

\bibitem[Voth and Yanagizawa-Drott, 2024]{Voth2024Images}
Voth, H.-J. and Yanagizawa-Drott, D. (2024).
\newblock Image(s).
\newblock CEPR Discussion Paper DP19219, Centre for Economic Policy Research.

\bibitem[Warrens, 2015]{warrens_cronbachs_2015}
Warrens, M.~J. (2015).
\newblock On {Cronbach}'s {Alpha} as the {Mean} of {All} {Split}-{Half}
  {Reliabilities}.
\newblock In Millsap, R.~E., Bolt, D.~M., van~der Ark, L.~A., and Wang, W.-C.,
  editors, {\em Quantitative {Psychology} {Research}}, pages 293--300, Cham.
  Springer International Publishing.

\bibitem[Williams and Kaufmann, 2012]{williams_reliability_2012}
Williams, B.~J. and Kaufmann, L.~M. (2012).
\newblock Reliability of the {Go}/{No} {Go} {Association} {Task}.
\newblock {\em Journal of Experimental Social Psychology}, 48(4):879--891.

\bibitem[Yinger, 1998]{yinger1998evidence}
Yinger, J. (1998).
\newblock Evidence on discrimination in consumer markets.
\newblock {\em Journal of Economic Perspectives}, 12(2):23--40.

\end{thebibliography}

\end{document}